\documentclass[11pt,preprint]{aastex}

\newcommand{\pks}{PKS\,2155$-$304}
\newcommand{\mkn}{Mrk\,421}

\newcommand{\lc}{light curve}
\newcommand{\lcs}{light curves}
\newcommand{\ts}{timescale}
\newcommand{\tss}{timescales}
\newcommand{\tcool}{$t_{\rm cool}$}
\newcommand{\tacc}{$t_{\rm acc}$}
\newcommand{\sy}{synchrotron}

\newcommand{\sax}{{\it Beppo}SAX}

\newcommand{\xr}{X-ray}
\newcommand{\xrs}{X-rays}

\newcommand{\etal}{et al.\,}

\newcommand{\beq}{\begin{equation}}
\newcommand{\eeq}{\end{equation}}

\slugcomment{ApJ, in press (Scheduled for the v572 no2, June 20, 2002
issue)}

\shortauthors{Zhang et al. }
\shorttitle{X-ray Variability of PKS~2155--304}

\begin{document}

\title{Four Years Monitoring of Blazar \pks\ with \sax : \\
	Probing the Dynamics of the Jet}

\authoraddr {Dipartimento di Scienze, Universit\`a dell'Insubria,
	   via Valleggio 11, I-22100 Como, Italy, 
		e-mail: youhong.zhang@uninsubria.it}

\author {Y.H. Zhang \altaffilmark{1},
A. Treves \altaffilmark{1},
A. Celotti \altaffilmark{2},
L. Chiappetti \altaffilmark{3},
G. Fossati \altaffilmark{4},
G. Ghisellini \altaffilmark{5}, \\
L. Maraschi \altaffilmark{6},
E. Pian \altaffilmark{7},
G. Tagliaferri \altaffilmark{5},
and F. Tavecchio \altaffilmark{6}
}
\email{youhong.zhang@uninsubria.it}

\altaffiltext{1}{Dipartimento di Scienze, Universit\`a dell'Insubria,
   via Valleggio 11, I-22100 Como, Italy}
\altaffiltext{2}{International School for Advanced Studies, SISSA/ISAS,
   via Beirut 2-4, I-34014 Trieste, Italy}
\altaffiltext{3}{Istituto di Fisica Cosmica G.Occhialini, IFCTR/CNR,
   via Bassini 15, I-20133 Milano, Italy}
\altaffiltext{4}{CASS, UCSD, 9500 Gilman Drive, La Jolla, CA 92093-0424,
   USA}
\altaffiltext{5}{Osservatorio Astronomico di Brera, via Bianchi
   46, I-22055 Merate, Italy}
\altaffiltext{6}{Osservatorio Astronomico di Brera, via Brera 28,
   I-20121 Milano, Italy}
\altaffiltext{7}{Osservatorio Astronomico di Trieste, via G.B. Tiepolo 
   11, I-34131 Trieste, Italy}

\begin{abstract}

PKS~2155--304 is one of the brightest blazars in the X-ray band. It
was repeatedly monitored with \sax\ during three long campaigns of
about 2 days each in November of 1996, 1997 and 1999. The source 
underwent different states of intensity and was clearly variable with
successive flares detected.
This paper presents temporal and spectral analysis to study the X-ray
variability trends for a blazar. The variability shows larger amplitude
and shorter \ts\ at higher energies. 
The power spectral densities have steep power-law slopes of $\sim$~2--3,
indicating shot noise variability.
Structure function analysis reveals the existence of ``typical'' \tss\
characteristic of the half duration of the flares. 
From the cross-correlation analysis we find that the values of soft
lags, i.e., delays of soft (0.1--1.5~keV) photons with respect to  
hard (3.5--10~keV) ones, differ from flare to flare, ranging from a
few hundred seconds to about one hour. There is a suggestion that
the flares with shorter duration show smaller soft lags. The soft lags
are also energy-dependent, with longer lags of lower energy emission 
with respect to the emission in the 4--10~keV.
The time-resolved X-ray spectral fits with a curved model show that
peak energies of the synchrotron component are located in the very
soft X-ray range or even below the \sax\ lower energy limit, 0.1~keV.
A correlation between peak energies and fluxes is marginal. Spectral
evolution during some flares shows clockwise loops in the spectral
index--flux plane, confirming the soft lags indicated by 
the cross-correlation analysis. Two flares, however, show evidence that
spectral evolution follows opposite tracks in the soft and hard
energy bands, respectively.
The rich phenomenology is interpreted in the context of a
model where relativistic electrons are accelerated through internal
shocks taking place in the jets. The most important parameter turns out
to be the initial time interval between the two shells ejected from the
central engine to produce the flare, which may determine the structure
of the shock and in turn the physical quantities of the emitting region
to reproduce the observed trends of the X-ray variability. 
\end{abstract}



\section{Introduction}\label{sec:intro}

The most remarkable property discriminating blazars from other Active
Galactic Nuclei (AGNs) is that they are strongly and rapidly variable
from radio to gamma-rays on different timescales. From the point of view
of the observations, there has been clear progress in the last
decade. EGRET onboard CGRO has detected about 60 blazars which are
(GeV) gamma-ray emitting sources (Hartman \etal 1999), and gamma-rays
from a few nearby sources have been detected up to the TeV energies
with ground-based Cherenkov telescopes.
Gamma-ray observations have revealed the remarkable feature that the
overall Spectral Energy Distribution (SED) of a blazar shows two
distinct components in the $\nu$--$\nu F_{\nu}$ representation,
typically characterized by their peak energies. The first (low
energy) component peaks from mm to the X-rays, while the second (high
energy) one peaks at GeV--TeV energies. In such a picture, the blazar
family could be unified according to the SEDs which are parameterized 
by the bolometric luminosity (e.g., Fossati \etal 1998). The
emission is believed to be produced by relativistic electrons tangled
with magnetic field in a relativistic jet through two 
processes. Synchrotron radiation is responsible for the low energy
component, while inverse-Compton upscattering by the same population of
electrons produces the high energy component (e.g., Ghisellini \etal
1998). The jet is supposed to be oriented close to the line of 
sight. Superluminal motions observed with VLBI suggest that the bulk
Lorentz factor of the jets is of the order of $\Gamma \sim 10 $ (e.g.,
Vermeulen \& Cohen 1994). Relativistic beaming is thus a marked feature
of blazars.

Constructing the overall SED and determining the relationships in
different energy bands can in principle constrain the physical parameters
and emitting mechanisms taking place in blazars. Moreover, the temporal 
evolution of both the SED and the inter-band relationships enables to
explore the dynamics and the structure of the jets, which ultimately give
clues on the physical properties of the central massive black hole
system. However, due to limitation of the observations, the earlier
studies of blazars were mainly based on the ``snapshot'' SEDs (most of
them are not simultaneous). Short time-coverage and undersampling of data
have prevented detailed studies of evolution of SED and of inter-band 
correlations. Relatively higher quality data from recent
monitoring of a few nearby TeV sources, especially in the X-rays and TeV
energies, has allowed to establish two important properties in these 
sources: correlated X-ray/TeV variability (e.g., Maraschi \etal 
1999) and energy-dependent time lags in the X-rays (e.g., Zhang
\etal 1999). These results have provided important clues on
the acceleration and cooling mechanism of relativistic electrons. 

Up to now, the X-ray energy band is still the best to perform the
detailed studies of variability in TeV-emitting sources, since the most
pronounced variations are expected in such band, which represents the
highest energy tail of the \sy\ component. Studies of the variability
expected in the highest energy end of Compton component, i.e., TeV
energies, are still limited at present because of undersampling.
Mrk~421 and Mrk~501 have been intensively monitored with \sax . 
The remarkable findings are the detection of significant up-shifts of
synchrotron peak energies in Mrk~501 (Pian \etal 1998; Tavecchio \etal
2001) and high energy photons lagging the low energy ones in Mrk~421
(Fossati \etal 2000a; Zhang 2000). 

\pks\ is a nearby TeV-emitting blazar ($z=0.116$; Falomo, Pesce, \&  
Treves 1993) with the synchrotron emission peaking in the UV-soft X-ray   
range. Because of the brightness, \pks\ is one of the best targets to
monitor across the whole electromagnetic spectrum. It has been repeatedly 
observed at optical (e.g., Zhang \& Xie 1996 and references 
therein; Xie \etal 2001;), UV (e.g., Pian \etal 1997; Marshall 2001),
X-rays (e.g., Treves \etal 1989; Sembay \etal 1993), gamma-rays
(Vestrand, Stacy, \& Sreekumar 1995), and TeV energies (Chadwick \etal
1999). These observations have demonstrated complex multiwavelength
variability in \pks\ (Edelson \etal 1995; Urry \etal 1997). Correlated
variability in different X-ray bands
and physical implications have been reported in Chiappetti \etal (1999),
Zhang \etal (1999), Zhang (2000), Kataoka \etal (2000), Edelson \etal
(2001). Mrk~421 and Mrk~501 have exhibited similar phenomenology as well 
(Takahashi \etal 1996; 2000; Fossati \etal 2000a; Zhang 2000; Tanihata
\etal 2001).

\pks\ was also monitored with \sax\ during three long campaigns (about
2 days each), allowing to explore its variability properties in different
brightness states. Therefore, these observations will provide us with
direct information on the evolution of the underlying physical processes,
which in turn can give some valuable clues on the dynamics and the
structure of the jets.
Temporal and spectral analysis of the 1996 and 1997 campaigns were   
presented in Giommi \etal (1998), Chiappetti \etal (1999) and Zhang
\etal (1999). In this paper we perform temporal and spectral analysis for
the third campaign, i.e., the 1999 data set. Since, for a direct
comparison, a homogeneous set of results is necessary, we re-analyze the
1996 and 1997 data sets, focusing on issues not
handled in the papers quoted above.

This paper is organized as follows. The observations are introduced in
\S\ref{sec:obs}. \S\ref{sec:timing} presents the light curves followed
by temporal analysis with various methods, including estimators of
variability (amplitude and doubling timescale), variations of hardness
ratios, power spectral density, structure function, and 
cross-correlation function. The time-resolved spectral analysis is
conducted in \S\ref{sec:spec} where we emphasize the evolution of
synchrotron peak energies and spectral indices. In \S\ref{sec:disc} we
summarize and compare the most important results presented in this and
other works, followed by a discussion of their implications to the
dynamics and the structure of the jet. Our conclusions 
are derived in \S\ref{sec:conc}.


\section{Observations}\label{sec:obs}

For a detailed description of the Italian/Dutch \textit{Beppo}SAX
mission we refer to Boella \etal (1997 and references therein).
The co-aligned Narrow Field Instruments (NFIs) onboard \sax\ consist of 
one Low Energy Concentrator Spectrometer (LECS; 0.1--10~keV), three 
identical Medium Energy Concentrator Spectrometers (MECS; 1.5--10 keV), 
a High Pressure Gas Scintillation Proportional Counter 
(HPGSPC; 4--120~keV), and a Phoswich Detector System (PDS; 13--300~keV). 
Due to the limited statistics of other instruments, in this work only 
data from LECS and MECS are considered to suite to our purpose of 
analysis. 
The MECS was composed at launch by three identical units. On
1997 May 6 a technical failure caused the switch off of the unit MECS
1. All observations after this date were performed with only two units
(i.e., MECS2 and MECS3).

The journal of the observations is summarized in Table~\ref{tab:obs}.
The observing efficiency, defined as the ratio of net exposure to the
observing time, is $\sim$~0.2 and $\sim$~0.5 for the LECS and the MECS
detector, respectively. These low efficiencies are mainly due to
periodic interruptions caused by the Earth occultation of low Earth
orbit satellite (with period of $\sim$~1.6 hours), and other reasons
such as passages through South Atlantic Anomaly (SAA), high background
region, etc. Moreover, the effective LECS exposure time is significantly
shortened, because LECS is exposed only during the part of the orbit in
the earth shadow cone, to avoid light leakage in the LECS instrument.

Our analysis is based on the linearized, cleaned event files for the LECS
and the MECS experiments. The events from the two
(1997 and 1999) or three (1996) units of MECS are merged together to
improve the photon statistics. These event files, together with
appropriate background event files, are produced at the \textit{Beppo}SAX
Science Data Center (SDC; {\scshape rev~0.2, 1.1 and 2.0}), and are 
available from the \sax\ SDC on-line archive.
The event files are further screened with a good time interval
(GTI) file (from SDC) to exclude events without attitude solution
(i.e., for these events it is impossible to convert detector to sky
coordinates, see Fiore, Guainazzi \& Grandi 1999).    

The photons of the source were accumulated from events within a circular
region centered on the position of the point source, and the extraction  
radii are 8 and 6 arcmin for LECS and MECS, respectively, which are  
typical radii applied to the bright and soft sources, like \pks\ 
(e.g., Fiore, Guainazzi \& Grandi 1999), and ensure that more than
95$\%$ of the photons are collected at all energies. No background light
curves were estimated and subtracted since the estimated background
count rates are of the order of about one percent of the source count
rates.

\section{Temporal Analysis}\label{sec:timing}

\subsection{Light Curves and Hardness Ratio} \label{sec:timing:lc}

\begin{figure}
\epsscale{0.9}
\plotone{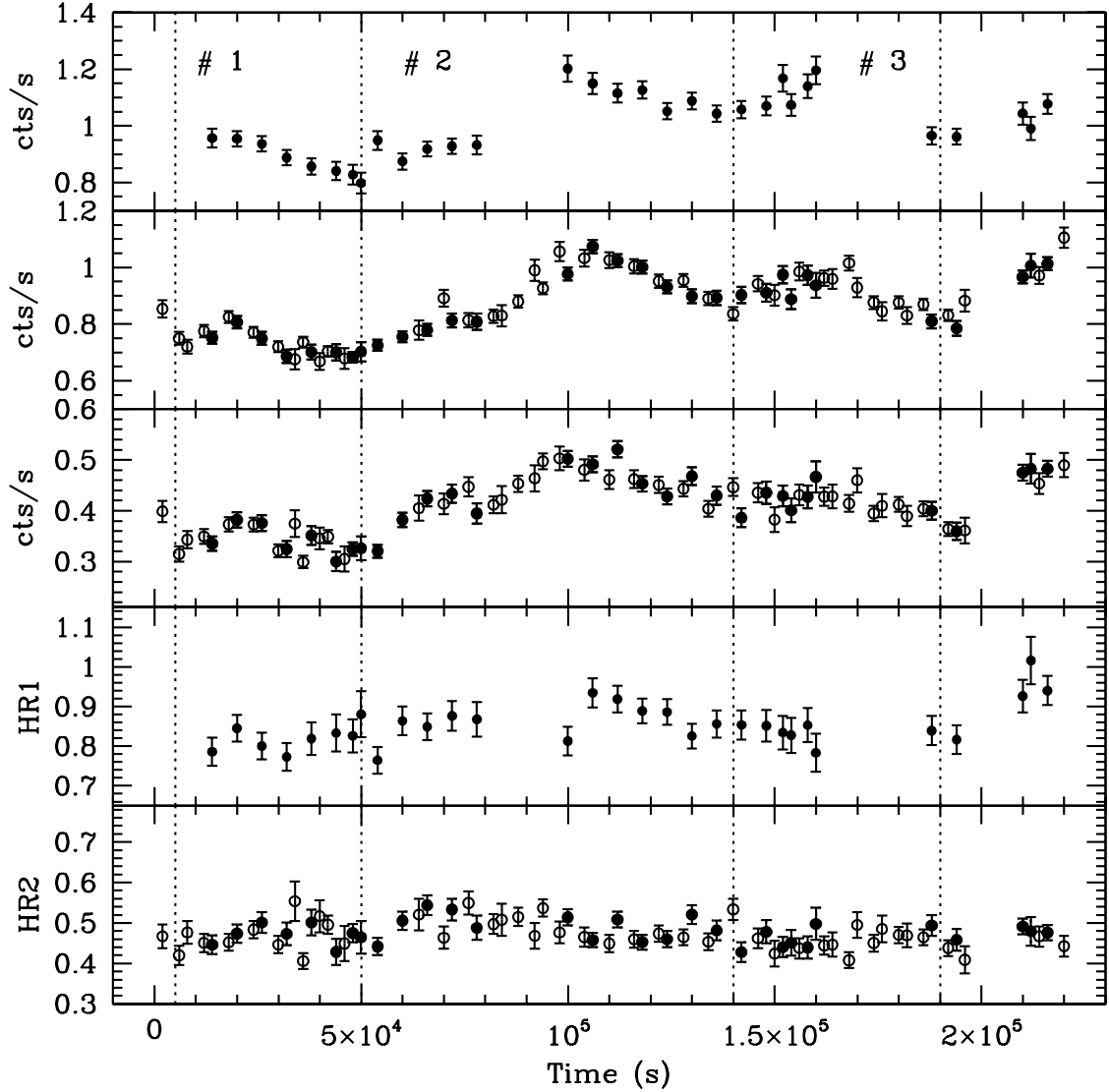}
\caption { \footnotesize Light curves and hardness ratios of 1996 November
20--22 observation. Data are rebinned over 2000~s. The reference time is
1996/11/20/ (TJD=10407) 00:00:00 UT. From top to bottom panel: light curve
in the 0.1$-$1.5, 1.5$-$3.5 and 3.5$-$10 keV bands, respectively, and
hardness ratio between the 1.5$-$3.5 and 0.1$-$1.5 keV bands (HR1) and
between the 3.5$-$10 and 1.5$-$3.5 keV bands (HR2). Note that the temporal
coverage of the LECS is much more sparse than that of the MECS. The
simultaneous LECS and MECS data points are indicated by filled
symbols. The whole data set is divided into three parts which in principle
contain single flares separated by the vertical dotted lines and numbered
as $\#1$, $\#2$ and $\#3$.  
}
\label{fig:lc:96}
\end{figure}

\begin{figure}
\epsscale{0.9}
\plotone{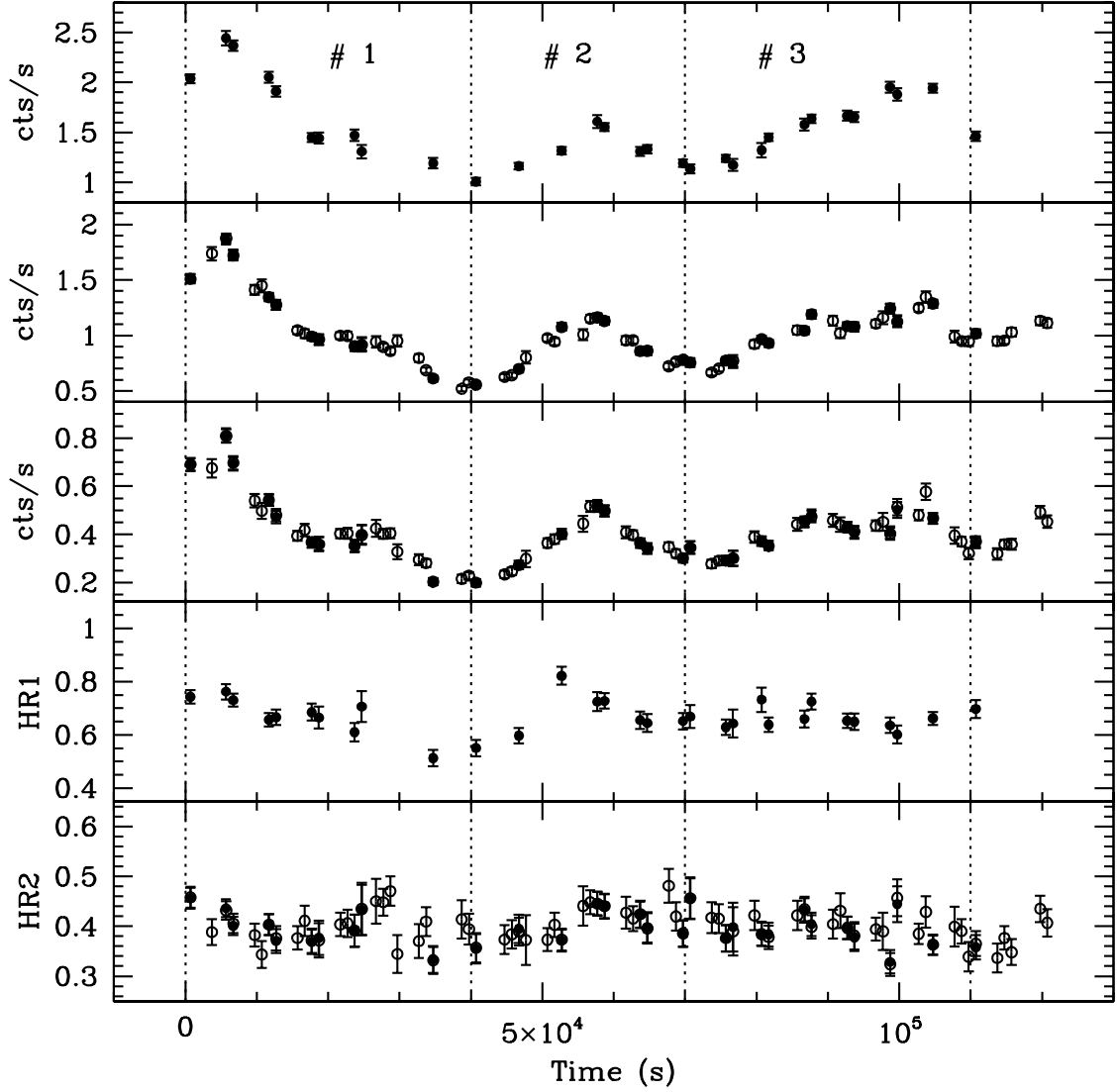}
\caption { \footnotesize Light curves and hardness ratios of 1997 November
22--24 observation. Data are rebinned over 1000~s. The reference time is
1997/11/22/ (TJD=10774) 16:00:00 UT. Panels and symbols have the same
meanings as those in Figure~\ref{fig:lc:96}. This data set is 
basically divided into three single flares numbered as $\#1$, $\#2$ and
$\#3$. 
}
\label{fig:lc:97}
\end{figure}

\begin{figure}
\epsscale{0.9}
\plotone{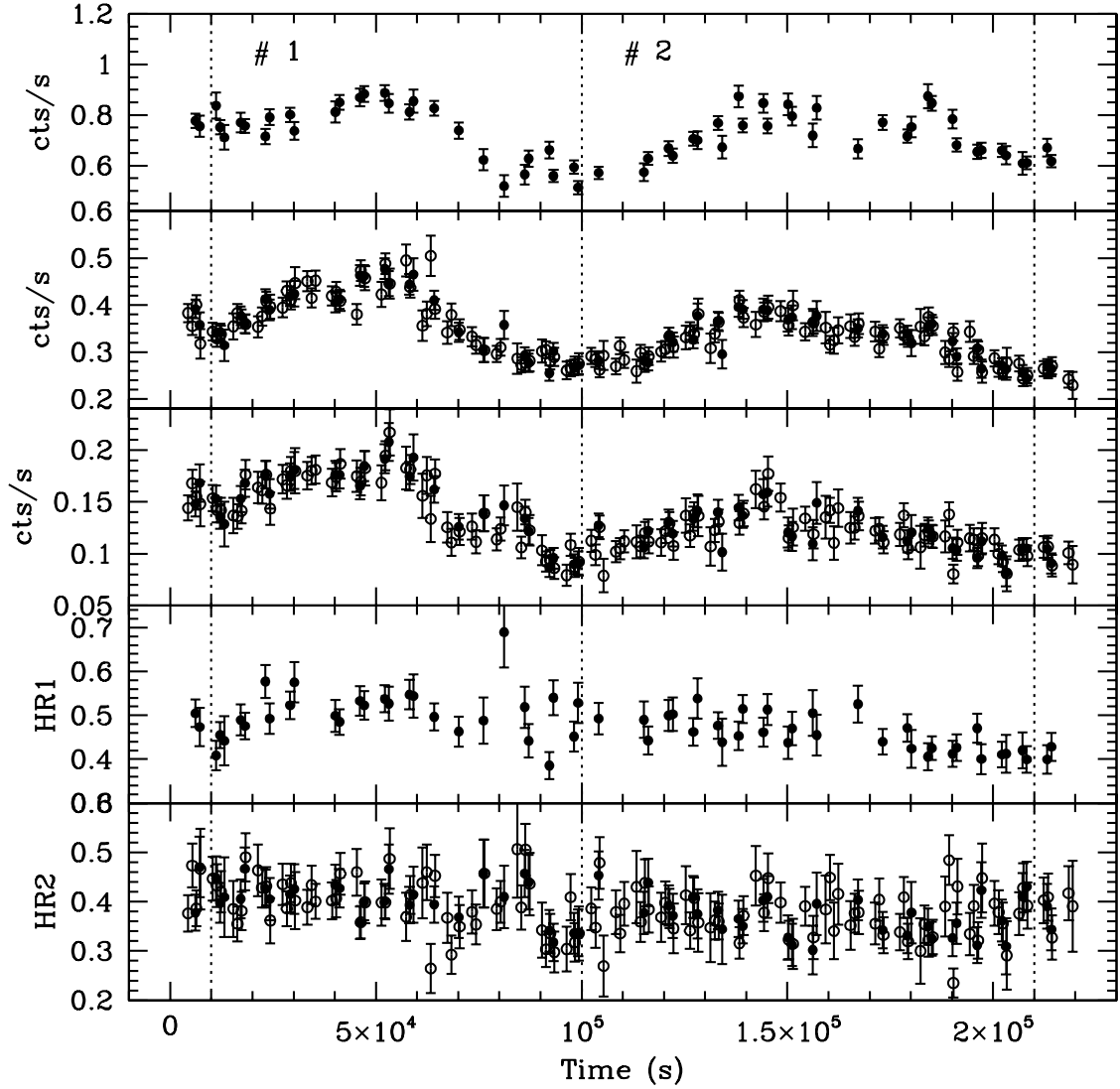}
\caption { \footnotesize Light curves and hardness ratios of 1999 November
4--6 observation. Data are rebinned over 1000~s. The reference time is 
1999/11/04/ (TJD=11486) 04:00:00. Panels and symbols have the same
meanings as those in Figure~\ref{fig:lc:96}. This data set is basically
divided into two single flares numbered as $\#1$ and $\#2$.
}
\label{fig:lc:99}
\end{figure}

In Figures~\ref{fig:lc:96}--\ref{fig:lc:99} are reported the \lcs\  
in the 0.1--1.5~keV (LECS), 1.5--3.5~keV and 3.5--10~keV (MECS), and the
hardness ratios between them. Each light curve has been divided into
several parts, coinciding with single flares considered in our
analysis. Each flare is numbered and separated by the dotted lines.
During all the three observation epochs, which refer to rather different
X-ray emission states of \pks , X-ray variability is clearly detected, 
showing recurrent flares and no quiescent state. In the following we
briefly describe the global properties of each light curve.

\subsubsection{1996} \label{sec:timing:lc:96}

\pks\ was in an intermediate state during this observation.
A flare of low amplitude, \#1, is visible at the beginning of the
observation. \#2 constitutes the major flare during this campaign.
This flare shows an asymmetric structure with different rising 
($\sim$~$5\times 10^4$~s) and decaying $\sim$~$3\times 10^4$~s) \tss 
. These \tss\ may be similar when extending the decay phase to the same
level of count rate as the beginning of the rising phase. Moreover, the
rising phase shows steeper slope than the decaying one does.
This major flare is followed, superimposed on its decay phase, by a lower
amplitude flare, \#3. Another flare occurs towards the end of the
observation, unfortunately only the rising phase was sampled. There are
still some small-amplitude flickers superimposed on the overall trend of
the flares. 

The hardness ratio HR1 shows a behavior similar to that of the light
curves, albeit rather marginal, in the sense that the spectrum becomes
harder with increasing intensity, while HR2 does not follow any clear
trend.  

The energy-dependence of this light curve is visible when one 
carefully examines the major flare, i.e., \#2. It appears that the
0.1--1.5~keV \lc\ leads
the 1.5--3.5~keV one, which in turn lags the 3.5--10~keV one, indicating  
that the inter-band time lags may change the sign across 0.1--10~keV
band. This interesting finding will be examined with the
cross-correlation (\S\ref{sec:timing:lag:lage}) and spectral analysis 
(\S\ref{sec:spec:lag}). 

\subsubsection{1997}\label{sec:timing:lc:97}

The source was in its brightest state during this observation.
At the beginning of the observation it exhibits the maximum
amplitude of variability (flare \#1) among the observations with \sax .
The rising phase of \#1 was not completely sampled. At the end of the
flare a fluctuation is significant which is most evident in the highest
energy band. In the middle of the observation, a relatively small
amplitude flare (\#2 ) presents quite similar rising and declining
timescales. The duration of this flare is the shortest among the
observations studied in this paper. The aspect of flare \#3 is different
from that of the first two, it is probably the result of convolution of
more than one ``flare'' event.

An overall trend similar to that of the count rates is apparent in
hardness ratio HR1, while HR2 does not show any trend.

\subsubsection{1999}\label{sec:timing:lc:99}

The source was in its faintest state during this observation.
The \lc\ is dominated by two ``isolated'' flares which show similar
durations of about one day. The shapes of the flares are quite different.  
Flare \#1, brighter and more pronounced than flare \#2,
shows faster decaying phase than the rising one. In contrast, flare
\#2 shows somewhat faster rising phase than the declining phase. The
energy-dependent trend is quite evident in both flares. For example,
during flare \#1, the decaying phase shows a convex shape in the
0.1--1.5~keV band, while concave shape is clear in the 1.5--3.5~keV band.
Flare \#2 is peculiar, it appears that the 0.1--1.5~keV light curve
consists of two ``flares'' which are invisible in the two higher energy
bands. 

It is difficult to discuss the correlation between the evolution of the
hardness ratio and that of the \lcs\ due to poor photon statistics in
the high energy bands, resulting in hardness ratios with large error
bars.

\subsection{Amplitude and Timescales of Variability}
\label{sec:timing:rms}

The normalized excess variance, $\sigma^2_{\rm rms}$ (see 
Appendix~\ref{sec:app:rms} for the definition), is intuitively 
utilized to quantify the amplitude of variability of a \lc . To
compare the amplitude of variability in different energy bands, the 
0.1--10~keV energy band is divided into four energy bands, i.e.,
0.1--0.5~keV, 0.5--2.0~keV (LECS), 2--4~keV, and 4--10~keV (MECS),
respectively. The \lcs\ are rebinned over 600~s. This integration time
is chosen so that it provides adequate statistical accuracy during the
low intensity state (e.g., 1999), especially for high energy bands where
photon counts are rather low. The bins which are less than $25\%$ exposed
are rejected, this assures that each bin has sufficient (more than
20) photons for the Gaussian statistics to be appropriate. The same
analysis is repeated with a rebinning over 5670~s (about one satellite
orbital period), and the bins which are less than $10\%$ exposed are
rejected to ensure adequate signal-to-noise ratio. This binning produces
comparable number of bins in the LECS and in the MECS.

The same \lcs\ (over bins and energies) used in estimating the excess 
variance are further utilized to quantify the ``minimum doubling \ts
'', $T_2$ (see Zhang \etal 1999 for detailed definition). We just
remind here that the doubling \tss\ that have uncertainties (formally 
propagated) larger than $20\%$ of the value itself are
rejected to minimize the risk of contamination by isolated data pairs
with large errors and insufficient fractional exposure time.

The results are tabulated in Table~\ref{tab:rms}.
Before discussing energy- and intensity-dependence of $\sigma_{\rm 
rms}^2$, it is important to know how different samplings affect
$\sigma_{\rm rms}^2$. This issue is discussed in detail in 
Appendix~\ref{sec:app:rms}. 
The largest uncertainty is the observing duration, $T$, since larger
$\sigma_{\rm rms}^2$ are expected for longer integration 
times. Therefore it is necessary to normalize different $\sigma_{\rm
rms}^2$ to the same $T$.
As discussed in Appendix~\ref{sec:app:rms}, we normalize $T$ of 1996
and 1997 to that of 1999. The amplification factor for $\sigma_{\rm
rms}^2$ is estimated to be $\sim
1.2$ and $\sim 2.5$ for the 1996 and 1997 observation, respectively. We
also discuss in Appendix~\ref{sec:app:rms} the uncertainties due to
binning size, the signal to noise ratio, and irregular gaps. The
determination of the correction factor due to these uncertainties
needs detailed simulations. In any case, it is not important for the
5670~s-binned light curves since they are evenly sampled and have high
ratio of signal to noise.

From Table~\ref{tab:rms}, one can see that the 1997 values of $\sigma_{\rm 
rms}^2$ are significantly larger, and the values of $T_2$ shorter, than
those of 1996 and 1999 in all the considered energy bands. 
Our results show a general trend that $\sigma_{\rm rms}^2$ increases,
and $T_2$ decreases, with increasing energy. There are however some
exceptions. It seems that the 1996 $\sigma_{\rm rms}^2$ calculated from 
the 600~s binned light curve keeps constant with energy.
The average count rates of 1996 are higher than those of 1999, but
show smaller $\sigma_{\rm rms}^2$ in the three high energy  
bands. However, in the 0.1--0.5~keV band, the count rates and $\sigma_{\rm 
rms}^2$ of 1996 and 1999 are statistically indistinguishable. 

In summary, \pks\ shows complex variability, and correlation of
$\sigma_{\rm rms}^2$ and $T_2$ with brightness is no longer clear, a
result which is different from that obtained by Zhang \etal
(1999). However, the results presented here may be biased by poor photon
statistics in some cases.  

\subsection{Variability of Hardness Ratio}
\label{sec:hardness}

\begin{figure}   
\epsscale{0.6}
\plotone{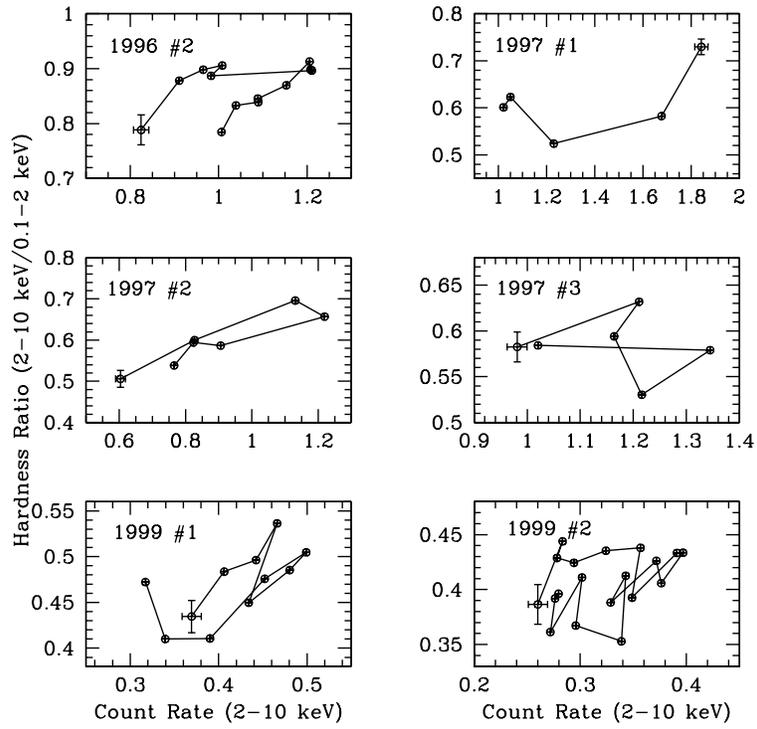}
\caption { \footnotesize Hardness ratio of the 2--10 keV to 0.1--2 keV
bands as a function of the observed count rate in the 2--10~keV. The
data are binned over 5670~s. The point with error bars represents the
starting point of the loop. The errors shown are typical. } 
\label{fig:hdnloop}
\end{figure}

Before spectral analysis (\S\ref{sec:spec}), spectral variability
versus flux can be simply studied using the hardness ratios as a 
function of count rates. In Figure~\ref{fig:hdnloop}, we
plot the hardness ratio of 2--10/0.1--2~keV versus the observed count rate
in the 2--10~keV band for each flare as numbered in
Figure~\ref{fig:lc:96}--\ref{fig:lc:99}. Each point is binned over 5670~s
and points with less than $10\%$ fractional exposure are rejected to
ensure sufficient photons in a bin as discussed in previous section. Six
flares are shown in the figure. The points with error bars indicate
the starting point of each loop, and the solid line tracks the time
sequences of development of each flare. In general they show clockwise
hysteresis as seen from Figure~\ref{fig:hdnloop}, indicating that
the low energy photons lag the high energy ones, so-called soft lag. 
It is worth noting that the flares are generally complicated with the
occurrence of smaller amplitude flickers overlapping on the major
flares, leading to a break of the clean and smooth loop of hardness
ratio versus count rate. For the 1997 flare \#1, the loop only starts
from the maximum point of the flare due to insufficient sampling
of the rising phase of this flare.

Spectral variability will be studied in a more accurate way with the 
time-resolved spectral analysis in \S\ref{sec:spec}. The feature of
soft lag deduced from the clockwise pattern of the loops of hardness
ratio versus count rate will be quantified with the  
cross-correlation techniques in \S\ref{sec:timing:lag}.

\subsection{Power Spectral Density Analysis}\label{sec:timing:psd}

Power spectral density (PSD) is the most common technique to
characterize the variability of a source. However, the \lcs\ obtained
with \sax\ are insufficient to perform a fully sampled PSD analysis
with the standard means because of the presence of the periodic data 
gaps (see \S\ref{sec:obs}).
We calculate the normalized power spectral density (NPSD) following
the method of Hayashida \etal (1998), which utilizes the
standard discrete Fourier transform (see Appendix~\ref{sec:app:psd}
for the definition).
To ensure this method to be appropriate, evenly sampled data sets are 
required. We therefore made 2--10~keV (MECS) \lcs\ with two kinds of bin
sizes, 256~s and 5670~s. The light curve binned over 5670~s is evenly
sampled for each observation, except for the last part of 1996 which is
rejected due to a long gap not due to the Earth occultation. 
The light curve binned over 256~s is divided into a series of individual
segments based on the orbital period. The segments that have no gaps
are used to ensure each segment being evenly sampled.   

The light curve binned over 5670~s produces the NPSD at low frequency
(less than $2\times 10^{-4}$ Hz) range. Similarly, each segment of the
\lc\ binned over 256~s yields one set of [f,p(f)] at high
frequency (larger than $2\times 10^{-4}$ Hz) range, which are first
sorted in frequency, and then averaged for the same frequency. NPSD of 
both the low and the high frequency range are then rebinned in a
logarithmic interval of 0.2 (i.e., a factor of 1.6) to allow 
determination of errors. The disadvantage of this technique is that it
introduces a large gap in the NPSD around $2\times 10^{-4}$ Hz. 

\begin{figure}
\epsscale{0.4}
\vspace{-1.6cm}
\plotone{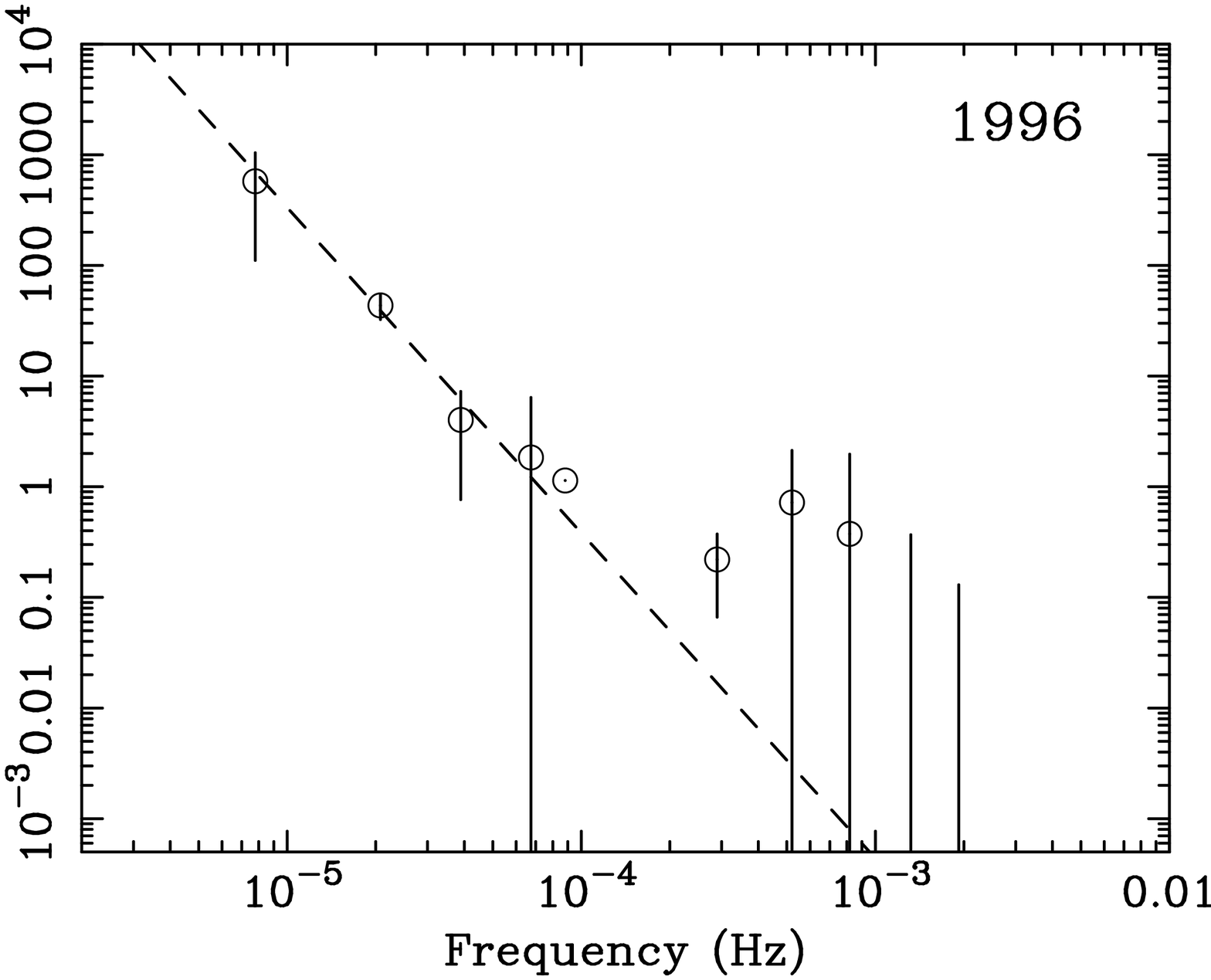}\\
\vspace{-1.6cm}
\plotone{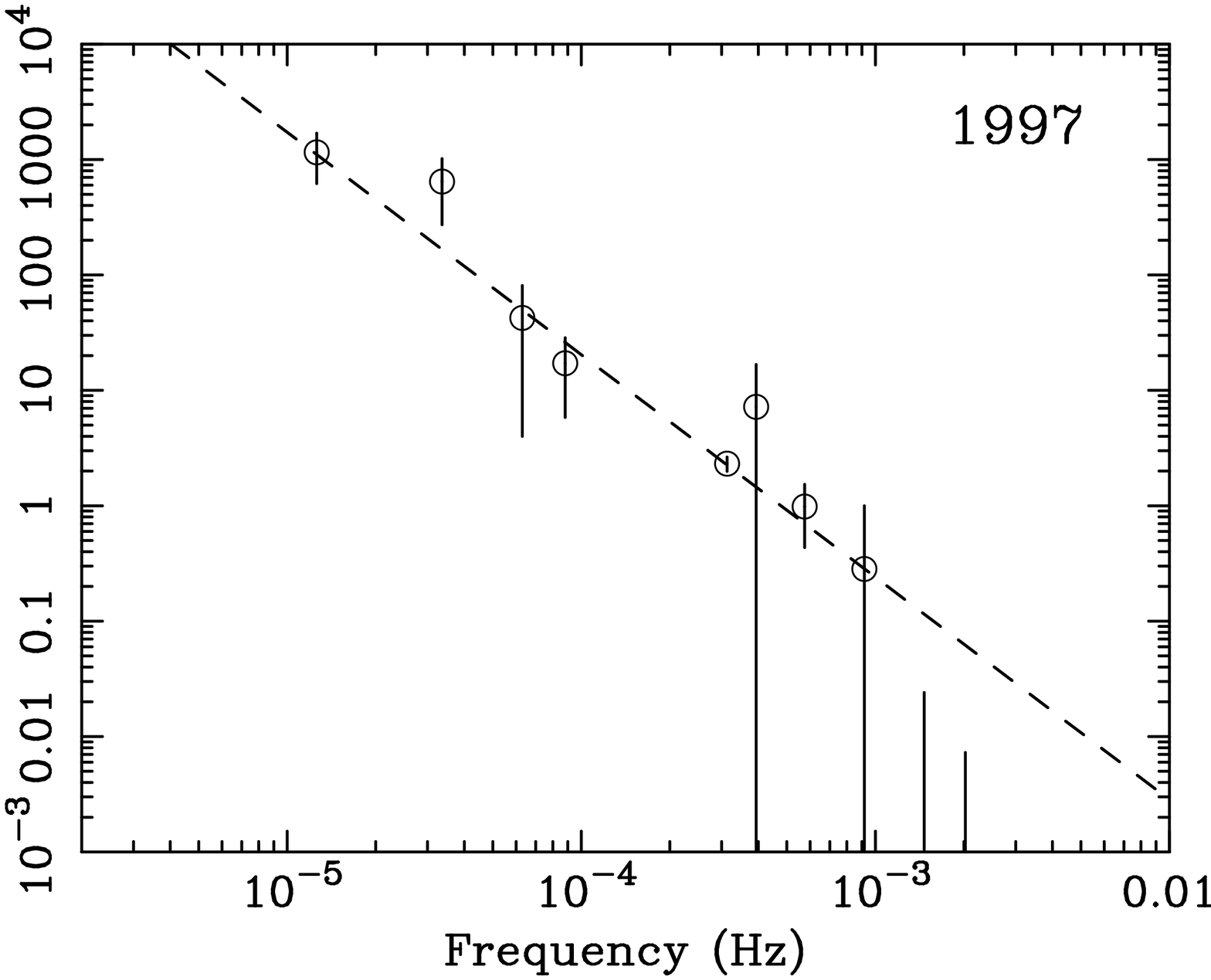}\\
\vspace{-1.6cm}
\plotone{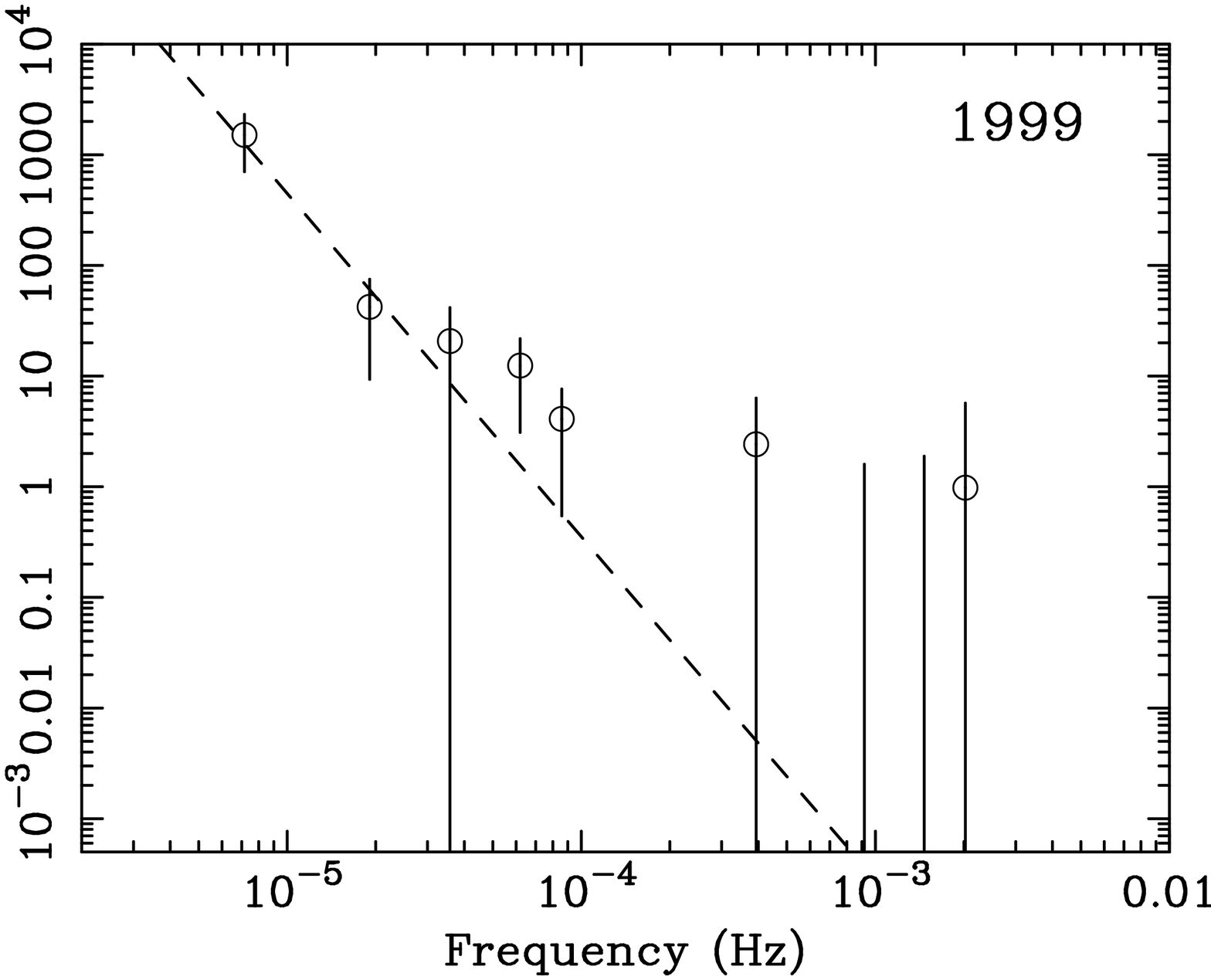}
\caption { \footnotesize Normalized power spectral density (NPSD) derived
from the light curves in the 2--10 keV band. The dashed line corresponds
to the best fit with a power-law model.
}
\label{fig:psd}
\end{figure}

The NPSDs obtained with this procedure are shown in 
Figure~\ref{fig:psd}. One can see that each NPSD generally follows 
a very steep power-law shape that quickly decreases with increasing
frequency in the low frequency range. It is important to note that
pronounced differences exist among the three NPSDs. The power law slope of
the 1997 NPSD may extend to $10^{-3}$ Hz, indicating rapid variability on
timescale of $\sim$~1000~s. However, power-law breaks of the 1996 and
1999 NPSDs are clear before $10^{-3}$ Hz, while this feature may be
overwhelmed by the poor photon statistics when the source is in the
fainter states.    

To quantify the slope, we fit each NPSD with a single power law
model, the best fit parameters are reported in Table~\ref{tab:psd}. 
The strong \textit{red noise} variability nature is indicated by
the steep slopes of $\sim$~2--3. The best fit power-law
slope shows that the NPSD of 1997 is flatter than those of 1996
and 1999, but the differences are statistically weak due to quite large
errors associated with the slopes of the 1996 and 1999 NPSDs. 

Finally, we compare the NPSDs presented here with those derived by
Zhang \etal (1999) where the gap filling technique was used to 
estimate the PSDs of the 1996 and 1997 observations. We found that the
1997 NPSD calculated with gap filling method are fully consistent with
that given here. For the 1996 observation, however, it seems that the
PSD slope derived with gap filling method is flatter than that
obtained here, but when the large errors of the NPSD are taken into
account, we believe that the 1996 slopes calculated with the two
methods are also consistent. Moreover, we note that ``red
noise leak'' could be reduced after linearly interpolating across the
gaps, but Poisson errors associated with these fitted points are still
uncertain. The true shape of PSD may be distorted by the sampling window
function. It is also necessary to point out that the PSDs derived in
Zhang \etal (1999) were averaged by dividing the whole light curves
into several intervals.  

\subsection{Structure Function Analysis}\label{sec:timing:sf}

With respect to PSD, the structure function (SF; see 
Appendix~\ref{sec:app:sf} for the definition) technique has advantages
in quantifying the time variability of a unevenly sampled \lc .
One of the most powerful features of SF analysis is the ability to
discern the range of the \tss\ that contribute to the variations of a
source. The most important one is $\tau_{\rm max}$, at which the SF
flattening occurs. 

\begin{figure}
\epsscale{0.71}
\plottwo{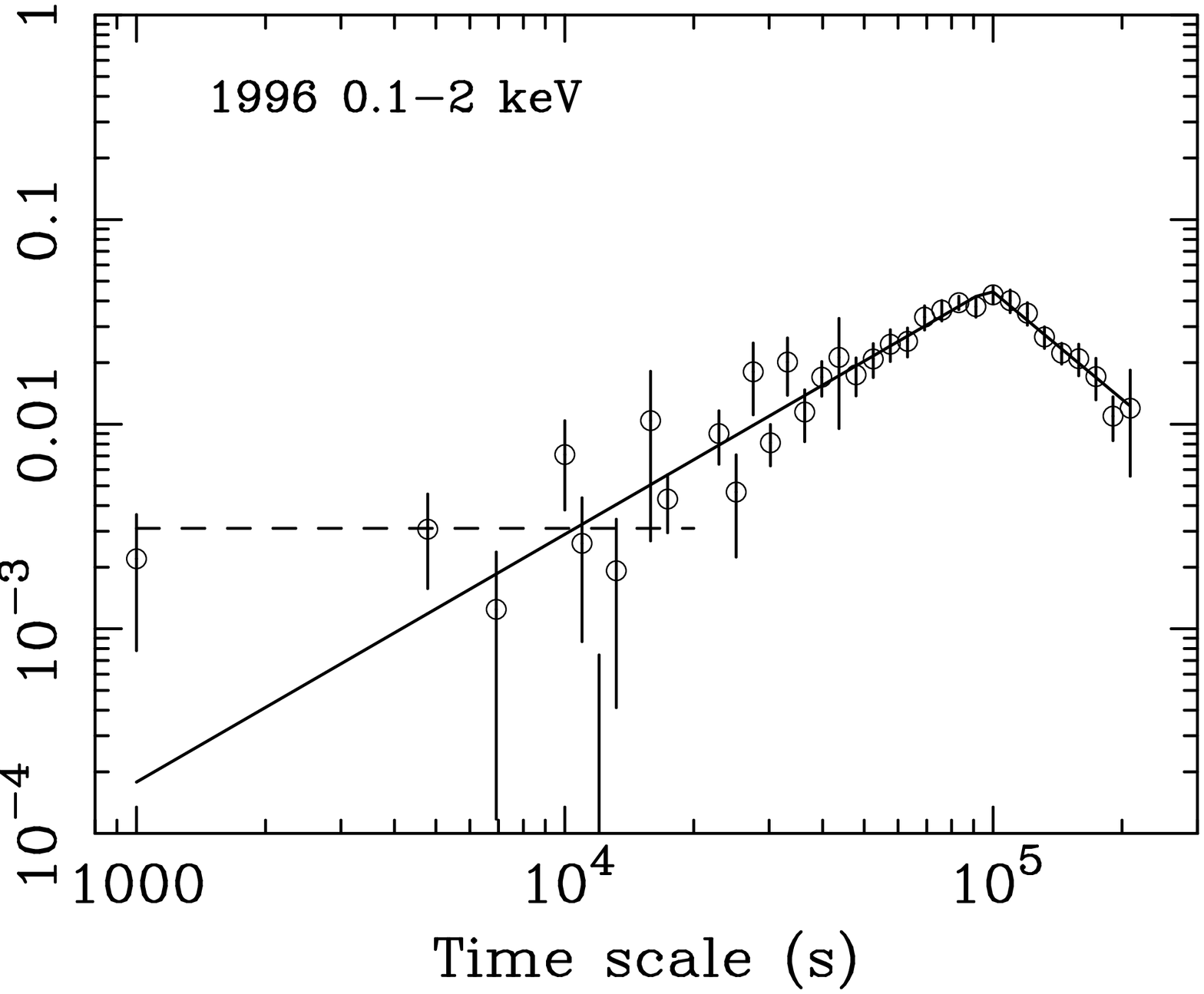}{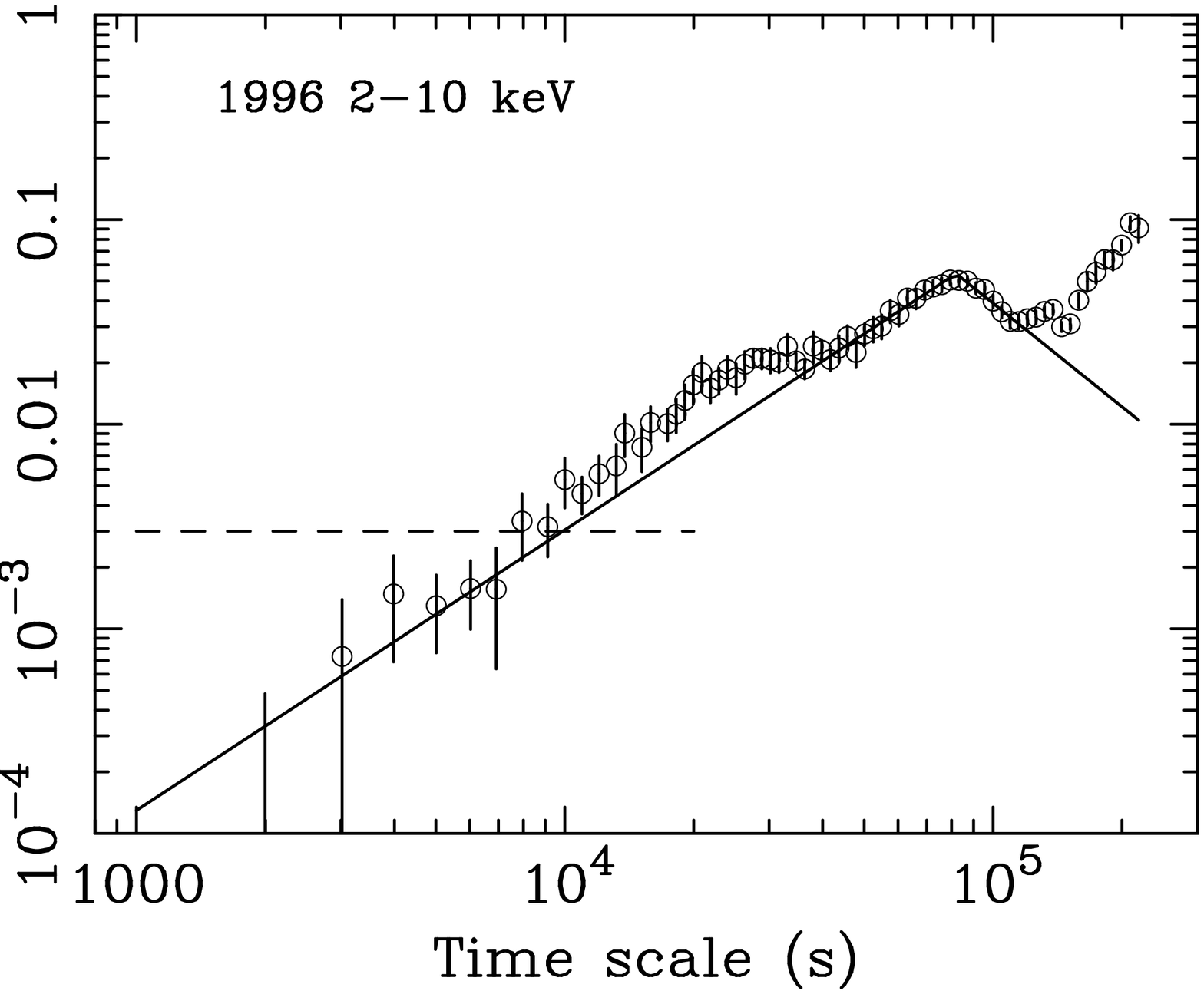}\\
\epsscale{1.6}
\plottwo{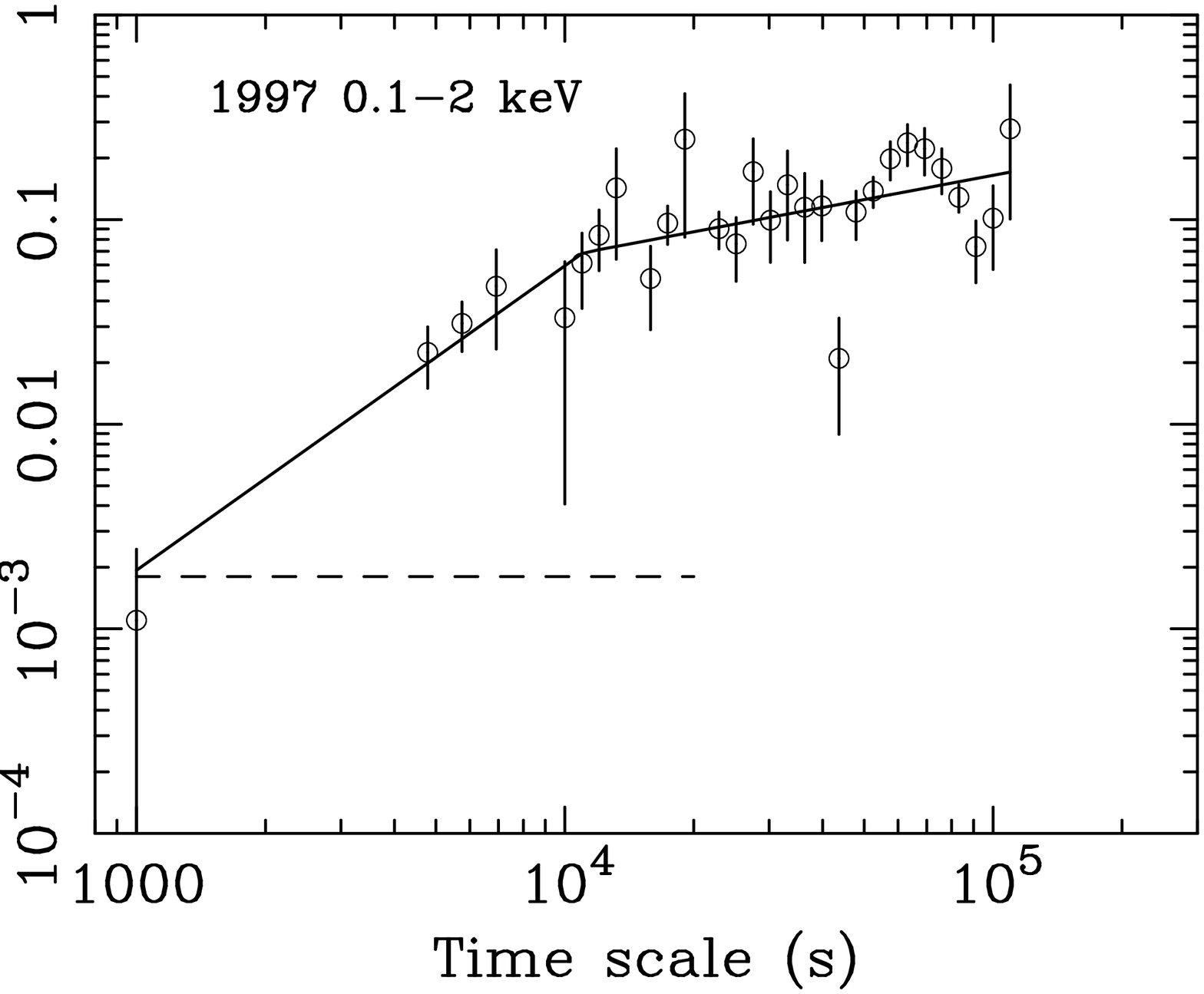}{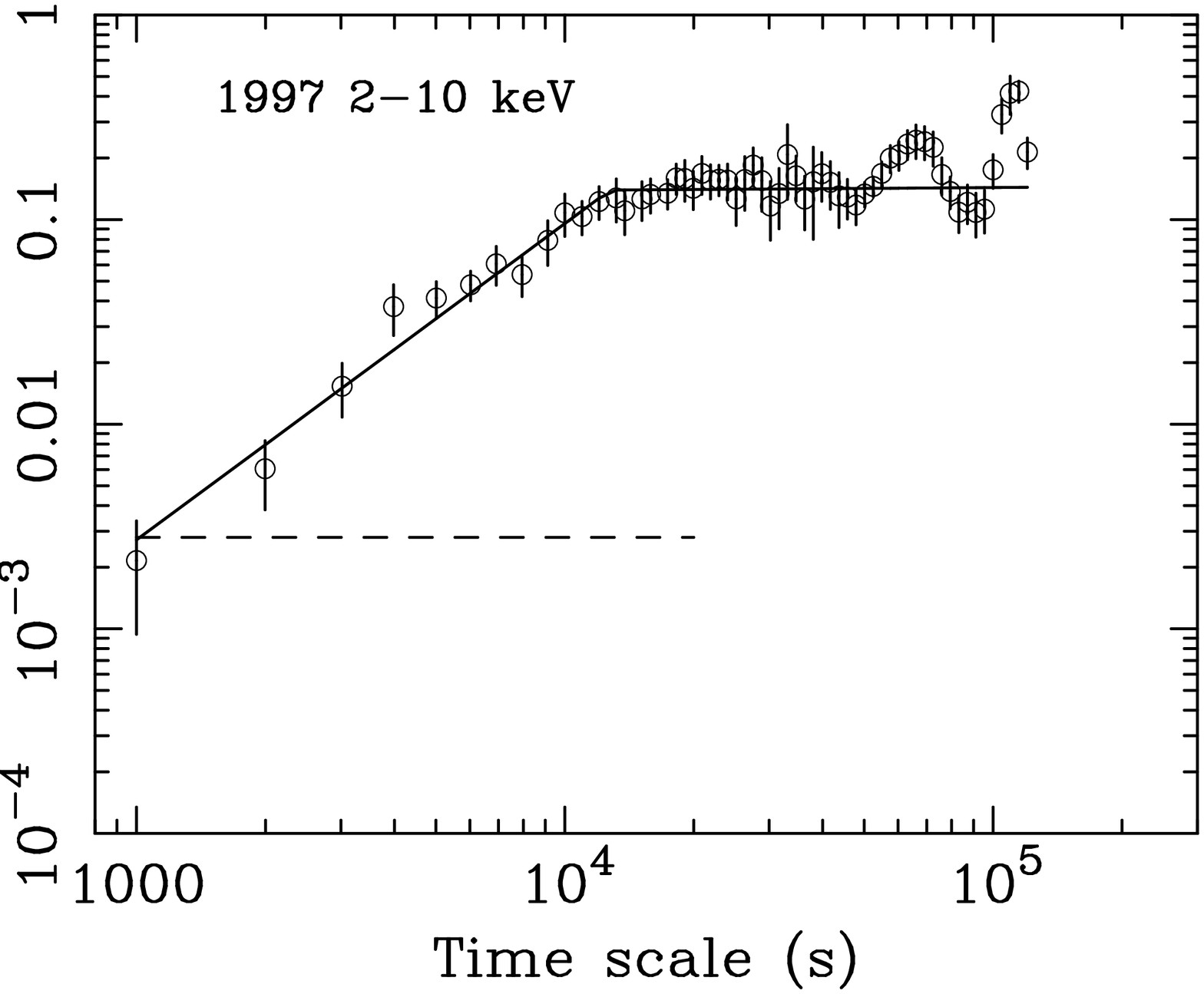}\\
\epsscale{3.6}
\plottwo{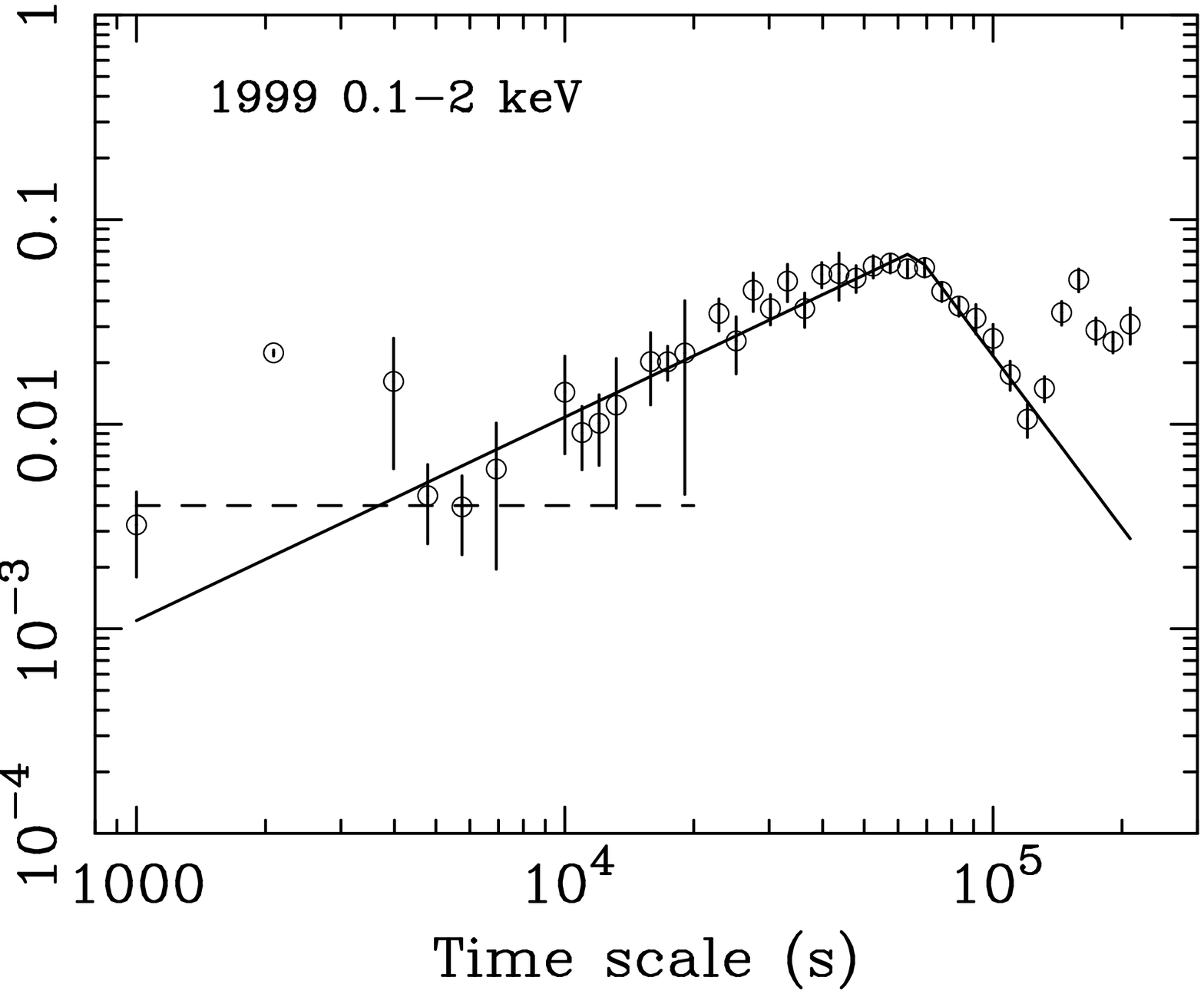}{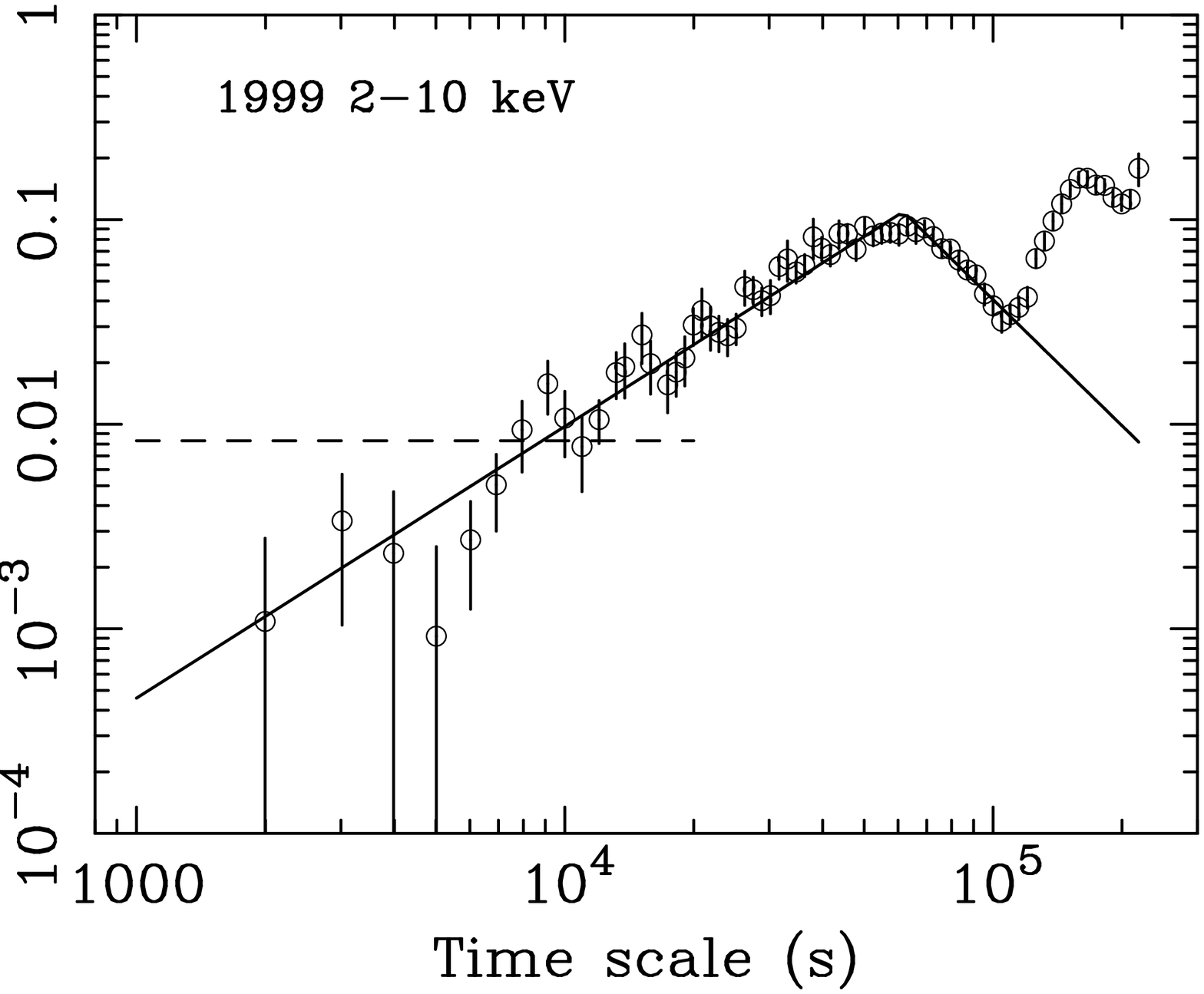}
\caption { \footnotesize Normalized structure function (SF) derived from
the 1996 (top), 1997 (middle), and 1999 (bottom) 1000~s binned \lcs\ in
the 0.1--2~keV (left) and 2--10~keV (right) bands, respectively.
Solid lines are the best fits with a broken power law model, and the
dashed lines are the level of measurement noise, $2\sigma_{\rm
noise}^2$, which is subtracted form the SFs. }
\label{fig:sf}
\end{figure}

The \lcs\ are binned over 1000~s and normalized by the mean count rate
before calculating SF. In this way, the SF is normalized by the squared
mean count rate of the \lc ,  allowing us to compare SFs in different
states of a source. The SFs are calculated for the 0.1--2~keV
(LECS) and 2--10~keV (MECS) energy bands to compare the
variability characteristics in the soft and hard \xr\ bands. The
contribution of the measurement (Poisson) noise to the SFs is
$2\sigma_{\rm noise}^2$, which is subtracted.  Figure~\ref{fig:sf} shows
the derived SFs. It is necessary to point out that the 0.1--2~keV SFs
are more poorly constrained than the 2--10~keV ones because of low
exposure efficiency of the LECS. 

The first piece of information we got from the SFs of the three campaigns
is that they show similar structures. The variations of \pks\ decrease
quickly with decreasing \tss\ as seen from the steep slopes of the SFs.
Similar to NPSDs, 1996 and 1999 SFs may show the evidence for
the shortest correlation \tss, $\tau_{\rm min}$, of around 6000~s,
while $\tau_{\rm min}$ could be $\sim$ 1000~s or even lower in
1997. This feature also indicates that \pks\ tends to be more rapidly
variable in high state, but, as for PSD, this difference may be 
overwhelmed by strong Poisson noise in its faint states.

With respect to the PSD, the most important inference from the SFs is
that one can determine the longest correlation \ts , $\tau_{\rm max}$,
identified by the first ``turn-over'' point at the end of the long \ts\
of the power-law shape. Importantly, $\tau_{\rm max}$ may reflect the
typical (half) duration time of the repeated flares. $\tau_{\rm max}$
thus may give an evaluation of the characteristic \ts\ of the
source. To accurately quantify $\tau_{\rm max}$ and the power-law
slope, a broken power-law model 
$$SF(\tau) = C(\tau/\tau_{\rm max})^{\beta}, \tau \le \tau_{\rm max} \,;$$
$$SF(\tau) = C(\tau/\tau_{\rm max})^{\beta_1}, \tau > \tau_{\rm max}$$  
is fitted to the SF between 1000~s and the \ts\ at which the first
minimum occurs. The best fit curves are plotted in
Figure~\ref{fig:sf} with solid lines, and the best fit parameters
with $90\%$ confidence region for one interesting parameter are
tabulated in Table~\ref{tab:sf}, where only the first power-law
slope, $\beta$, and $\tau_{\rm max}$ are shown. The results show that
the power law slopes of the SFs are $\sim$~1--1.5, this corresponds to a
PSD power law slope of $\sim$~2--2.5, suggesting that the \xr\
variability of \pks\ is always dominated by shot noise. Moreover, our
results suggest that $\tau_{\rm max}$ differs by a factor of $\sim 6$
from epoch to epoch. This may shed light on some clues on the dynamics
and the structure of the jet (see \S\ref{sec:disc:implication}).

\subsection{Cross-correlation Analysis}\label{sec:timing:lag}

The main goal of the cross-correlation analysis is to determine the
degree of correlation and the time lags between the variations in 
different energy bands. Before constructing the 
cross-correlation function (CCF), light curves are normalized to zero
mean and unit variance by subtracting the mean count rate and dividing by
the rms of the \lcs\ (e.g., Edelson \etal 2001).
CCFs are measured using two techniques suited to unevenly sampled time
series: the Discrete Correlation Function (DCF, Edelson \& Krolik
1988) and Modified Mean Deviation (MMD, Hufnagel \& Bregman 1992). 
In addition, model-independent Monte Carlo simulations taking into
account ``flux randomization'' (FR) and ``random subset selection''
(RSS) of the data sets (Peterson \etal 1998) are used to statistically
determine the significance of time lags derived from DCF and MMD. 
FR/RSS is based on a computationally intensive statistical
``bootstrap'' method. This will build a
cross-correlation peak distribution (CCPD; Maoz \& Netzer 1989). For 
details of this procedure we refer to Zhang \etal (1999).
In most cases, the bin sizes of light curves and DCF/MMD are chosen to
be about 3 times smaller than the possible lag after a series 
of experiments. DCF/MMD function is fitted with a Gaussian
function plus a constant, and the time lags are evaluated with the
Gaussian centroid rather than the DCF/MMD peak/dip to reduce the
possibility of spurious estimates of time lags due to a particular
DCF/MMD point that could originate from statistical errors and data
gaps (see Zhang 2000 for details). The accuracy of a cross-correlation
result can be better than the typical binning as long as the DCF/MMD 
function involved are reasonably smooth. Throughout the paper, a
positive lag indicates that the lower energy photons lag the higher
energy ones (soft lag), while a negative lag represents the opposite
(hard lag).

We first comment on some issues that may affect the evaluations of 
time lags. The detection of a time lag is dependent not only on the 
sampling characteristics of the \lcs , but also on the specific variations
and the measurement uncertainties of the \lcs . The increasing level of
the measurement uncertainties with respect to real variability will
broaden the CCPD built from Monte Carlo simulations, and this will
decrease the confidence of the detection of a time lag. The detection of
a time lag is thus mainly sensitive to a pronounced flare in which
relative level of the measurement uncertainties tend to be minimum. In
addition, a light curve generally shows multiple flares which can be
dominated by different timescales, as discussed in
\S\ref{sec:disc:implication}, different inter-band time lags are then
expected from flare to flare. 
Cross-correlating the entire \lc\ may prevent the clear detection of a
time lag as it only gives a mean value of more than one time lags
corresponding to the respective flares. In the worst case the average of
the positive and negative time lags may lead to non-detection of a
time lag. To obtain a proper determination of a time lag, from the
point of view of physical processes, one must analyse each flare
individually rather than the whole light curve. 
However, the statistical confidence of a cross-correlation result
decreases because the duration of a single flare is generally much
shorter than that of the whole light curve.    
Since the results for the 1996 and 1997 data sets presented in Zhang
\etal (1999) were based on the whole light curves, we re-analyze them
along with the 1999 data set following this criterion. As we will show in 
the following, the values of time lags depend significantly on the 
different parts of the light curve which are cross-correlated.

The time lags derived with DCF/MMD for the six flares indicated in
Figure~\ref{fig:lc:96}--\ref{fig:lc:99} are reported in 
Table~\ref{tab:lag}. Only the time lags between the 0.1--1.5~keV and the  
3.5--10~keV energy bands are shown. 
It is worth noting that the differences in the
results obtained using the DCF and MMD techniques are significant  
in some cases (e.g., for the 1999 \#2 flare), these differences are 
mainly caused by the irregularities of the DCF/MMD function which
results in large uncertainties when fitting the DCF/MMD function with 
a simple Gaussian function.
It can be seen that the source shows
soft lags in all flares, while the values of soft lags are different
from flare to flare, ranging from a few hundred seconds to about one
hour. If one compares the values of soft lags with the duration of the
flares, or more precisely with $\tau_{\rm max}$ inferred with the SFs (see
Table~\ref{tab:sf}), one can find that there may be a correlation between
them, i.e., smaller soft lags seem to be associated with the flares of
shorter duration.

\subsubsection{1996} \label{sec:timing:lag:96}

We found a soft lag of $\sim$~4 hours (between the 0.1--1.5~keV and
3.5--10~keV energy band) evaluated with the whole 1996 observation
(Zhang \etal 1999). We re-calculate DCF/MMD for the three parts of the
light curve as indicated in Figure~\ref{fig:lc:96}. The
time lags of \#1 and \#3 parts can not be well determined because
there are no proper DCF/MMD function defined, possibly due to sparse 
sampling. The \#2 part, the major flare during this campaign,
shows clear soft lag: 
$6.16^{+1.38}_{-1.33}$~ks (DCF) and 5.09$^{+4.98}_{-3.95}$~ks (MMD). 
The confidence range of this value at 1$\sigma$ level based on FR/RSS
simulations is [1.7, 9.3]~ks (DCF) and [0.6,11.8]~ks (MMD). It is
important to note that the soft lag derived from this major flare is
just about half of the soft lag derived from the whole observation,
demonstrating strong dependence of the estimate of the time lag on
the examined part, basically containing a single flare, of the observed
light curve. 

\subsubsection{1997} \label{sec:timing:lag:97}

The whole 1997 observation is characterized by a soft lag of $\sim$~1~ks
between the 0.1--1.5 and 3.5--10~keV bands (Zhang \etal 1999). This
light curve can be divided into three flares with duration of $\sim$
40~ks each (Figure~\ref{fig:lc:97}). Thanks to their brightness, we
can make a proper estimate of the time lags for each flare. 
The first two flares show quite small soft lags of a few hundred seconds,
with lower limit down to zero lags. The soft lag of the third flare is
significant when determined with the DCF ($\sim$~1600~s), but not with the
MMD (no proper MMD function exists). FR/RSS simulations are not performed
because of short duration of the flares, but we refer to Zhang \etal
(1999) for the simulations based on the whole light curves.   

\subsubsection{1999} \label{sec:timing:lag:99}

With respect to the previous two campaigns, \pks\ is in its faintest state
during this observation. Of interest is that two ``isolated'' flares
were completely sampled, allowing us to study in detail the inter-band
correlations and the time lags occurring in the faint state of the
source. 

\begin{figure}
\epsscale{0.71}
\plottwo{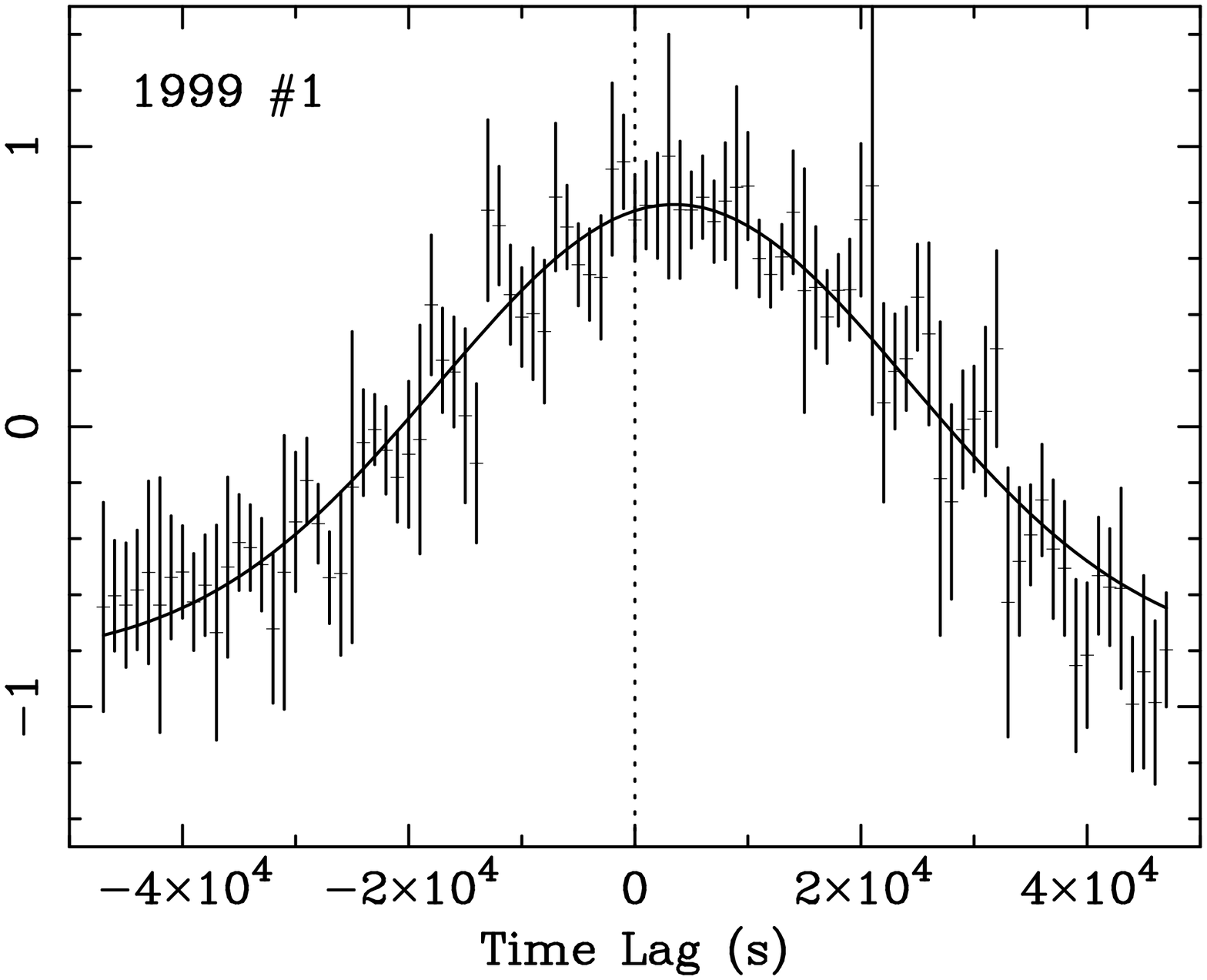}{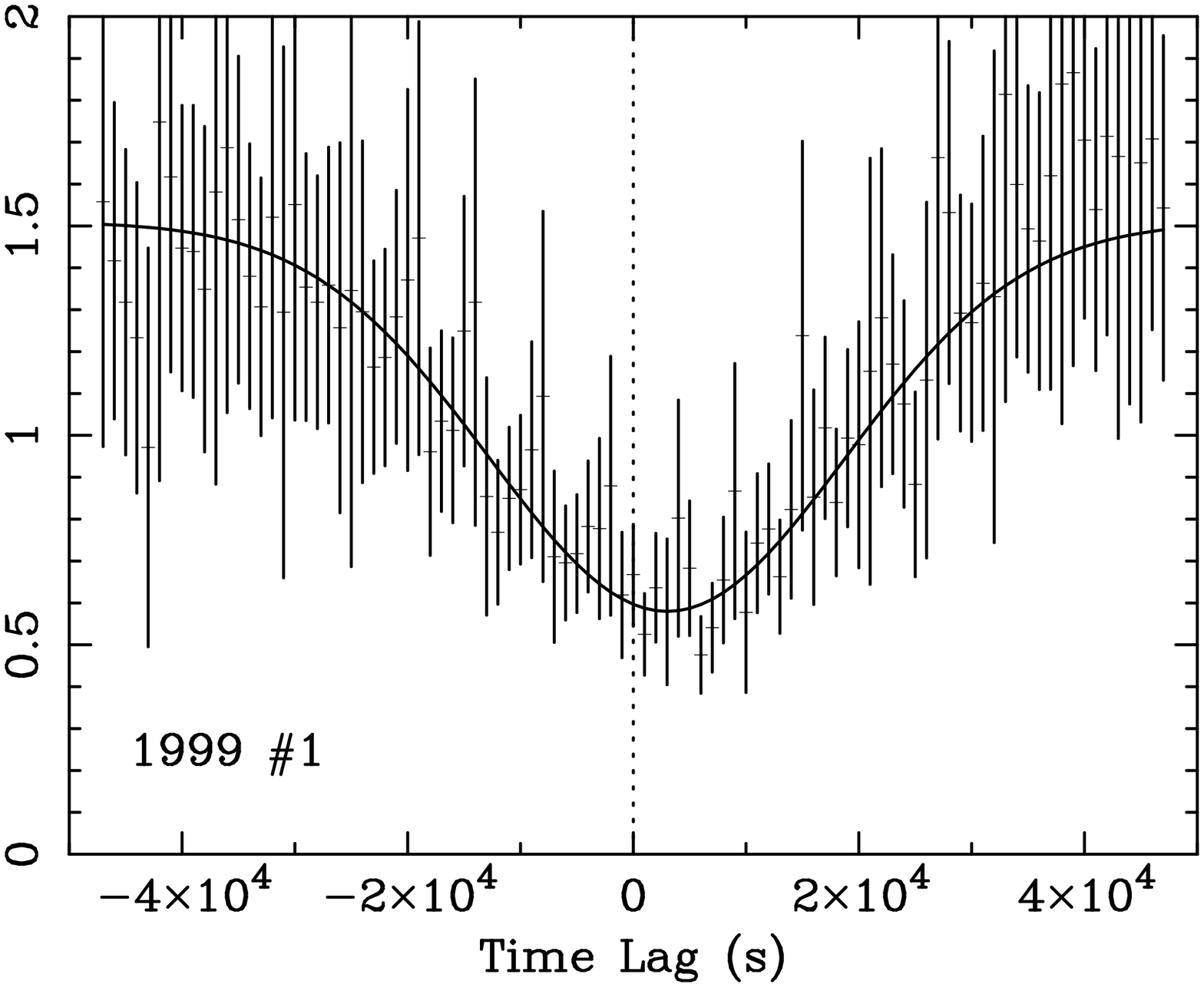}\\
\epsscale{1.6}
\plottwo{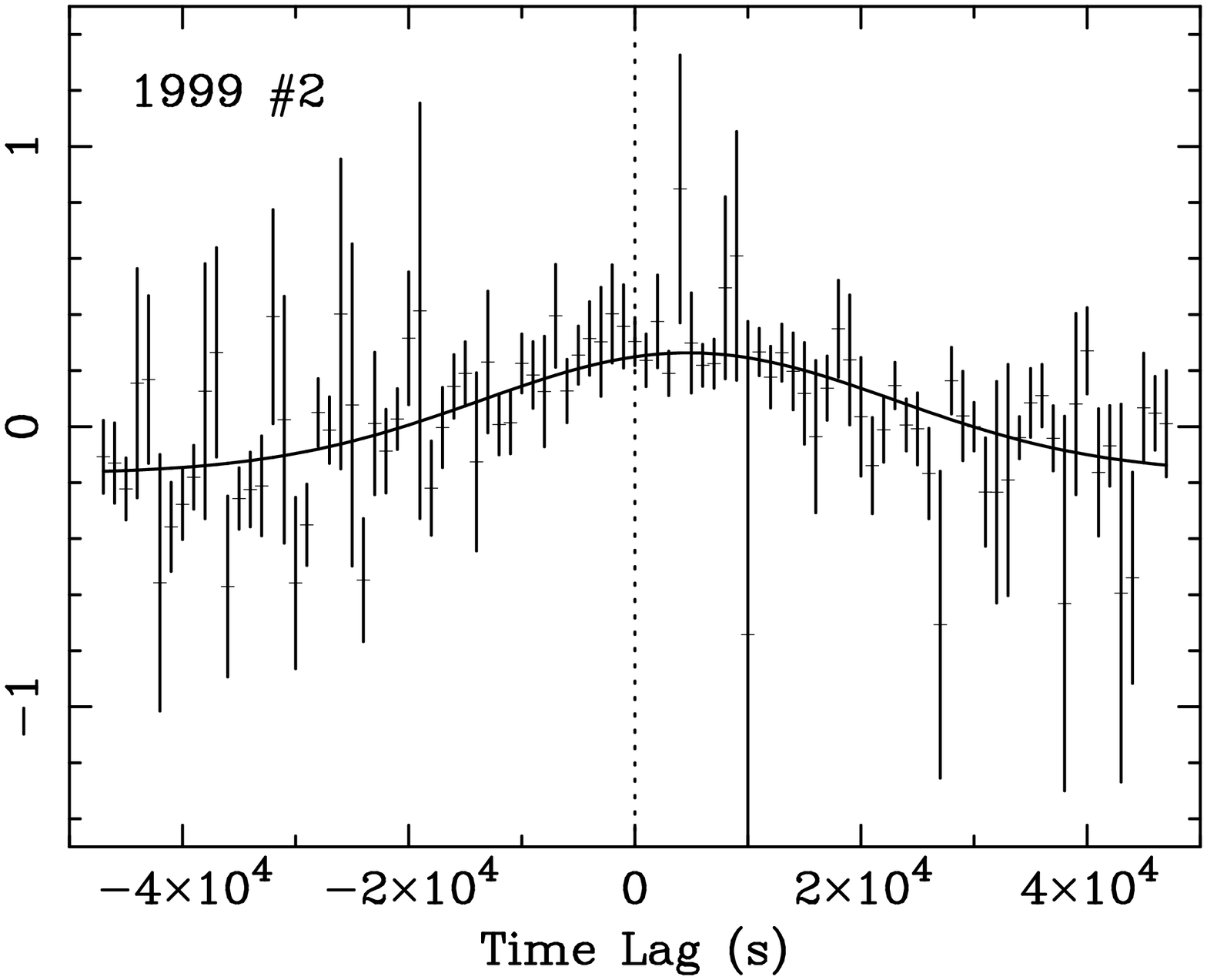}{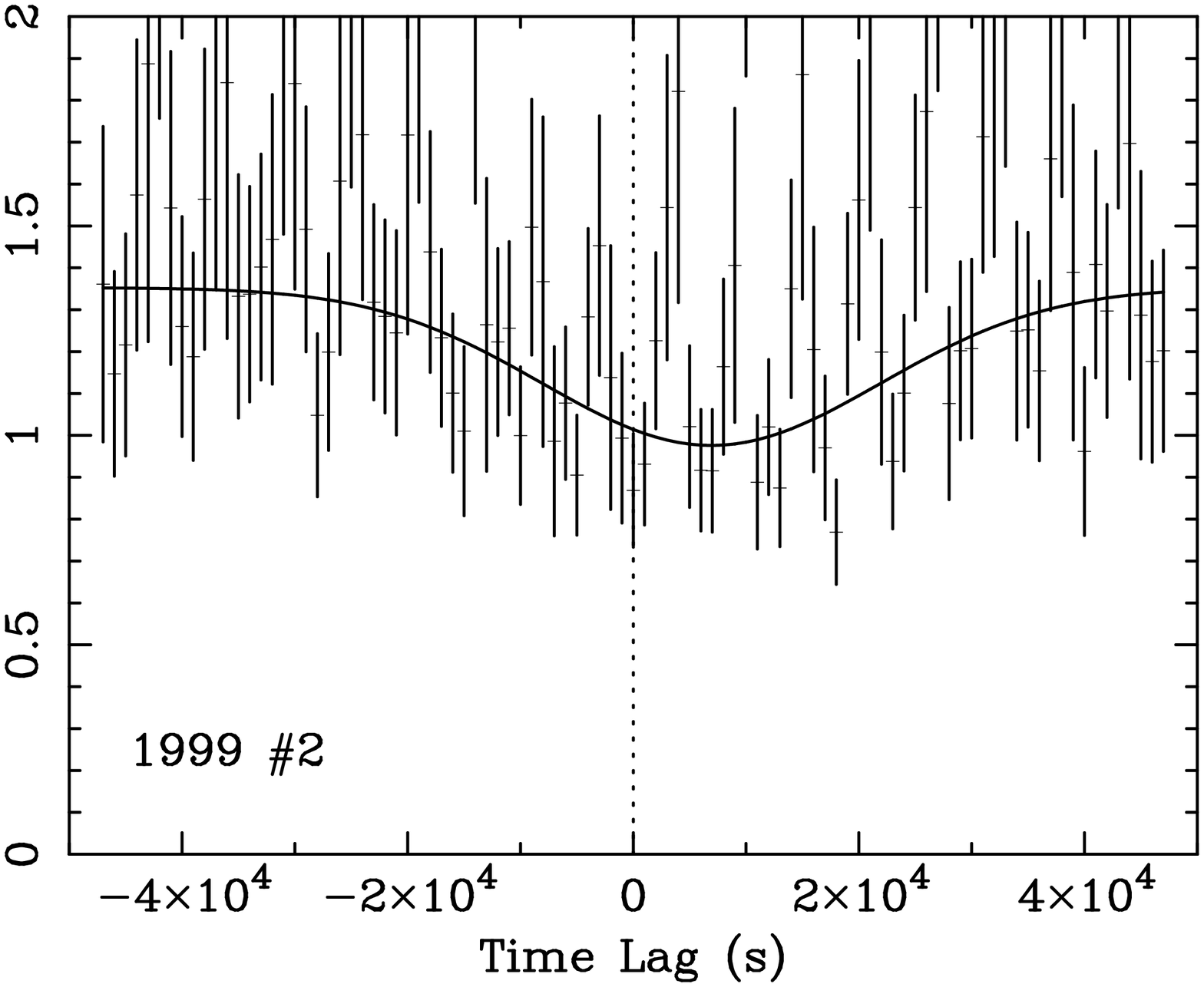}
\caption{ \footnotesize DCF (left) and MMD (right) between the
0.1--1.5~keV and the 3.5--10~keV bands derived from the 1999 \#1
(top) and \#2 (bottom) flare, respectively. The solid curve indicates the
best fit consisting of a Gaussian function plus a constant. }
\label{fig:ccf:99}
\end{figure}

The DCF/MMD functions between the 0.1--1.5~keV and the 3.5--10~keV bands
are shown in Figure~\ref{fig:ccf:99}.
The \#1 flare has relatively well-defined DCF/MMD function, and the
correlation between the two energy bands is high, $\sim$ 0.8, as seen
from the centroid of Gaussian fit to the DCF function. Both DCF and
MMD suggest a soft lag of about 3~ks. FR/RSS simulations indicate that the
confidence (1$~\sigma$) range of this soft lag is [2.7,11.6]~ks (DCF) and
[1.3,12.2]~ks (MMD).  

DCF/MMD of the \#2 flare does not present a well-defined function. The
DCF suggests a very low correlation of about 0.2, indicating that the
variations of the soft and hard photons are weakly correlated. This
might be expected already from the light curves
(Figure~\ref{fig:lc:99}): in the 0.1--1.5~keV band there might be two
flares, which are not seen in the 3.5--10~keV band. Therefore, the
cross-correlation analysis can only give an indication of the
inter-band time lags, which does not necessarily give a clean and
unambiguous measure of the relationship between the two time series. 
FR/RSS simulations result in broader CCPDs of this soft lag,
which is [0.3, 11.0]~ks (DCF) and [3.7, 13.7]~ks (MMD) at 1$\sigma$
confidence level.

Finally, as we commented above, the CCPDs are strongly broadened due to
weak statistics of photons. This gives estimates of the soft lags with
low confidence.

\subsubsection{Energy-dependence of Soft Lags} 
\label{sec:timing:lag:lage}

We further quantify the energy-dependence of soft lags by dividing
the 0.1--10~keV energy band into 6 narrow energy bands, i.e., 
0.1--0.5, 0.5--1.0, 1.0--1.5 (LECS), 1.5--2 (LECS and MECS), 2--4 and
4--10~keV (MECS), and measure the soft lags for each flare with
respect to the 4--10~keV energy band. This is an important issue to
test synchrotron cooling mechanism.  

\begin{figure}
\epsscale{0.4}
\vspace{-1.6cm}
\plotone{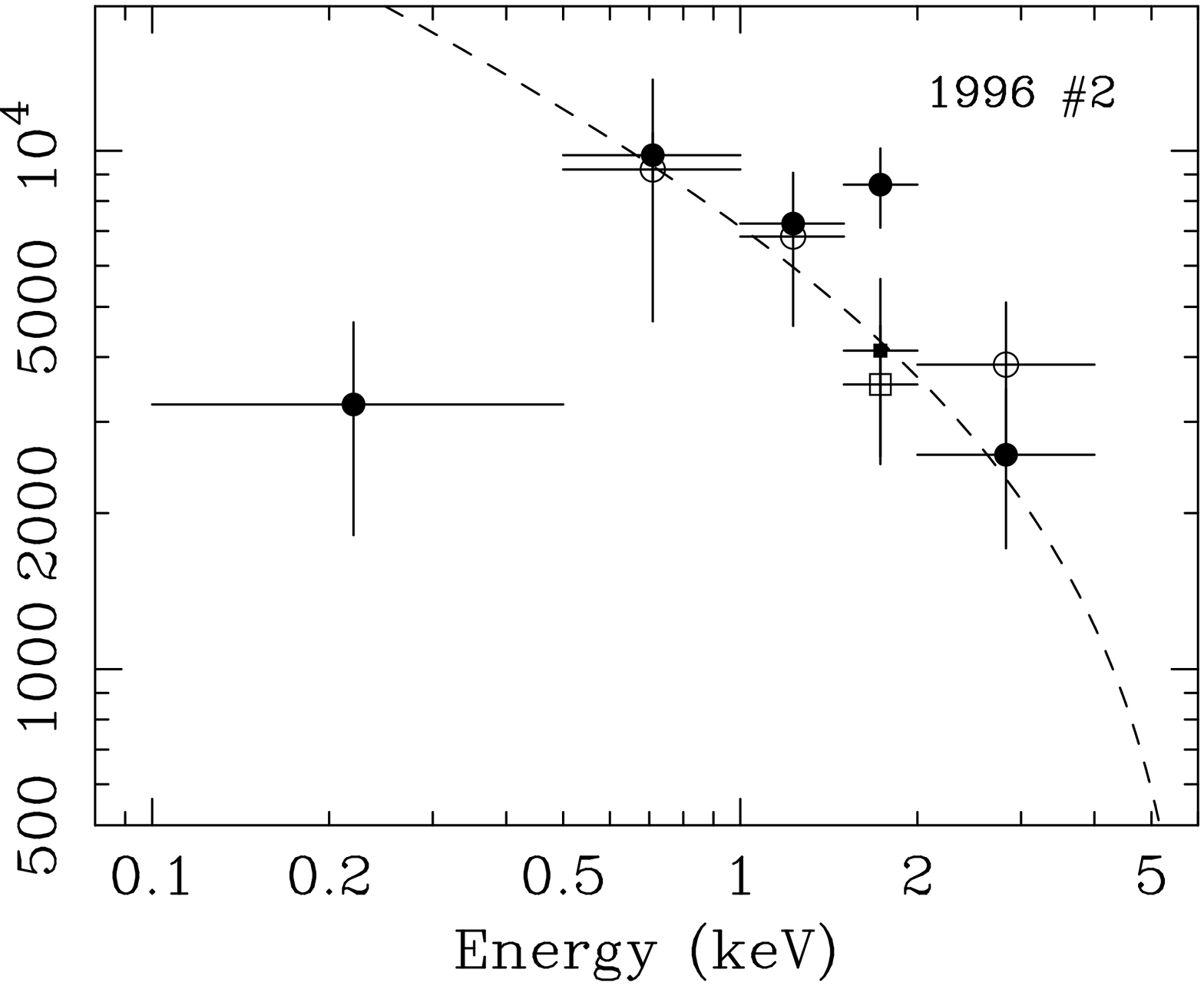}\\
\vspace{-1.6cm}
\plotone{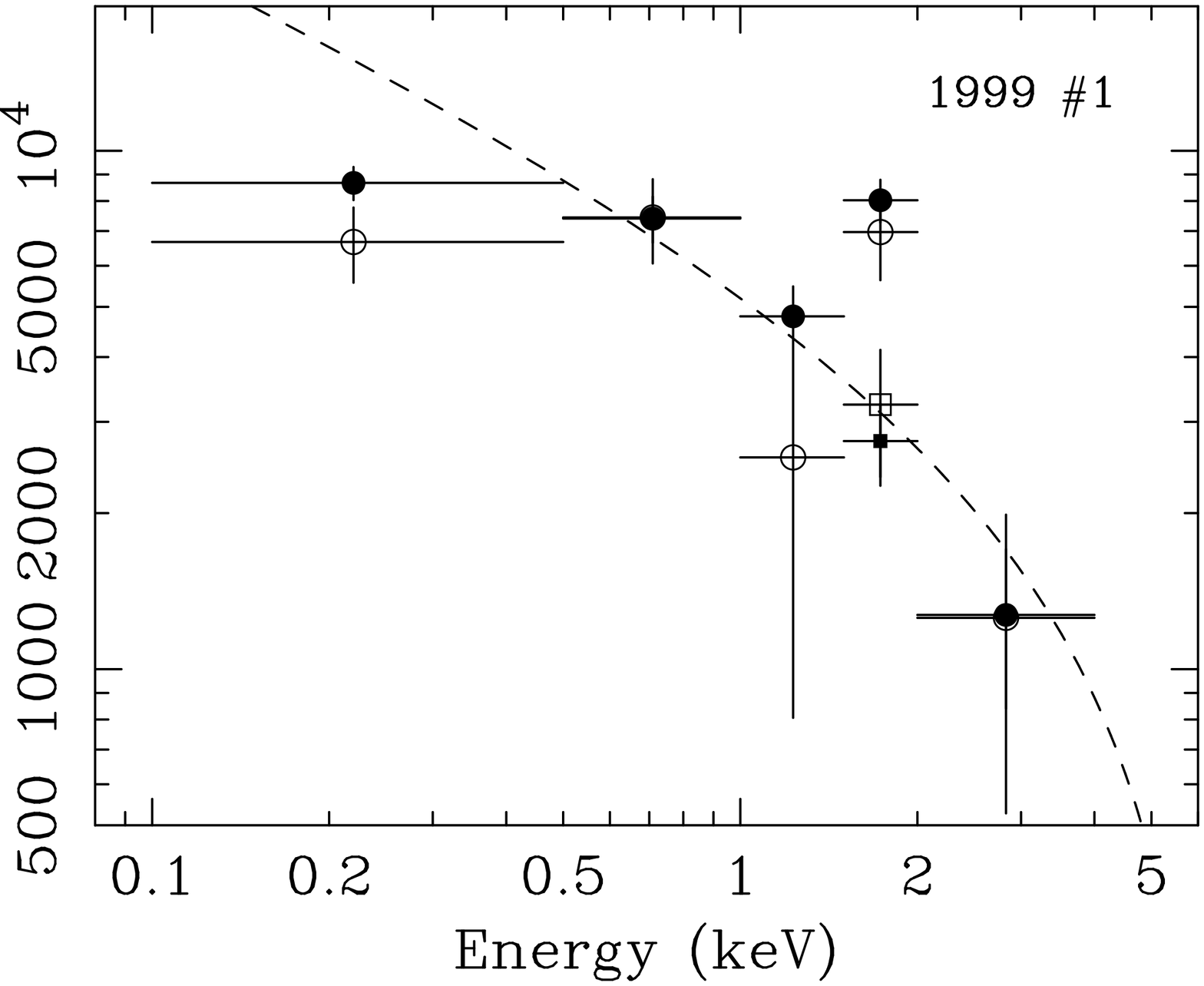}\\
\vspace{-1.6cm}
\plotone{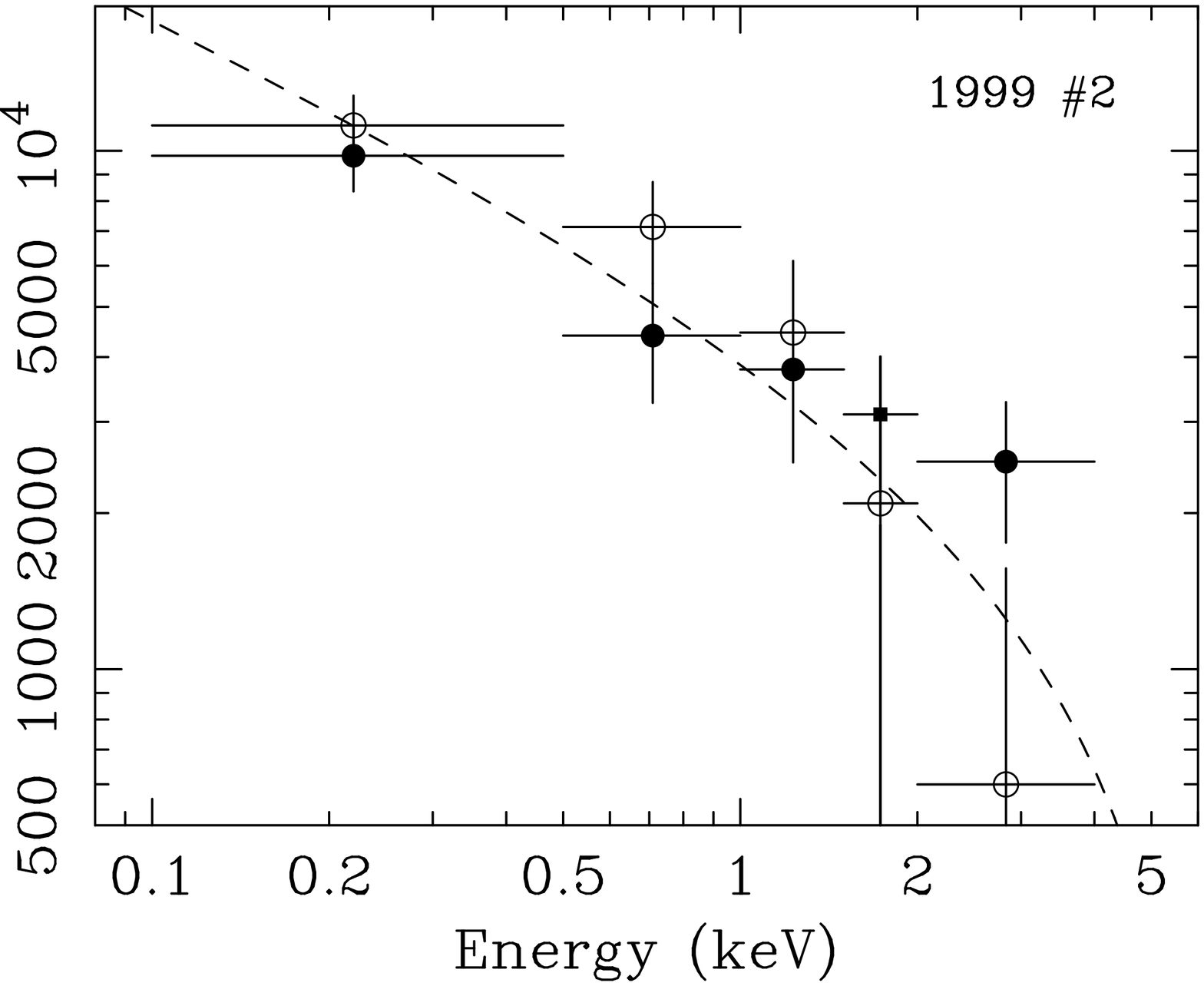}
\caption{ \footnotesize Soft lags, measured with respect to the 4--10~keV 
band, as a function of photon energies. The dashed curve is the best
fit with the energy dependence of \sy\ cooling \ts . The results with
DCF and MMD are indicated with the solid and open symbols,
respectively. Note that the soft lags in the 1.5--2~keV are derived
from both the LECS (circles) and the MECS (squares). }
\label{fig:lage}
\end{figure}

The results are shown in Figure~\ref{fig:lage} for the 1996 \#2, 1999
\#1 and 1999 \#2 flare. The energy-dependence of soft lags for the
three flares of 1997 can not be determined because of the short 
duration and the small soft lags. Further observations with higher
time-resolution may determine the energy-dependence of very small time 
lags which probably associate with the flares of short duration.  

The soft lags of the variations of the soft \xrs\ with respect to
those of the hard \xrs\ are thought to be due to the energy-dependence
of synchrotron cooling time, $t_{\rm cool}$, of relativistic
electrons responsible for the emission in the studied energy bands,
which results in the dependence of $t_{\rm cool}$ on the emitting
photon energies (see \S\ref{sec:disc:implication}). Therefore, the
observed energy-dependence of the soft lags can be fitted with
Equation~[\ref{eq:tcoolsykev}], which will give a constraint to the
magnetic field $B$ and the Doppler factor $\delta$ of the
emitting regions in the form of $B\delta^{1/3}$ (see also 
Equation~[\ref{eq:softlag}]). The best fits are shown
in Figure~\ref{fig:lage} with dashed lines and give $B \delta^{1/3} =
0.40$, 0.37 and 0.40 Gauss for the 1996 \#2, 1999 \#1 and \#2 flares,
respectively. 
 
For the 1996 \#2 flare, our analysis reveals that, with respect to the
4--10~keV band, the soft lag of the 0.1--0.5~keV band is significantly
smaller than those of other low energy bands (see
Figure~\ref{fig:lage}a), this suggests that the 0.1--0.5~keV photons
may lead the other low energy photons. This point has already been
noted in the \lcs\ (see \S\ref{sec:timing:lc:96}). Therefore, in this
flare \pks\ may show evidence of hard lag in the soft X-rays, and 
soft lag in the hard X-rays. This is the first time that a blazar
shows opposite behavior of the inter-band time lags in the different
energy bands, which demonstrates the complexity of the variability of
the source. 

\section{Spectral Analysis}
\label{sec:spec}

  Spectral analysis for the 1996 and 1997 data sets was presented
in Giommi \etal (1998) and Chiappetti \etal (1999), respectively, 
but both paper did not examine detailed spectral evolution. 
To reveal in detail the spectral variability of \pks , in particular
during individual flares and between the flares, we perform again a
time-resolved spectral analysis for the 1996 and 1997 data sets, along
with the new 1999 data set. To do so, we divide each data set into  
sub-segments on the basis of single \textit{Beppo}SAX orbit or a
multiple of it to reach sufficient statistics for each segment. 
The main goal of such analysis is to produce homogeneous spectral
information of \pks\ to study flare-dependent
spectral evolution, which is, together with temporal variability,
essential to understand and constrain the jet physics of the source.

Due to remaining calibration uncertainties, in the spectral analysis
LECS and MECS data have been considered only in the ranges 0.1--3~keV and
1.6--10~keV, respectively. These ``good'' spectral channels (i.e.,
channels 11--285 for the LECS, and 36--220 for the MECS) are further
rebinned using the grouping templates available at \textit{Beppo}SAX
SDC. Because of the very steep spectral shape of this source, we
modified the MECS grouping template above $\sim 7$~keV in order to have
sufficient photons in each new bin.

Spectral analysis is performed with XSPEC version 11.1 package,
using the latest available calibration files (i.e., response matrices
and effective areas) and blank sky fields. The background spectra
for each observation have been evaluated from these blank field event
files by applying the same extraction region as the source spectra,
i.e., the extraction position and radii are identical for both the source
and the background spectra.

The Galactic absorption column in the direction of \pks\ ($N_{\rm H}
= 1.36 \times 10^{20}$ cm$^{-2}$; Lockman \& Savage 1995) is used during
the spectral fits.  
Due to a small mismatch in the cross-calibration between the LECS and
the MECS, we add a free multiplicative constant factor in the fit
procedure to allow a variable normalization between the LECS and
the MECS data. The acceptable range of this constant is 0.7--1.0
which depends mainly on the source position on the detectors
(see Fiore, Guainazzi \& Grandi 1999). 

\subsection{Spectral Models} \label{sec:spec:model}
 
The observed X-ray spectra of blazars are usually thought to be
represented by a single power law model.
However, since \pks\ (and other TeV sources) tends to show
continuously downward curved spectra, a model consisting of a single
power law plus a free low energy absorption or fixed at Galactic value
generally gives a very poor fit to
the LECS$+$MECS spectra of most segments. Therefore, we will not discuss in
detail spectral fits with a single power law. We then test a spectral
model consisting of a broken power law with free absorbing column
density free or fixed at Galactic value. This model gives statistically
acceptable minimum reduced $\chi^2$ in each of the fits. 

We further apply a continuously curved model (e.g., 
Tavecchio, Maraschi, \& Ghisellini 1998) as formalised by Fossati
\etal (2000b). This model is described as 
$$
F(E) \;=\; K\; E^{-\alpha_{-\infty}}\; \left[ 1 + \left(\frac{E}{E_{\rm
B}}\right)^f \right ]^\frac{\alpha_{-\infty}-\alpha_{+\infty}}{f} \,,
$$
where $\alpha_{-\infty}$ and $\alpha_{+\infty}$ are the asymptotic values 
of spectral indices for $E \ll E_{\rm B}$ and $E \gg E_{\rm B}$, 
respectively, while $E_{\rm B}$ and $f$ determine the scale length of
the curvature. This model is constrained by four parameters
($\alpha_{-\infty}$, $\alpha_{+\infty}$, $E_{\rm B}$, $f$).

This function is better understood in terms of the local spectral indices 
at finite $E$ instead of the asymptotic ones.
The available parameters of this spectral model are then re-expressed as
$(E_1, \alpha_1, E_2, \alpha_2, E_{\rm B}, f)$ instead of
($\alpha_{-\infty}$, $\alpha_{+\infty}$, $E_{\rm B}$, $f$), where
$\alpha_i$ is the spectral index at energy $E_i$ $(i=1,2)$. 
The relationship between these two sets of parameters
can be obtained by differentiating the above function to obtain the local
spectral indices (for more details see Fossati \etal 2000b). As there 
are two more parameters in the new description of this model we have to
fix one for each of the pairs $(E_1, \alpha_1)$ and $(E_2, \alpha_2)$ in
order to have a meaningful use of this spectral description. It is
interesting to note that it is this degeneracy that determines some
advantages of this improved model.
In particular, it allows us to derive a local spectral index by
setting $E_{\rm i}$ at a preferred value, or the energy of a certain
spectral index by setting $\alpha_{\rm i}$ at the desired value. 
The most important aspect is the possibility of estimating the position
of the peak (as seen in $\nu$--$\nu F_{\nu}$, i.e., $E$--$EF(E)$, 
representation) of the 
synchrotron component, $E_{\rm peak}$ (if it falls within the observed
energy band ): one of the crucial physical quantities in blazar 
modeling. This can be obtained by setting one spectral index, i.e., 
$\alpha=1$, and leaving the corresponding energy $E$ free to vary in the
fit: the best fit value of $E_{\alpha=1}$ gives $E_{\rm peak}$. 

\subsection{Results}\label{sec:spec:results}

As the curved model presents more useful information than the broken
power-law model does, our following discussions focus on the results
from the former model although the latter also gives statistically
acceptable fits.

Following Fossati \etal (2000b), to estimate the really interesting
parameters, i.e., spectral index at preferred energies and
peak energy, we first estimate proper value of parameter $f$ for
each epoch. This is done by allowing all parameters but $N_H$ to
freely vary in the fit of the time-averaged spectra extracted from each
data set. The value of $f$ corresponding the best fit is $2.34\pm0.52$,
$1.82\pm0.37$, and $0.91\pm0.86$ for the 1996, 1997 and 1999 data sets, 
respectively. So $f=2$ is
fixed for the 1996 and 1997 data sets, and $f=1$ for the 1999 one. 
With $f$ fixed, the spectrum of each segment has been fitted a few
times in order to derive spectral indices at desired energies and 
E$_{\rm peak}$ (The errors are derived with $68\%$ confidence level
for one parameter of interest). The fluxes have been computed in the
0.1--2~keV and the 2--10~keV energy ranges and corrected for Galactic
absorption. 

\subsubsection{Peak Energies and Spectral Variability}

\begin{figure}
\epsscale{1.0}
\plottwo{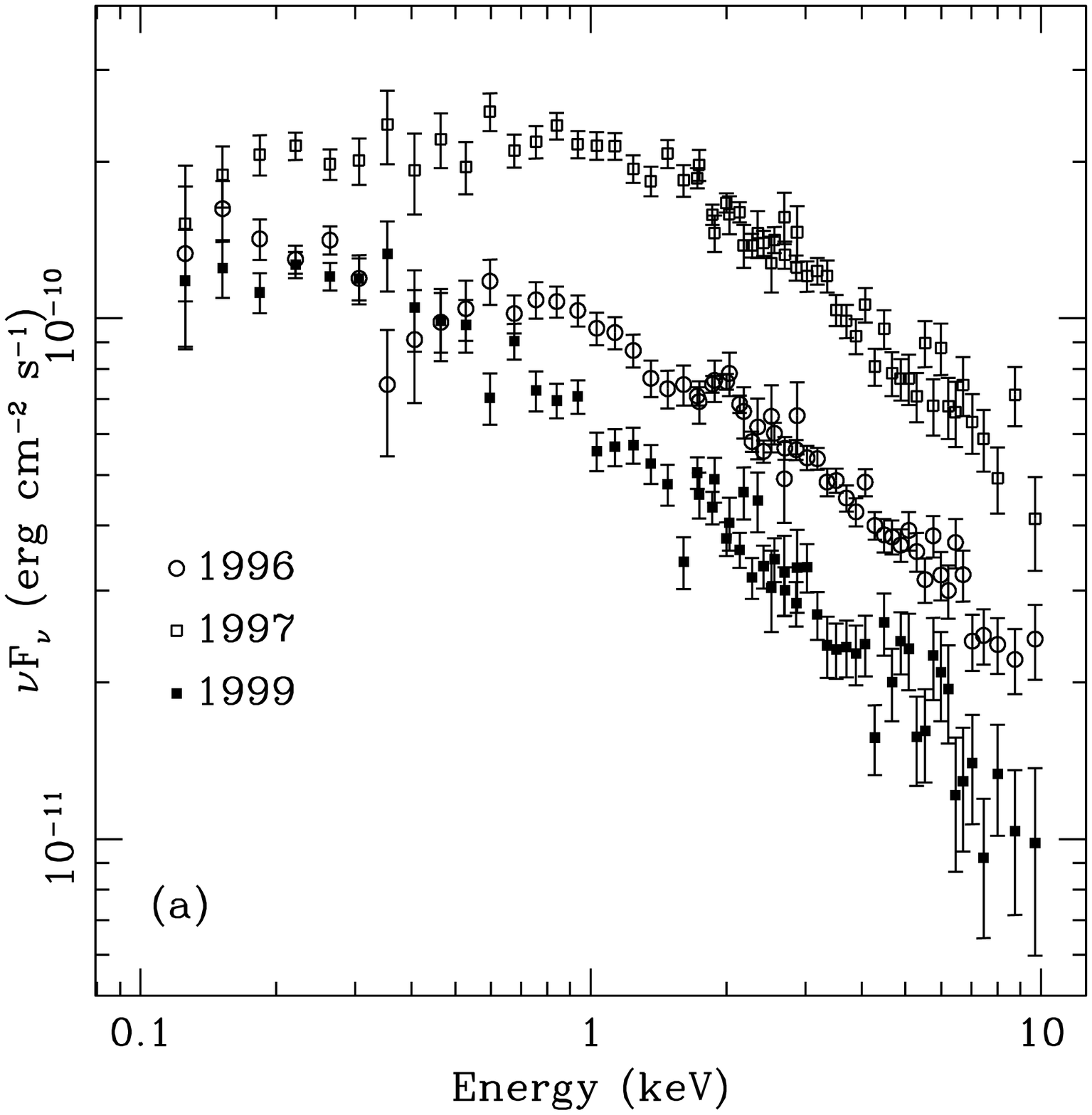}{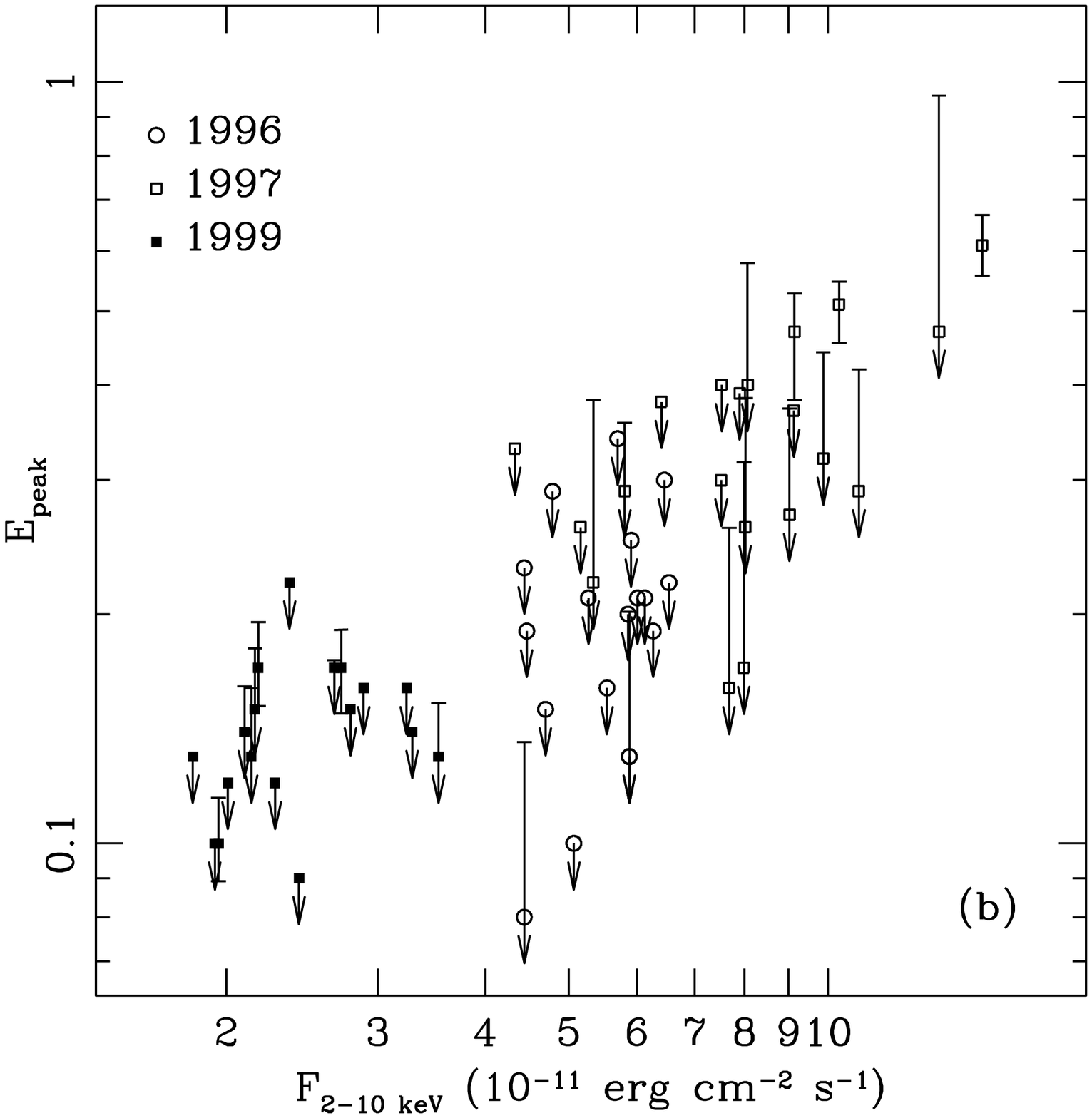}
\caption {\footnotesize (a) Deconvolved $\nu F_{\nu}$ spectrum derived
from the segment with maximum flux among each data set, which
corresponds to segment \#10 of 1996, \#2 of 1997, and \#5 of 1999,
respectively. (b) Peak energies of synchrotron component versus the
2--10~keV fluxes. A correlation between them is seen, albeit dominated
by the upper limits of $E_{\rm peak}$ (see text). These plots
demonstrate spectral evolution characterized by the shifts of 
synchrotron peak energies.  
}
\label{fig:spec:xsed}
\end{figure}

The source revealed a large flux variability, a factor $\sim$~10 in
the 2--10~keV energy band, ranging from $15.1 \times 10^{-11}$ (1997
maximum) to $1.8 \times 10^{-11}$ (1999 minimum) erg cm$^{-2}$
s$^{-1}$. Such flux changes were accompanied by changes of 
spectral curvature characterized by the shifts of the peak energy of 
synchrotron component as seen in the $\nu - \nu F_{\nu}$ diagram.
Figure~\ref{fig:spec:xsed}a shows the spectral energy distribution derived
from the segment with the maximum flux of each
data set. Unfolded spectra have been corrected for low energy
absorption assuming $N_{\rm H}$ equal to the Galactic value. Spectral
convexity and shift of peak energy are apparent, while $E_{\rm
peak}$ of the 1996 and 1999 spectra may be below the X-ray band. It is
worth noting the difference between the 1996 and 1999 spectra: both 
show similar fluxes in the soft energy band, while the 1996 spectrum
shows higher flux than the 1999 one at energies higher than 0.5~keV.

We show in Figure~\ref{fig:spec:xsed}b $E_{\rm peak}$ versus the 
2--10~keV flux. There is a suggestion that $E_{\rm
peak}$ shifts to higher energy with increasing flux, but this trend
is clearly dominated by the upper limits of $E_{\rm peak}$. $E_{\rm peak}$ 
carries direct information on the source's physical properties, however,   
they are very weakly constrained due to two facts: the signal-noise
ratio is not good enough and, more importantly, $E_{\rm peak}$ of \pks\ is
around or below the \sax\ lower energy limit, 0.1~keV. A proper
determination of $E_{\rm peak}$ in \pks\ will depend on the combined UV
and soft X-ray spectrum together.
In fact, among the 56 data segments, in only three cases can the curved
model determine $E_{\rm peak}$ with reasonable accuracy, these 
correspond to the brightest segments in 1997. In most cases only upper
limit on $E_{\rm peak}$ could be determined. 

\begin{figure}
\epsscale{1.0}
\plottwo{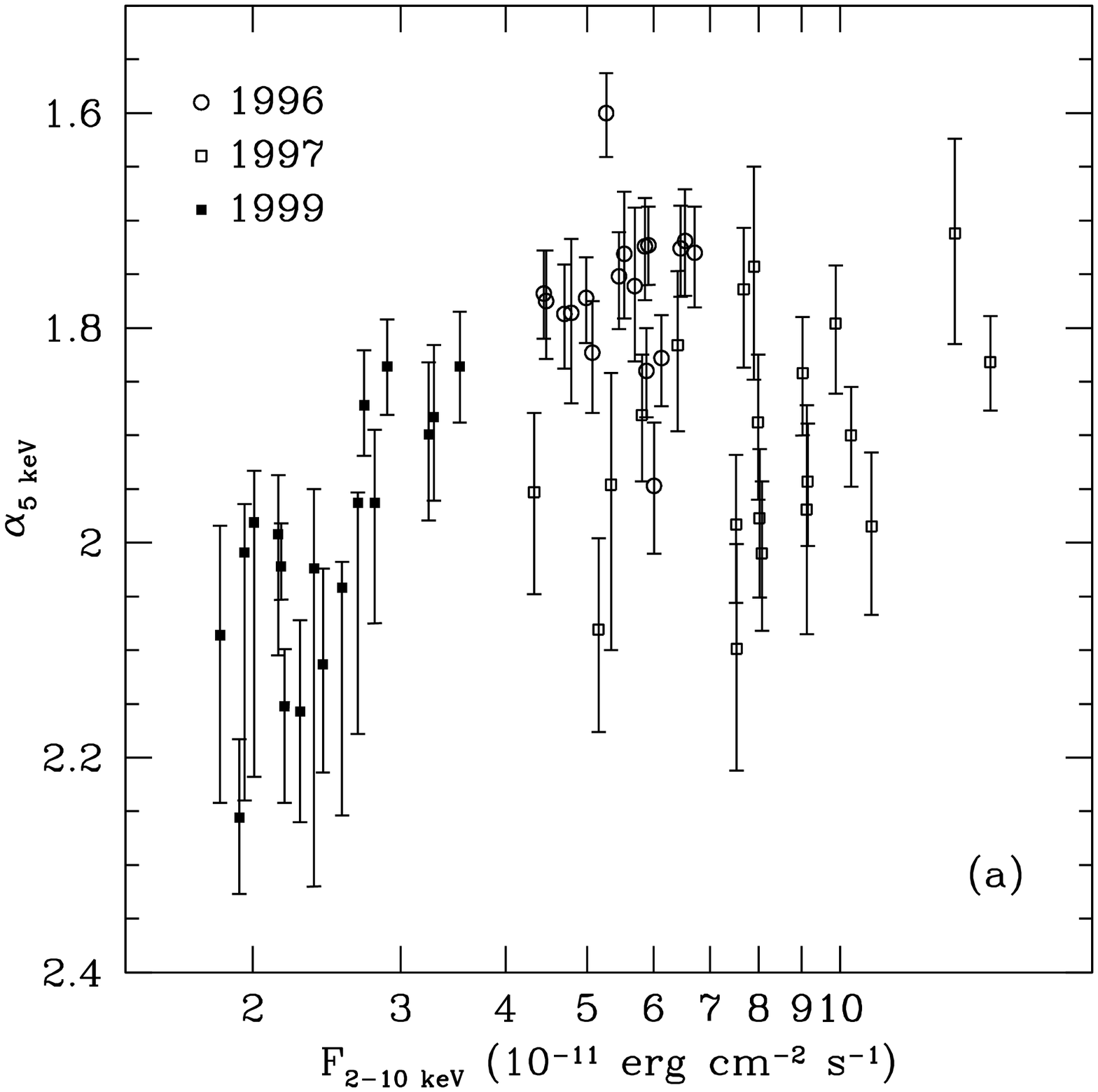}{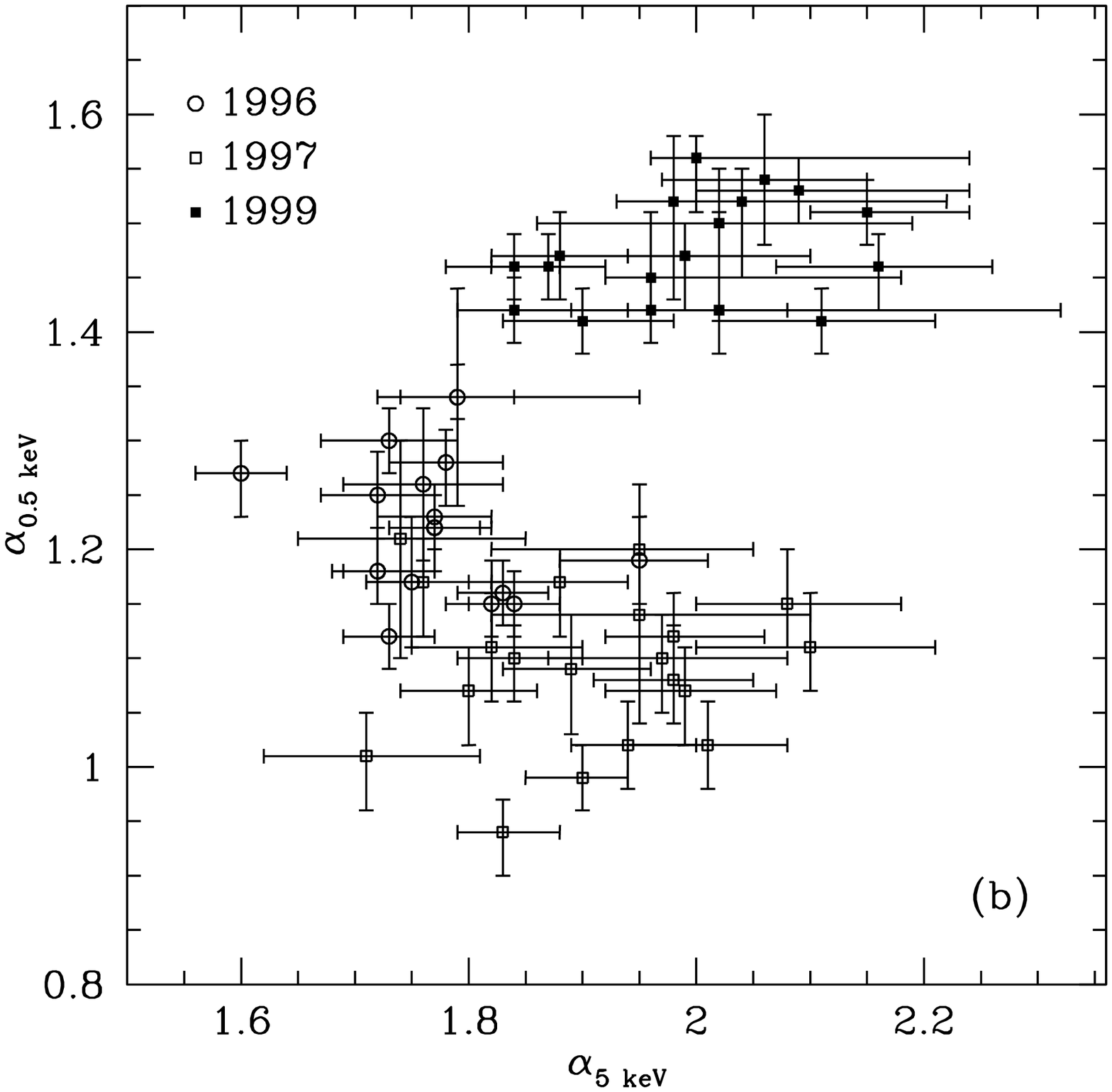}
\caption {\footnotesize (a) Spectral index at 5~keV is plotted against
the 2--10~keV flux. (b) Spectral index at 0.5~keV versus spectral index at
5~keV. }
\label{fig:spec:index}
\end{figure}

In Figure~\ref{fig:spec:index}a we show the relationship of the
spectral index $\alpha$ at 5~keV and the flux. One can see that there
is little correlation between them. 
In Figure~\ref{fig:spec:index}b we have plotted the spectral index at
0.5~keV against the spectral index at 5~keV. Again this plot does not
show any clear correlation, in particular each data set is
clustered. These results indicate that, on the long epochs, the spectral
changes in the soft and hard energy range do not depend on each
other. However, for each data set, this inference could be biased 
because of large errors on the spectral index. 
As a comparison, we compute the unweighted average spectral index at
0.5 and 5~keV for the three data sets:  
$<\alpha_{\rm 0.5keV}>^{1996} = 1.22\pm0.07$,
$<\alpha_{\rm 5keV}>^{1996} = 1.77\pm0.07$;
$<\alpha_{\rm 0.5keV}>^{1997} = 1.09\pm0.07$, 
$<\alpha_{\rm 5keV}>^{1997} = 1.91\pm0.11$;
$<\alpha_{\rm 0.5keV}>^{1999} = 1.47\pm0.05$; 
$<\alpha_{\rm 5keV}>^{1999} = 2.00\pm0.12$.
The average spectral curvature from 0.5~keV to 5~keV is: 
$\Delta\alpha^{1996}=0.55$,
$\Delta\alpha^{1997}=0.82$, 
$\Delta\alpha^{1999}=0.53$.
The large spectral steepenings inferred in the X-ray band
should be ascribed to the large steepenings of the (radiating) electron
energy distribution because of fast cooling of high energy electrons.

\subsubsection{The Sign of Time Lags} \label{sec:spec:lag}

\begin{figure}
\epsscale{0.4}
\plotone{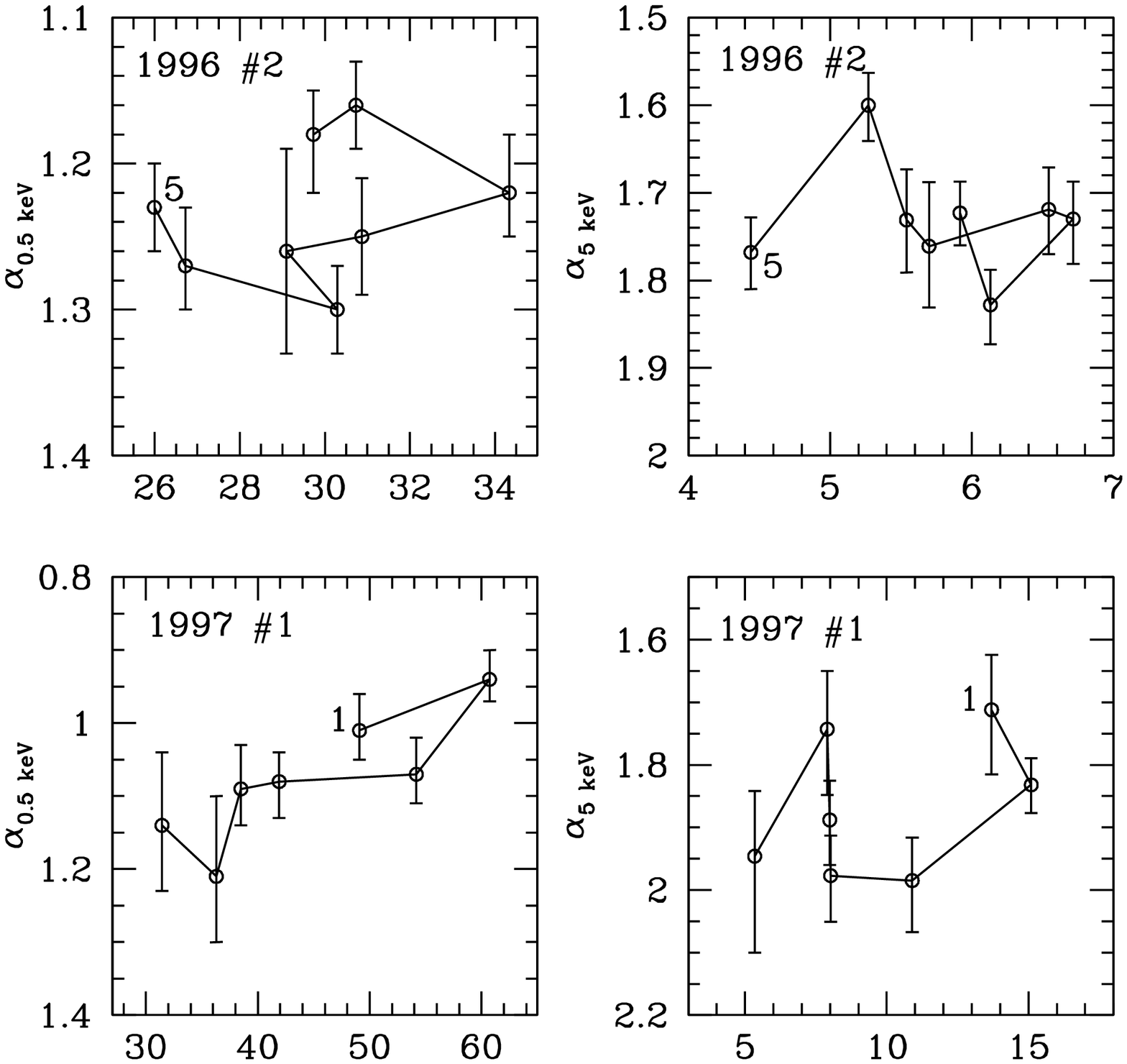}\\
\vspace{0.5cm}
\plotone{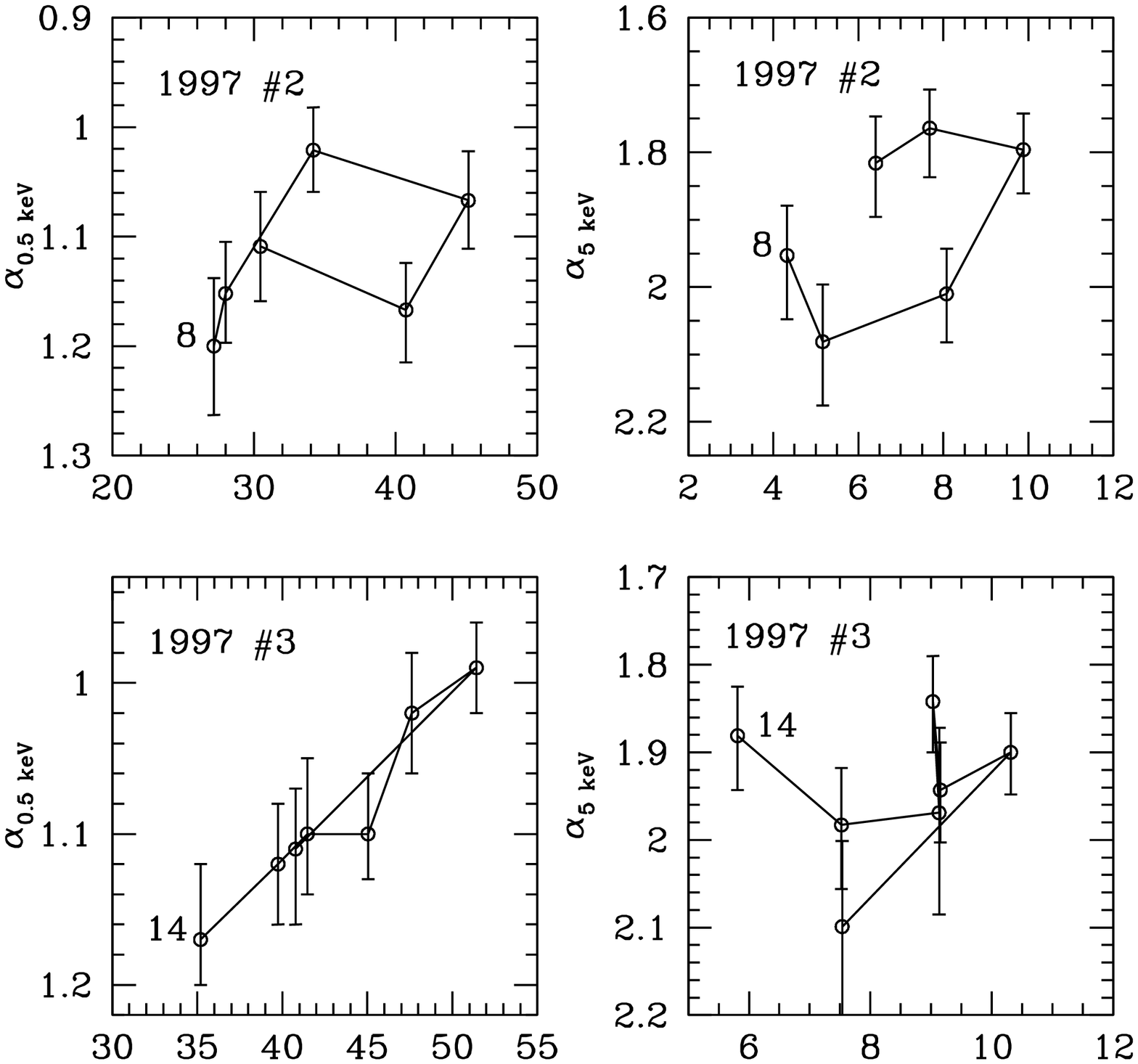}\\
\vspace{0.5cm}
\plotone{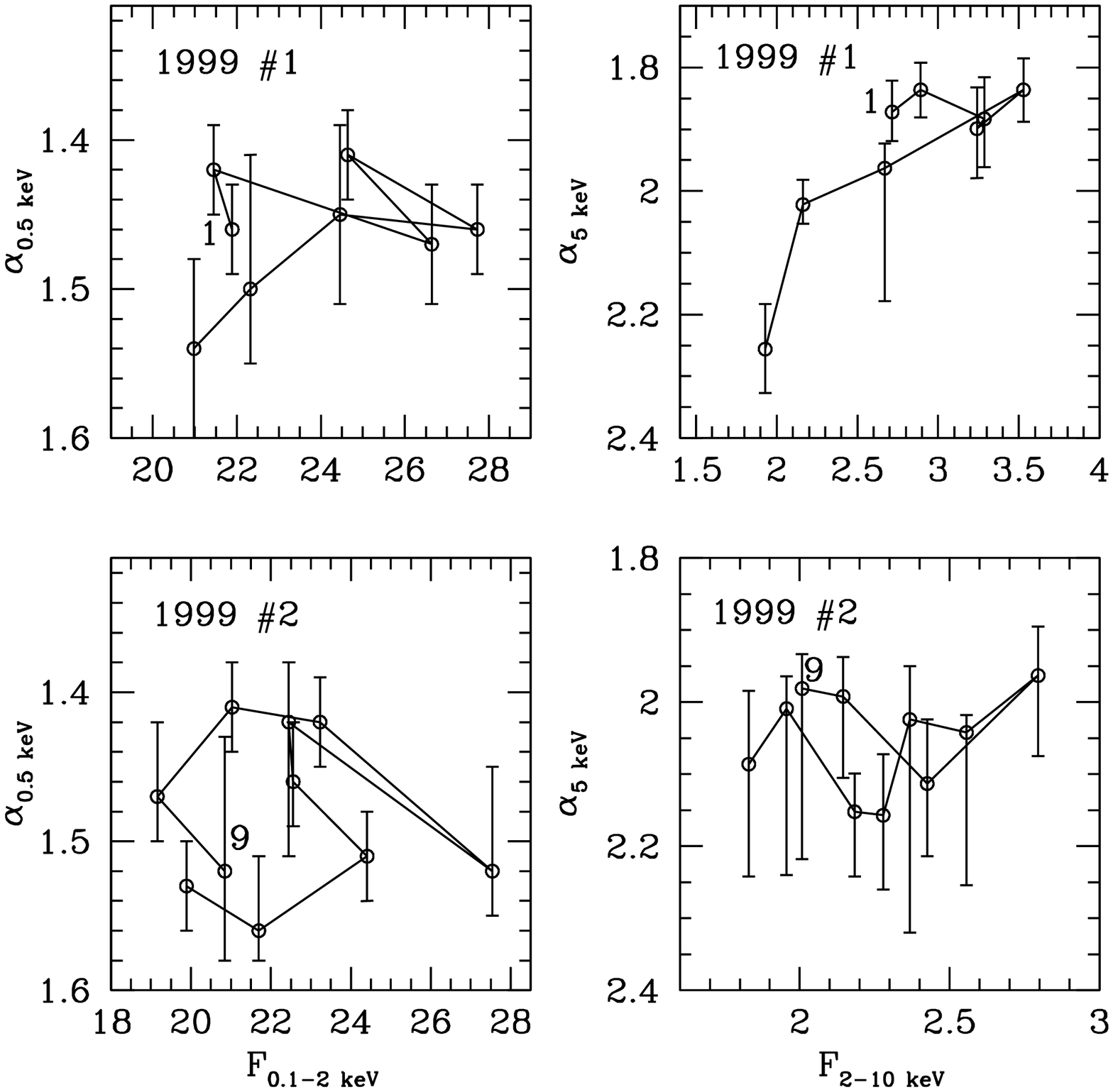}
\caption { \footnotesize Spectral index at 0.5~keV (left) and 5~keV
(right) as a function of the 0.1--2~keV and 2--10~keV flux (in unit of
$10^{-11}$~erg~cm$^{-2}$~s$^{-1}$), respectively, to illustrate the sign
of time lags. The starting point of each loop is indicated with the
segment number of each observation, and the evolutionary direction follows
the connected line. }
\label{fig:spec:loop}
\end{figure}

For each flare, spectral variability versus flux is
further studied with the diagram of $\alpha$--$F$ whose property is
that a soft and hard time lag, during a flare, can be qualitatively
determined with a clockwise and anti-clockwise loop, respectively.
In Figure~\ref{fig:spec:loop} we have plotted the spectral index at
0.5~keV and 5~keV versus the 0.1--2~keV and the 2--10~keV
absorption-corrected flux, respectively. The first point of each loop, 
representing the segment number of each data set, has been numbered. 
As a further aid to clarity, points are sequentially linked with a solid
line in its original time order. The overall behavior of each flare is
briefly summarized below:

1996 \#2---A normal clockwise loop, corresponding to soft lag, is
indicated by the $\alpha_{\rm 5keV}$--$F$ loop, but showing small changes
of spectral index, $\Delta \alpha_{\rm 5keV} \simeq 0.1$, apart from the
second point which hardens from $\alpha \simeq 1.8$ to $\alpha \simeq
1.6$. However, an anti-clockwise loop, corresponding to hard lag, is
indicated by the $\alpha_{\rm 0.5keV}$--$F$ relation. This is the
first time that $\alpha$--$F$ loop of a TeV source shows opposite
behavior in the soft and hard X-ray bands. 
Note that such behavior is in agreement with the results obtained
from the cross-correlation analysis (\S\ref{sec:timing:lag:lage}) and the
visual inspection of the 1996 \lcs\ (\S\ref{sec:timing:lc:96} and
Figure~\ref{fig:lc:96}).

1997 \#1---Since the increasing phase was not sampled, the spectral
behavior is mainly determined by the flare peak and decaying phase,
but a clockwise track is clear in both loops, indicating soft lag
during this flare. The change in spectral slope is relatively large,
with $\Delta \alpha_{\rm 0.5keV} \simeq 0.2$, and $\alpha_{\rm 5keV}
\simeq 0.3$.

1997 \#2---The 2--10~keV flux doubled and spectral index at 0.5 and
5~keV change by $\Delta \alpha_{\rm 0.5keV} \simeq 0.2$ and $\Delta
\alpha_{\rm 5keV} \simeq 0.3$, respectively. The loops are quite 
well-defined with a quasi-circular form. The rise and decay phases
follow clearly different tracks. It is noticeable that we again find
opposite behavior of the loops, but in the opposite sense to the
behavior found in the 1996 \#2 flare. The $\alpha_{\rm 0.5keV}$--$F$
relation follows a normal clockwise, while $\alpha_{\rm 5keV}$--$F$
follows an anti-clockwise loop. 

1997 \#3---This might not be a single flare, so both loops might not
have direct meanings. The spectral changes are about 0.2 in both
energies. 

1999 \#1---Apart from the second point in the $\alpha_{\rm
0.5keV}$--$F$ loop, which corresponds to an abrupt rise in the soft
energy but not in the hard energies (see Figure~\ref{fig:lc:99}), the
$\alpha$--$F$ relation might follow clockwise loops in both
energies. The spectral index changes little in the rising phase then
softens dramatically by $\Delta \alpha_{\rm 5keV} \simeq 0.4$ during
decreasing phase, the maximum changes of spectral index obtained in
this work. This feature agrees with the flare aspect: a slow rising  
phase followed by a rapidly decaying phase.

1999 \#2---The relation of index and flux again follows a clockwise
loop, albeit with less significance. The spectral index with large
errors changes by $\Delta \alpha_{\rm 5keV} \simeq 0.2$ during this
flare. However, both loops are not well-defined due to swings feature
of the \lc\ during the peak phase. For example, the visible softening
in the third point of $\Delta \alpha_{\rm 5keV}$--$F$ loop corresponds
to a drop of flux during the rising phase of this flare.

Finally, we notice, in most cases, that the changes of $\alpha_{\rm
0.5keV}$ are smaller than those of $\alpha_{\rm 5keV}$, and in 
general, that clockwise loops are in agreement with the fact that the
DCF/MMD gives soft lags.

\section{Discussion}\label{sec:disc}

In this section, we summarize and compare the most important aspects
about the variability of \pks\ emerged from this and other
works (\S\ref{sec:disc:summary}). The variability of Mrk~421 and Mrk~501
are also compared. The implications of our results are
explored on the basis of the internal shock taking place in the jet
(\S\ref{sec:disc:implication}). 

\subsection{Summary and Comparison with Other Observations}
\label{sec:disc:summary}

\subsubsection{Power Spectral Density}\label{sec:disc:summary:psd}

Power spectral density or structure function can identify the nature
of the variability of a source. The information carried by the PSD has 
not yet been well explored in blazars. In the X-rays, PSD has been studied
recently in some details for the three TeV sources. 

The PSDs (\S\ref{sec:timing:psd}) and SFs 
(\S\ref{sec:timing:sf}) derived in this work with the 3 \sax\ data 
sets indicate that \pks\ shows \textit{red noise} variability. The
steep power-law slopes (PSD) of $\sim$ 2--3 further suggest that
the source shows shot noise variability.
PSD or SF of this source was previously studied in the X-ray and
optical bands. Tagliaferri \etal (1991) firstly analyzed the 3 EXOSAT
short observations (length of $\sim$ 10 hours each). In the 1--6~keV
energy band, an average PSD power-law slope of $\alpha \simeq 2.5\pm0.2$
was found, which, however, reduced to $\alpha \simeq 1.9\pm0.4$ after the
removal of the linear trend in the \lcs . The SF derived from the
EXOSAT data was shown in Paltani (1999). Hayashida \etal
(1998) derived an average PSD of \pks\ using 4 GINGA observations
(length of about 1--2 days each; \lcs\ were presented in Sembay \etal
1993). The PSD slope was $2.83^{+0.35}_{-0.24}$. Very recently,
Kataoka \etal (2001) analyzed the ASCA and the RXTE data sets. They
reported steep power-law slopes of $\sim$~2--3. They also reported a break
frequency of $(1.2\pm0.4) \times 10^{-5}$~Hz for the PSD derived with
the about 12 day long RXTE data set. In the optical, Paltani \etal
(1997) studied the short-term variability of \pks\ and
derived the SFs based on the dense-sampled 15 nights data. They found
the optical PSD (from SFs) was well described by a power-law slope of
2.4. In summary, the slopes of PSDs of \pks\ obtained by other
authors are essentially compatible with those of the present analysis,
that is, this source shows the power-law PSD with slopes of
$\sim$~2--3 in the studied frequency range.

PSDs of Mrk~421 and Mrk~501 were recently studied by Kataoka \etal 
(2001) with the ASCA and the RXTE data sets. These two sources also show
power-law PSDs with slopes of $\sim$~2--3. For the long data set, Mrk~421
and Mrk~501 also show a break frequency of $(9.5\pm0.1) \times
10^{-6}$~Hz and  $(3.0\pm0.9) \times 10^{-5}$~Hz, respectively. 
Cagnoni, Papadakis \& Fruscione (2001) studied the EUV variability of
\mkn . The EUV PSD is well represented by a power-law slope of 
$2.14\pm0.28$, breaking at $\sim$~3 days. 

Finally, we point out that the PSDs of the TeV sources are different
from those of Seyfert galaxies and Galactic black hole candidates 
(GBHCs). The PSDs of the latter can be represented by a
power-law model in the high frequency range, while the slopes are
rather distinct from those of the TeV sources. It is noticeable that the
TeV sources tend to show steeper slopes ($\alpha \sim$~2--3) than Seyfert
galaxies and GBHCs do ($\alpha \sim$~1--2, see, e.g., Hayashida \etal 
1998). This difference presumably
indicates the different origins and/or the sites for the X-ray
production. The X-ray emission of Seyfert galaxies and GBHCs are thought
to be due to thermal Comptonization of the soft photons (from the
accretion disk) by the hot, thermal electrons (from coronae) (e.g., Nowak
\etal 1999; Nandra \etal 2001), while the synchrotron emission from a
beamed jet is the most probable origin of the X-rays in the TeV blazars.

\subsubsection{Timescales}\label{sec:disc:ts} 

The X-ray observations have revealed that the TeV sources exhibit
successive flares. A natural question is whether there is a characteristic
timescale related to this phenomenology. However, it is important to
first  note that most of the observed flares show complicated 
behaviors, e.g., different durations and amplitude, multiple 
sub-structures with shorter timescales.    

Timescales of the X-ray variability are important parameters because
they carry direct information about the physical processes in the
vicinity of the central black holes. PSD technique is basically used
to detect these timescales which can be inferred from the presence of
the low and the high frequency ``breaks'' or ``periodicities'' of the PSD. 
However, because some issues, such as irregularly sampled astronomical
data and inefficiently sampled length of the observations, plague the 
Fourier-based PSD analysis, the inference of \tss\ from PSD becomes 
impossible in most cases. The SF method, partially minimizing these 
issues, may have these advantage to give an estimate of the \tss . SF 
analysis (\S\ref{sec:timing:sf}) showed that the X-ray variability of
\pks\ has two common features: a steep slope ($\sim$~1.4) at the 
short \ts\ range and a ``turn-over'' 
\footnote {Although the presence of this
``turn-over'' should in principle correspond to the low frequency
``knee'' of the PSD, our analysis does not show this correspondence,
probably due to the limitation of the data (see
Appendix~\ref{sec:app:sf}). }
at $\tau_{\rm max}$ which is indicative of the \ts\ characteristic of
the durations of the individual flares. $\tau_{\rm max}$ measures the
maximum correlated \ts\ of the system, i.e., whatever the origin and
the nature of the variations, the \ts\ during which ``memory'' of the
variability can be maintained. 

It is not yet clear what determines the observed (rising and
decay) \tss\ and the shape of a flare. There are several physical \tss\
involved, including cooling and acceleration \ts\ of relativistic
electrons, $t_{\rm cool}$ and $t_{\rm acc}$, electron injection (shock
crossing) timescale, $t_{\rm inj}$, and light crossing timescale, $t_{\rm
crs}$. These timescales may play a joint role (see, e.g., Chiaberge \&
Ghisellini 1999). If $t_{\rm cool}$ is shorter than $t_{\rm crs}$, the
variability of the flare is dominated by $t_{\rm crs}$, and the rising and
the decaying phase of the observed flare should have symmetric profiles
whose \ts , $\tau_{\rm f}$, reflects the size of the emitting region
(\S\ref{sec:disc:implication}; Equation~[\ref{eq:esize}]). In such a
case, we have the relationship of $\tau_{\rm max} = \tau_{\rm f} =
t_{\rm crs}$. This scenario is plausible to apply      
to the TeV sources in the hard X-ray energy band which corresponds
to the shortest $t_{\rm cool}$ range of relativistic electrons 
responsible for the emission.  Within the jet geometry 
(\S\ref{sec:disc:implication}; Equation~[\ref{eq:edistance}]), the site
of the emitting region can be estimated. This idea is supported by the 
observed quasi-symmetric profiles of some flares, but this feature is
usually broken by unclear factors. However, if the system is observed
at the energy where $t_{\rm cool}$ is larger than $t_{\rm crs}$, the flare
will show slower decay phase than the rising phase. In such a case,
$\tau_{\rm f}$ of the rising phase, which is smaller than $\tau_{\rm
f}$ of the decaying phase, reflects the size of the emitting 
region. These two cases are viable if $t_{\rm inj}$ is smaller than
$t_{\rm crs}$. If $t_{\rm inj} > t_{\rm crs}$, a plateau should appear
during the flare. 

However, the determination of $\tau_{\rm max}$ from an observed \lc\ is 
rather ambiguous. Apart from the underlying emission component which
may exist, the obvious scenario is that an observed flare can be a
superposition of several individual flares. These complexities 
prevent us from getting physical meaning of $\tau_{\rm max}$. Moreover, 
each \lc\ consists of successive flares that show different amplitude
and duration. SF analysis, by definition, is dominated
by the flares with larger amplitude and longer duration, and the more
rapid shots are suppressed. Therefore, $\tau_{\rm max}$ only
carries information on the flares with larger amplitude.

In the literature, $\tau_{\rm max}$ is assumed to be a constant
as a common feature of a source (e.g., Kataoka \etal 2001). In fact,
this statement is oversimplified, because it means that the site
and the size of the emitting region is constant with time. Of course,
we believe that $\tau_{\rm max}$ should be in a limited range as
suggested by our results, which show that $\tau_{\rm max}$ varies
by almost an order of magnitude, ranging from $\sim 10^4$~s to about
one day (\S\ref{sec:timing:sf}; Table~\ref{tab:sf}). This is in
agreement with the visible differences of the flare duration between
different observations. We also notice similar phenomenology in \mkn\
observed with \sax , $\tau_{\rm max}$ ranges from $\sim 2 \times
10^4$~s to $\sim 7 \times 10^4$~s (Zhang 2000). 

$t_{\rm crs} > t_{\rm cool}$ at the energies under consideration
also implies that the symmetric profile of the observed X-ray flares 
traces the time-dependence of the electron acceleration mechanism
(modulated by light-crossing delays). The importance of the
light-crossing effects suggest the manner in which the electrons are   
accelerated and the density profile of the electrons are both important
factors in determining the shape of the flare (Lawson, M$^{\rm c}$Hardy 
\& Marscher 1999). The flare, corresponding to the variation in the
number of relativistic (radiating) electrons, could correspond either
to the variation in the efficiency of acceleration of relativistic
electrons or to the density of electrons that are accelerated. As an
example of a reasonable physical situation, Lawson, M$^c$Hardy \&    
Marscher (1999) simulated a flare via synchrotron self-Compton 
(SSC) emission, caused by a square-shock excitation passing through a
spherical emitting region with electron density and magnetic field
falling exponentially with distance from the center. The simulated
light curve fits quite reasonably the symmetric X-ray flare of 3C~279
observed with RXTE. Thus the flare can be reproduced with a relatively
simple geometry and the cooling timescales do not dominate the shape of
the light curve.

\subsubsection{Time Lags} \label{sec:disc:summary:lag}

The cross-correlation analysis focusing on the ``single'' flares 
(\S\ref{sec:timing:lag}) showed that the inter-band soft lags of \pks\  
differ from flare to flare. The value of the soft lag may relate
with the duration of the flare, in the sense that a flare with longer
duration may show larger soft lag. The energy-dependence of the soft
lags was suggested for the three flares with longer duration 
(Figure~\ref{fig:lage}). It is necessary to point out that,  given the
large uncertainties on the determination of the time lags, these
results are still indicative and yet to be further tested with higher
quality data. However, in most cases such uncertainties might be
intrinsic to the complicated behaviors of the observed flares 
themselves rather than to the sampling problems and photon statistics,
and to the analysis techniques. 

It is important to compare our results with others, especially those  
obtained with other satellites. A long look of \pks\ with ASCA seems
to reveal larger soft lags (Tanihata \etal 2001) in two flares. If the
correlation between the soft lag and the duration of the flare were
real, the results with ASCA would be explained because the flares show
larger duration. The exception is the third flare that is much more
complicated.  Observations of Mrk~421 and Mrk~501 with \sax\ and ASCA
also showed that the time lags are different from flare to flare. The
results for Mrk~421 with \sax\ and ASCA are essentially consistent with  
each other (Zhang 2000; Takahashi \etal 2000), both the soft and hard
lags were found. A hard lag of about 2 hours was reported for
Mrk~501 (Tanihata \etal 2001). Energy dependence of time lags has also
been reported for Mrk~421 and Mrk~501 (Zhang 2000; Tanihata \etal 2001). 

The latest results obtained with XMM--Newton are questioning the
detection of the time lags with \sax\ and ASCA. Due to the highly
eccentric $\sim$~48 hours orbit, uninterrupted data (of one flare) can
be obtained. Therefore, it is believed that the results derived from
such data are more reliable. Edelson \etal (2001) and Sembay \etal
(2002) reported that no inter-band time lags exist in two (\pks ) and
three (\mkn ) data sets observed with XMM--Newton. They further
proposed that the soft or hard lags detected with ASCA and \sax\ could
be biased by the periodic interruption due to the  
Earth-occultation. We, however, notice that there is an important 
difference between the XMM-Newton data and those of ASCA and \sax
: the available XMM-Newton observations did not show flares with
large duration comparable with those of \sax\ and ASCA. It is necessary to
point out that the \sax\ and the ASCA data also showed time lags
close to zero. Furthermore, as mentioned in \S\ref{sec:timing:lag},
CCFs calculated from more than one flare together may produce spurious
zero lag. We have noticed this issue for the long look of \mkn\ with 
\sax\ in May 2000 which involved several significant flares: the CCF
of the entire observation suggested zero lag. However, when each
single flare was cross-correlated, different results were derived and
both soft and hard lags were found (Zhang 2000). Variable time lags
are physically expected if the parameters of the emitting region change
from flare to flare (see \S\ref{sec:disc:implication}). 
Finally, we point out that, due to the complexities of the observed light
curves, there exist ambiguities in defining a flare.

\subsubsection{Synchrotron Peak Energies}
\label{sec:disc:summary:epeak}

Being $E_{\rm peak}$ close to the \sax\ lower energy limit, the 
present analysis can not give reliable results about the evolution
of $E_{\rm peak}$ of \pks . However, an indicative correlation between
$E_{\rm peak}$ and the flux (Figure~\ref{fig:spec:xsed}b) is still 
meaningful, indicating that $E_{\rm peak}$ may shift to higher
energies with increasing flux.  

Such a correlation was found with higher confidence in Mrk~421 and 
Mrk~501. Fossati \etal (2000b) found that Mrk~421 showed a correlation
of $E_{\rm peak} \propto F^{0.55\pm0.05}$, and Tavecchio \etal
(2001) reported that Mrk~501 showed a relation of $E_{\rm peak}
\propto F^{\sim 2}$. The significant difference of the slopes was
interpreted by Tavecchio \etal (2001) with the dependence of the slope
of $E_{\rm peak}$--$F$ relation on the position of the peak, being
steeper when the peak is at higher energies. The reason for the change
of the slope is that the flux is evaluated in a limited energy
band. When $E_{\rm peak}$ is located at the lower energy boundary of
the studied energy band, a small shift of $E_{\rm peak}$ can cause a
large change of the flux, and hence the slope of the $E_{\rm
peak}$--$F$ relation will be smaller than 1. Once $E_{\rm peak}$ moves
toward the upper limit of the studied energy band, the increase of the
flux is less rapid, and hence the $E_{\rm peak}$--$F$ relation will be
steeper. This phenomenology has also been found in Mrk~421 itself: the
2000 \sax\ data (in a very high state; Fossati \etal in 
preparation) may show steeper $E_{\rm peak}$--$F$ relation than that
found in the 1997 and 1998 \sax\ data (Fossati \etal 2000b).

\subsubsection{Spectral Variability} \label{sec:disc:summary:spec}

Our analysis has revealed that \pks\ showed complex X-ray spectral
variability, with different variability modes detected. Our results
suggest the following basic facts: (1) on \ts\ of hours (for each
observation), there is no correlation of
the spectral slopes and the fluxes (Figure~\ref{fig:spec:index}a). If
the rising and decaying phase of each flare follows different tracks
on the $\alpha$--$F$ plane, there will be large scatter
for the correlation between them. Moreover, there might be no such
correlation over long \ts\ (from epoch to epoch). This phenomenology
was also found in Mrk~421 with the XMM-Newton observations (Sembay
\etal 2002); (2) It seems that spectral indices in the soft and hard
X-rays do not correlate with each other 
(Figure~\ref{fig:spec:index}b). This could be
interpreted if there were time delay of the spectral changes at different 
energies; (3) our analysis indicates that the spectral curvature
(from soft to hard X-rays) of the 1997 data is larger than the curvature 
of the 1996 and 1999 spectra. This can be explained if $E_{\rm peak}$ of
the former is located in the soft X-rays, while $E_{\rm peak}$ of the 
latter may move down to the UV band, from which smaller spectral
curvature is expected in the X-ray band. Note, however, that Sembay
\etal (1993) found tight correlation between the spectral index in
the 3.8--17.9~keV and in the 1.7--3.8~keV ranges with GINGA 
observations, and the higher energy index is systematically steeper by
$\Delta\alpha \sim$~0.2, indicating that the degree of curvature of
the spectrum is constant. The two behaviors are not in contradiction,
because the results of Sembay \etal do not cover the soft energy
part of spectrum.

We have entered into details to study the $\alpha$--$F$ relation of each
flare of \pks\ (Figures~\ref{fig:hdnloop} and \ref{fig:spec:loop}). With
this technique, Sembay \etal (1993) found both the soft and hard lag
behaviors in this source. Kataoka \etal (2000) reported a clockwise loop
with ASCA data. We further found more complicated evidence that two flares
track opposite directions in the soft and hard bands, indicating changes
of the sign of the time lags from the soft to hard energy band. 

On the $\alpha$--$F$ plane, Mrk~421 showed an anticlockwise loop during a
large flare detected by \sax , indicating a hard lag (Fossati \etal
2000b; Zhang 2002). 

\subsection{Implications for the Dynamics and the Structure of the 
Jet}\label{sec:disc:implication}

Due to rapid cooling, relativistic electrons responsible for the X-ray
emission of the TeV blazars must be repeatedly accelerated 
(injected) in order to account for successive flares. It is now
believed that relativistic electrons are most likely accelerated at
shock fronts taking place in the jets. Therefore, the rich
phenomenology of the X-ray variability of \pks\ emerged from this work ,
in particular, characteristic \ts\ (if any; we identify it as the half
duration of the flare), time lags, peak energies of synchrotron component,
and spectral variability, can impose important clues on the dynamics and
the structure of the jet and its central engine.

Two kinds of shocks have been proposed: internal shocks versus external
shocks. The basic difference between them is the site and the way of shock
formation in blazar jets. Both have been extensively applied to gamma-ray
burst (GRB) models.

In the scenario of external shock, the shocks are formed when the
relativistic plasma ejected from the central engine sweeps up
material from the surrounding medium, where the jet plasma decelerates and
its bulk kinetic energy is converted into non-thermal particle energy
(e.g., Dermer \& Chiang 1998).   

The idea of internal shock was first proposed to explain some features
of the optical jets in M87 (Rees 1978). It was applied to GRB
modelling and reconsidered in blazar jets (e.g., Ghisellini 
2000; Spada \etal 2001) inspired by the close similarities between
them. The key idea of this scenario is to assume that the central engine
of a blazar is working in an intermittent rather than in a continuous way
to expel discrete shells of plasma with slightly varying velocities. The
shock will be formed due to collision when a later faster shell
catches up an earlier slower one. The dissipation of bulk kinetic
energy carried by the shells during the shock is used to accelerate
particles and generate the magnetic field, from which the radiation
is produced by synchrotron and inverse Compton mechanisms.      

The observed large amplitude and rapid variability of blazars,
particularly the recurrent flares, must be accounted for by a non
steady-state pulse (jet) model. The internal shock model can be thought to
be the simplest way, in which a series of shell-shell collisions can
naturally produce successive flares. In this scenario there should
be a ``typical'' \ts\ (at least for the first collisions) related to
the initial time interval between the two colliding shells and their 
widths. Inhomogeneities within the shells and smaller scale
instabilities can cause variability on shorter \tss\ to account for
small amplitude flickers superimposed on the global trend of the
flare. Extremely rapid variability could be due to shocks where the
width of the shock is smaller than the width of the jet. There are
still other points in favor of the internal shock model (see, e.g.,
Ghisellini 2001).

Relativistic electrons in the jets of the TeV sources are thought to
be accelerated to the maximum energies. The first collisions of the 
shells have the most efficient energy conversion. Therefore, for 
simplicity, it is reasonable to assume that an X-ray flare of the TeV
sources is produced during the first collision of two shells which have 
never collided with other shells before, while the more realistic
situation is that the central engine would produce a series of shells
involving multiple collisions to produce multiple flares.  

We assume that the two relativistic shells have bulk Lorentz factor
$\Gamma_1$ and $\Gamma_2$ ($\Gamma_2 > \Gamma_1$), and that the initial
time interval between them at their departure from the central source is
$t_0$. The shell width is initially of the same order of the initial
shell-shell separation. The expected distance where these two shells
collide is
\begin{equation}
R \simeq \frac{2a^2}{a^2 - 1} \Gamma^2 c t_0 \,,
\end{equation}
where $a=\Gamma_2 / \Gamma_1 > 1$ is the ratio of bulk Lorentz
factor of the two shells, and $\Gamma$ is the bulk Lorentz factor
of the merged shell which is usually identified as the emitting region 
(blob). 

The emitting region is assumed to be a ``one-zone'' homogeneous
spherical blob with radius $r/2$ tangled with magnetic field $B$,
moving at relativistic velocity $\beta c$ at a small angle $\theta$
with respect to the line of sight. The bulk Lorentz factor of the blob
is described as $\Gamma = (1-\beta^2)^{-1/2}$, and relativistic 
effects are described by the Doppler factor
$\delta=\Gamma^{-1}(1-\beta \cos\theta)^{-1}$. 
If $\Gamma \gg 1$, the open angle of the jet, $\theta$, is 
approximately equal to $1/\Gamma \simeq 1/\delta$.
The blob is filled with ``cold'' electrons which are
being accelerated and, in turn, suffering cooling to produce the
emission we are observing.

If one assumes that $t_{\rm crs}$ dominates the system (this condition
is thought to be valid in the X-rays of the TeV sources), $\tau_{\rm
crs}$ can be observationally identified as $\tau_{\rm max}$ inferred
from the SF. Therefore, $\tau_{\rm max}$ gives a constraint on the
size of the emitting region
\begin{equation}
r \simeq \Gamma c\tau_{\rm max} 
\label{eq:esize}
\end{equation}
If $\Gamma \sim 10$ is assumed, the inferred $\tau_{\rm max}$ 
(Table~\ref{tab:sf}) suggests that the size of the emitting blob is
$\sim 3\times 10^{15}$~cm and $\sim 2\times 10^{16}$~cm for the 1997 and
1996/1999 observations, respectively. 
With the jet geometry, this in turn constrains the distance (from the
central engine) where the acceleration and emission takes place
\begin{equation}
R_{\rm e} = \frac {r}{{\rm sin} \theta} \simeq \frac{r}{\theta} \simeq
          r\Gamma \simeq \Gamma^2 c \tau_{\rm max} \,. 
\label{eq:edistance}
\end{equation}
It is reasonable to assume that the emission takes place just after
the collision, i.e., $R_{\rm e} \simeq R$, then we have
\begin{equation}
\tau_{\rm max} \simeq \frac{2a^2}{a^2 - 1} t_0 \,.
\end{equation}
Thus $\tau_{\rm max}$ is a measurement of the initial time interval of
the two shells ejected from the central engine if $a$ is not too small
which is also constrained by the efficiency of the energy
dissipation. In general, $a \simeq 2$, then $\tau_{\rm max}$ is
between two to three times $t_0$. 

The minimum value of $t_0$ is constrained by light crossing
time across the central source. For a black hole of $M = 10^9 M_\odot$,
this timescale is $\sim 10^4$~s. Therefore the initial time
intervals of the shell pairs to produce the X-ray flares in 1997 is
comparable to the minimum timescale constrained by the central
engine. The flare durations of 1997 may represent the shortest
timescales which can be ever detected in \pks .  

The acceleration mechanism remains open. In this work the widely
discussed diffusive shock acceleration (e.g., Drury 1983; Blandford \&
Eichler 1987) is assumed to be the mechanism responsible for electron
acceleration in blazar jets.
In the comoving frame, the diffusive shock acceleration timescale,
$t'_{\rm acc}(\gamma)$, of a relativistic electron with energy $\gamma
m_{\rm e}c^2$ , is approximated (e.g., Kusunose, Takahara, \&
Li 2000) as
\begin{equation}
t'_{\rm acc}(\gamma) = \frac{20 \lambda(\gamma) c}{3 u_s^2} 
\sim 3.79 \times 10^{-7} \xi B^{-1} \gamma \quad {\rm s} \,,
\label{eq:tacc}
\end{equation}
where $\gamma$ is the Lorentz factor of the electron, $m_{\rm e}$
electron mass, $c$ light speed, $u_s \approx c$ is the speed of
relativistic shock, and
$\lambda(\gamma) = \gamma m_{\rm e} c^2 \xi / (e B)$ is the mean free path 
assumed to be proportional to the electron Larmor radius, and $\xi$ is
described as the acceleration parameter. $B$ is the magnetic field in
Gauss.

In the case of TeV sources, X-ray emission is primarily due to
\sy\ radiation. In the comoving frame, the corresponding \sy\ cooling
\ts\ of a relativistic electron with energy $\gamma m_{\rm e} c^2$ ,
$t'_{\rm cool}(\gamma)$, is (e.g., Rybicki \& Lightman 1979)
\beq 
t'_{\rm cool}(\gamma) = \frac{6\pi m_{\rm e} c}{\sigma_T}
	B^{-2}\gamma^{-1}
	 \sim 7.74\times 10^8 B^{-2}\gamma^{-1} \quad {\rm s} \,.
\eeq

One can see from the above two equations that $t'_{\rm acc}$ and
$t'_{\rm cool}$ depend on electron energy, in the sense that
lower energy electrons can be accelerated to the radiative window in
shorter time but cool during a longer time. It is convenient to express 
$t'_{\rm acc}(\gamma)$ and $t'_{\rm cool}(\gamma)$ in terms of the
observed photon energy because the typical \sy\ emission frequency, 
averaged over pitch angles, of an electron with energy $\gamma m_{\rm
e}c^2$ peaks at photon energy, $\nu \sim 3.73 \times 10^6 B\gamma^2$
Hz. In the observer's frame, we have 
\beq
t_{\rm acc}(E) = 9.65 \times 10^{-2} (1+z)^{3/2} \xi 
	B^{-3/2} \delta^{-3/2} E^{1/2} \quad {\rm s} \,,
\label{eq:tacckev}
\eeq
\begin{equation} 
t_{\rm cool}(E) = 3.04 \times 10^{3} (1+z)^{1/2} B^{-3/2}
	\delta^{-1/2} E^{-1/2}  \quad {\rm s} \,,
\label{eq:tcoolsykev}
\end{equation}
where $E$ is the observed photon energy in unit of keV for the
convenience of X-ray studies.  
It is obvious that $t_{\rm cool}$ and $t_{\rm acc}$ have 
the ``same'' degree of dependence on the photon energies (square
root of the energy) but in the \textit{opposite} way, i.e., the lower
energy photons have shorter accelerating but longer cooling \tss\ than
the higher energy photons do.
It is then reasonable to assume that the soft and hard lag of the
response of the low energy X-ray variations compared with those of the
high energy X-ray ones reflect the difference of $t_{\rm cool}$ and
$t_{\rm acc}$ of relativistic electrons responsible for the emission in
the studied energy bands, respectively. If $t_{\rm acc} \sim t_{\rm
cool}$, the observed hard lag is expected to be $\tau_{\rm
hard} = t_{\rm acc}(E_{\rm h}) - t_{\rm acc}(E_{\rm l})$, from which
physical parameters of the emitting blob can be constrained 
\beq
B\delta \xi^{-2/3} = 0.21 \times (1+z) E_{\rm h}^{1/3} 
    \left [ \frac{1 - (E_{\rm l}/E_{\rm h})^{1/2}}
                 {\tau_{\rm hard}} \right ]^{2/3} \quad {\rm Gauss} \,,
\eeq
where $E_{\rm l}$ and $E_{\rm h}$ refer to the low and high energy
band in unit of keV, respectively. Similarly, if $t_{\rm acc} \ll 
t_{\rm cool}$, the observed soft lag is expected to be $\tau_{\rm soft}
= t_{\rm cool}(E_{\rm l}) - t_{\rm cool}(E_{\rm h})$, from which
physical parameters of the emitting blob can be constrained 
\beq
B\delta^{1/3} =209.91 \times \left (\frac{1+z}{E_{\rm l}}\right
)^{1/3}
	\left [\frac{1 - (E_{\rm l}/E_{\rm h})^{1/2}}
        {\tau_{\rm soft}} \right ]^{2/3} \quad {\rm Gauss}  \,.
\label{eq:softlag}
\eeq
Note that, in practice, $E_{\rm l}$ and $E_{\rm h}$ are taken as the
logarithmic mean energies in the corresponding energy bands used in the
cross-correlation analysis taking into account power-law decrease of
the X-ray flux with increasing energy in the TeV sources.
It is interesting to note that \tacc\ and \tcool\ have the same
dependence on the magnetic field $B$, the ratio of \tacc\ to \tcool\
is thus independence of $B$
\beq
t_{\rm acc}(E)/t_{\rm cool}(E) 
  = 3.17 \times 10^{-5} (1+z) \xi \delta^{-1} E
  = 0.32 \times (1+z) \xi_5 \delta_1^{-1} E \,,
\label{eq:tratio}
\eeq
where $\xi_5$ and $\delta_1$ are in unit of $10^5$ and $10$, 
respectively. 

Since $\delta$ is thought to be of the order of $\sim$~10--25, one can see
that $\xi$ is the key parameter to modulate the observed behavior of the
variability, i.e., the observed inter-band relationship of a flare
depends on not only the value of $t_{\rm acc}/t_{\rm cool}$ but also
its energy dependence. The value of $\xi$ is poorly known, but changes 
of $\xi$ provide clues on changes of parameters of shock formation and 
acceleration. There are three different behaviors expected for an
observed energy range:
(1) if $t_{\rm acc} \ll t_{\rm cool}$ across the studied energy band, 
which corresponds to instantaneous injection of radiating electrons, 
the cooling process dominates the system, and information on the
acceleration is suppressed by cooling. Then information about the
emission propagates from higher to lower energy, higher energy
photons will lead lower energy ones, and clockwise loop of spectral   
index against the flux will be observed, this corresponds to the soft  
lag pattern (\S\ref{sec:timing:lag} and \S\ref{sec:spec:lag}); 
(2) in contrast, if $t_{\rm acc}$ is comparable to $t_{\rm cool}$ 
under the considered energy range, the acceleration process dominates
the system, then information about the emission propagates from lower
to higher energy, higher energy photons will lag lower energy ones, and 
anticlockwise loop of spectral index against flux will be observed,
this corresponds to the hard lag behavior. Therefore, information about
particle acceleration is observable only if the source is observed at
energies closer to the maximum (characteristic) synchrotron radiating
energy, $E_{\rm max}$, where $t_{\rm acc} = t_{\rm cool}$ (Kirk, 
Rieger, \& Mastichiadis 1998), indicating a relatively low rate of 
acceleration. The value of $E_{\rm max}$ emitted by the maximum energy
of electrons ($\gamma_{\rm max}$) which can be accelerated is 
determined by the balance between acceleration and cooling (e.g., Kirk, 
Rieger, \& Mastichiadis 1998);  
(3) the most interesting behavior may occur if $E_{\rm max}$ where
$t_{\rm cool} = t_{\rm acc}$, is inside the high energy end of
the considered energy band. $E_{\rm max}$ ($= 3.7\times 10^6 \delta B
\gamma_{\rm max}^2$~Hz) can be estimated to be a few keV with the 
general parameter ranges ($\gamma_{\rm max} \sim 10^5$, $B \sim 
0.1$~Gauss, and $\delta \sim 10$) of the emitting region inferred from 
the SEDs of the source (e.g., Kino, Takahara, \& Kusunose 2002). The
observed time lag will gradually evolve from soft lag to hard lag at
an energy ($E_{\rm app}$) where $t_{\rm acc}$ approaches $t_{\rm
cool}$, say $t_{\rm acc} \sim 0.9t_{\rm 
cool}$. Equation~\ref{eq:tratio} implies that $E_{\rm app}$
is $\sim  2.5$~keV if $\xi_5=1$, $\delta_1=1$, and $z=0.116$ are
assumed. This is an alternative and the simplest explanation for the
behavior of the 1997 \#2 flare which shows soft and hard lag in the
soft and hard energy band, respectively (see 
Figure~\ref{fig:spec:loop}).
This is the first time that the interplay between acceleration and cooling
time is simultaneously observed through the behavior that soft lags
evolve to hard lags from low to high energies. $E_{\rm max}$
determines the position of the switching point from the soft to
hard lag. This phenomenology demonstrates the importance of $t_{\rm
acc}$ being energy-dependent. However, the current time-dependent
models involving acceleration have not yet considered this effect. For
the convenience of numerical simulations, $t_{\rm acc}$ is assumed to
be energy-independent (e.g., Kirk, Rieger, \& Mastichiadis
1998; Kusunose, Takahara, \& Li 2000). 

However, the opposite behavior observed in 1996 \#2 flare can not be
accounted for by this interpretation, but could be explained with a two
components model, of which one component (lower energy) dominates at soft
energies and one (higher energy ) dominates at hard energies. If the
former comes first, hard lag in the soft energies and soft lag in the hard
energies are expected if the latter is dominated by soft lag.
It seems that this idea is consistent with the fact that $\sigma_{\rm
rms}^2$ in the 0.1--0.5~keV band may be larger than that in the 0.5--2~keV
band (see \S\ref{sec:timing:rms} and Table~\ref{tab:rms}) as long as
the synchrotron peak energy of the first (lower energy) component is
smaller than that of the second (higher energy) component. 

The internal energy dissipated from the bulk kinetic energy of the
two colliding shells is shared among protons, electrons and magnetic
field. In the case of magnetic field generated only from the energy
dissipation in each collision, and neglecting any seed magnetic field 
which can be amplified by shock compression, the value of magnetic
field $B$ strongly anti-correlates with collision distance $R$
approximately as (Spada \etal 2001) 
\begin{equation}
B \propto R^{-3/2} \,.
\end{equation}
This relation should be the case for the TeV sources because the X-ray
flares are thought to be produced from the first collisions of the
shells which have never collided before (i.e., shells without internal 
energy before collisions).

Equation [\ref{eq:softlag}] gives the following relation
\begin{equation}
B\Gamma^{\frac{1}{3}} \propto \tau_{\rm lag}^{-2/3} \,.
\end{equation}
Then we have 
\begin{equation}
\tau_{\rm lag} \propto \Gamma^{-1/2} R^{9/4} \,.
\end{equation}
Substituting $R$ (i.e., $R_{\rm e}$) with equation
[\ref{eq:edistance}], we can get
\begin{equation}
\tau_{\rm lag} \propto \Gamma^{4} \tau_{\rm max}^{9/4} \,.
\end{equation}
One can see that the soft lag increases steeply with $\tau_{\rm max}$
since the change of $\Gamma$ is very small. Therefore, the changes
of soft lags from flare to flare inferred from the cross-correlation
analysis are in qualitative agreement with this prediction, which
suggests that smaller soft lags are related to the flares with shorter
duration.

In the simplest ``one-zone'' SSC model, the electron distribution is
described by a broken power-law with cutoff energy, $\gamma_{\rm max}$,
either on the basis of the pure phenomenology or on the basis of the
kinetic equation. $\gamma_{\rm max}$ is determined by a detailed
balance between acceleration and cooling. The break energy,
$\gamma_{\rm break}$, is observationally recognized as the $E_{\rm
peak}$, through the relation: $E_{\rm peak} \propto \delta B
\gamma_{\rm break}^2$. Therefore, an increase of $E_{\rm peak}$ from a
low state (similar to that observed in 1996 and 1999) to a high state
(like that of 1997) can be ascribed to an increase of $\gamma_{\rm
break}$, $B$ and $\delta$, of which $\gamma_{\rm break}$ is the key
parameter regulating the observed variability. 

The simulation by Spada \etal (2001) showed that $E_{\rm peak}$ of
synchrotron radiation pulses decreases with the increasing collision
distance along the jet (see their figure~5). The values of $E_{\rm peak}$ 
at each distance have been obtained by averaging over all of the collisions
occurring at that distance. This is in agreement with our findings
that $E_{\rm peak}$ in 1997 is higher because of smaller collision
distance (i.e., smaller duration of the flares). Note that the parameters 
of their simulations are adopted typically for FSRQs like 3C~279, which
has very low peak energy with respect to the TeV sources.  

The variations of $\gamma_{\rm break}$ ($E_{\rm peak}$) from
flare to flare can be explained by the variations in the structure of
the shock (the key parameter is $t_{\rm acc}$) which may depend on the
colliding distance of the two shells. The initial separation time
$t_0$ of the two shells is the most critical parameter which determines 
the distance where the collision happens. 

Finally, it is necessary to point out that our interpretation for
the complex variability patterns is based on the simplest scenario,
i.e., the interplay between acceleration and cooling processes. It is
worth noting other possible
effects that may influence the observed flux behavior. The X-ray
emission in PKS~2155--304 (and other TeV sources) originates in the
fast-cooling regime of the relativistic electron energy distribution,
which causes the fast-changing of the SED. Therefore the light curve
(flare) at a fixed-frequency corresponds to the different parts of the
SED of the source (as in the case of GRBs, e.g., Sari et
al. 1998). This uncertainty may change the relation of light curves at
different energies and the evolution of spectral indices with the
flux. Furthermore, the flare's profiles depend on such parameters as
shock/dissipation lifetime, electron injection time profile,
adiabaticity, and half-opening angle of the jet (Sikora \etal 2001). In
the context of the radiative hydromagnetic shocks, the interplay
between the cooling and the compression may also play an important
role. Granot \& K\"{o}nigl (2001) showed that, when synchrotron
radiation dominates the cooling, this effect becomes more pronounced on
account of the feedback effect between the field amplification and the
emission process: a strong magnetic field increases the emissivity,
which in turn induces a larger compression that further amplifies the
field. However, the determination of a specific mechanism responsible
for a flare pattern (e.g., energy-dependent time lags and local
spectral evolution) requires the comparison between the flare behavior
and the numerical simulation of the radiative shocks. We also note that 
the theoretical study of the radiative shock structure and evolution in
the jets of blazars has so far been little developed.


\section{Conclusions}\label{sec:conc}

\sax\ continuously monitored \pks\ for about two days in November of 1996,
1997 and 1999, respectively. We have studied and interpreted the joint
behavior of temporal and spectral evolution in the X-ray range. During
these observations, the source was in different states of brightness, and
showed a variety of variability behavior. This provides us with rich
information about this source, which in turn reveals some interesting
clues on the physical processes operating in jets. We summarize our main
results as follows:

\begin{enumerate}

\item The amplitude of variability is larger, and doubling \ts\ shorter
at higher energy. Both quantities do not correlate with the brightness
of the source, and complex behaviors are detected.

\item The PSDs (and the SFs) are characterized by a featureless red
noise steep power-law of $f^{-2\sim -3}$, indicating that the 
variability of the source can be ascribed to the shot noise.

\item The SFs show evidence for the presence of ``turn-over'' 
characteristic of the timescales of the repeated flares, i.e., the
average value of the half duration of the flares. More importantly,
our analysis revealed that this \ts\ changes over a factor of a few,
which could determine the overall behavior of the flares.

\item Detailed cross-correlation analyses revealed that the soft X-ray
photons lag the hard ones. Interestingly, the values of soft lags 
seem to correlate with the duration of the flares. 

\item Time-resolved spectral fits with a curved model suggested that the 
peak position of synchrotron emission moves to higher energy with 
increasing flux. 

\item Spectral changes are complicated without any clear correlations
of spectral slope versus flux and between spectral slopes at different
energies. 

\item The tracks between the spectral indices and the fluxes showed a
variety of modes. The changes of the sign of the time lags from the
soft to hard energy were found in two flares. The other flares show
only soft lags across the studied energy band. 

\end{enumerate}

As a possible interpretation to the observed variability of \pks , the
internal shock was discussed in line with the homogeneous synchrotron 
cooling mechanism. It turns out that our analyses are qualitatively
consistent with the predictions of such a scenario:

\begin{enumerate}

\item The (half) duration of the flare indicates the initial time
interval of the two shells ejected from the central engine to produce
the flare. This time interval determines the distance where collision
of the shells takes place, which in turn may fix the structure of
the shocks and physical parameters of the  emitting region. 

\item Shot noise mode of the variability indicates recurrent
flares. A series of shells should be intermittently ejected in order to
form successive flares.  

\item The changes of the durations, the soft lags and synchrotron peaks 
from flare to flare are fully consistent with predictions of the internal
shocks if they form at different distances. 

\item The discovery of the inter-band time lags switching from
soft to hard lag during the 1997 \#2 flare emphasizes the importance of
electron acceleration being energy-dependent.

\item The variability behavior of the 1996 \#2 flare suggests that the
flare may consist of more than one emission component with different
spectral energy distribution.

\end{enumerate}

In the scenario of internal shock, the initial conditions (mainly the time
interval) of the two shells which collide to produce the flare may 
determine at the first order of approximation, the main properties of the
observed flares in \pks .

\acknowledgments

Part of this work was conducted at SISSA during the Ph.D. project of 
Y.H.Z.. The anonymous referee is thanked for constructive comments.
We greatly thank the \sax\ Science Data Center (SDC) for providing the
standard event files archive. We acknowledge the Italian MUIR for
financial support.   

\appendix
\section{Excess Variance}\label{sec:app:rms}

The normalized excess variance, $\sigma^2_{\rm rms}$, is defined as
\beq 
\sigma^2_{\rm rms} = \frac{1}{N\bar{x}^2} \sum_{i=1}^{N} 
	\left [(x_i - \bar{x})^2 - \sigma^2_{i} \right ] \,,
\eeq
where $x_i$, with quoted error, $\sigma_{i}$, is the count rate, and
$\bar{x}$ is the unweighted arithmetic mean count rate over $N$ points
of a \lc . One can see that $\sigma^2_{\rm rms}$ is defined as
the difference between total (standard) variance, 
$\sigma^2_{\rm total}= \frac{1}{N} \sum_{i=1}^{N} (x_i -\bar{x})^2$,
and noise variance, $\sigma^2_{\rm noise}= \frac{1}{N}
\sum_{i=1}^{N} \sigma_i^2 $, which in turn is normalized by
$\bar{x}^2$ to compare $\sigma^2_{\rm rms}$
between different light curves. 
The error on $\sigma^{2}_{\rm rms}$, is estimated by 
$s_{\rm D}/(\bar{x}^2\sqrt{N})$ (Turner \etal 1999), where
\beq 
s_{\rm D}^{2}=\frac{1}{N-1}\sum_{i=1}^{N}\left \{ 
[(x_i - \bar{x})^2 - \sigma_{i}^{2}] - \sigma_{\rm rms}^2\bar{x}^2
\right \}^2 \,, 
\eeq
i.e., the variance of the quantity $(x_i - \bar{x})^2 - 
\sigma_{i}^{2}$. The fractional variability parameter, $F_{\rm var}$,
used in Zhang \etal (1999) is the square root of the excess
variance: $F_{\rm var} = (\sigma^{2}_{\rm rms})^{1/2}$.

Power spectral density analysis of \pks\ performed in
\S\ref{sec:timing:psd} have revealed larger amplitude variability at
lower frequencies than at higher frequencies. The PSD can been 
parameterized by a power-law form (see also Appendix~\ref{sec:app:psd}) 
\begin{equation}
P(f) = Cf^{-\alpha} \,,
\end{equation}
where $C$ is the normalized amplitude of the PSD, and $\alpha$ is the
slope of PSD. The excess variance, $\sigma_{\rm rms}^2$, can be
obtained by integrating the PSD 
\begin{eqnarray}
\sigma_{\rm rms}^2 & = & 2C\int_{f_{1}}^{f_{2}}P(f)df 
\label{eq:rmspsd} \\
& = & \frac{2C}{1-\alpha}(f_{2}^{1-\alpha}-f_{1}^{1-\alpha})\\
& = & \frac{2C}{\alpha -1}\left [T^{\alpha-1}- \left (\frac{\Delta 
t}{2}\right ) ^{\alpha-1} \right ]  
\end{eqnarray}
The $\sigma_{\rm rms}^2$ will thus depend on the exact range of
temporal frequencies, i.e., observing length ($T=1/f_1$) and binning
($\Delta t = 1/(2f_2)$) of the light curve. Different sampling pattern can
be corrected by normalizing variance to the same frequency range.

If $T \gg \Delta t$, we have
\begin{equation}
\sigma_{\rm rms}^2 \propto T^{\alpha-1}
\end{equation}
for $\alpha > 1$ and no changes of $\alpha$ across the frequency  
interval $[1/T,1/(2\Delta t)]$. 

The $\sigma_{\rm rms}^2$ of 1996 and 1997 is normalized to the
observing length of 1999, the longest one, and $\alpha \sim 2.5$ derived
in  \S\ref{sec:timing:psd} is applied. The amplified factor is $\sim$~1.2
and $\sim$~2.5 for 1996 and 1997, respectively. However,the real values
may decrease significantly if there is knee or break of PSD at frequency
larger than $1/T$, especially for 1997 as relative short characteristic
timescale is suggested by its SFs.

The second important effect on the observed variance is different binning.  
For the same length of observation, the variance from the smaller bin size
should be no smaller than those from larger bin size since larger binning
will tend to reduce the observed variance by integrating out high
frequency power. However, in fact, the situation is not
always like this. As seen from Table~\ref{tab:rms}, $\sigma_{\rm rms}^2$
derived from 600~s bin size are significantly smaller than those derived
from 5670~s bin in some cases. We consider this effect as being due to 
lower signal-noise ratios associated with 600~s bin, and particularly  
in some bins noise variance will be larger than the count rate variance,
which inevitably reduces source variance. Therefore, smaller binning
has a strong tendency to reduce the observed variance by integrating
negative high frequency power caused by insufficient signal-noise 
ratios. When the signal-noise ratio is high enough, this will not
happen. As seen from Table~\ref{tab:rms}, $\sigma_{\rm rms}^2$ derived
from 600~s bin size are indeed larger than those derived from 5670~s
bin size in the case of 0.5--2~keV and 2--4~keV band of 1997, while the
statistics is not significant. 

Quantifying the effects caused by binning, signal to noise ratio and
gaps together would be analytically impossible, and needs detailed
simulations. However, the observed excess variance estimated from light
curves binned over 5670~s will not be affected by these uncertainties, and
the only uncertainty is the observing length. This is because the 5670~s
binned light curves are evenly sampled, and signal-noise ratio is high. 
There are still uncertainties caused by low exposure time of the LECS, but
it may be not significant. 

Therefore, in \S\ref{sec:timing:rms} we use $\sigma_{\rm rms}^2$
derived from 5670~s binned light curves to discuss the energy and
intensity dependence of amplitude of variability.

\section{Power Spectral Density}\label{sec:app:psd}

The normalized power spectral density (NPSD) at frequency $f_k$ is
defined as
\begin{eqnarray} 
P(f_k) =\frac{[c^2(f_k)+s^2(f_k)-\sigma^2_{\rm noise}/N]T}{\bar{x}^2} \,,
 \\
c(f_k) =\frac{1}{N}\sum_{i=1}^{N} x_i \cos (2\pi f_k t_i) \,,
\\
s(f_k) =\frac{1}{N}\sum_{i=1}^{N} x_i \sin (2\pi f_k t_i) \,,
\\
\end{eqnarray}
where $x_i$ is the count rate at time $t_i=i\Delta t$ ($i=1, 2,...,N$,
$\Delta t$ is the binning size), $\bar{x}$ the unweighted arithmetic mean
count rate, and $T$ the observing length. $c(f_k)$ and $s(f_k)$ ($f_k =
k/T$ is the Fourier frequency, $k=1, 2, ..., N/2$) represent the finite
cosine and sine components of Fourier transform of a light curve. The
power due to the white noise, $\sigma_{\rm noise}^2$, is subtracted. With
this definition, the integration of the source power over the positive
frequencies yields half of the excess variance of the same light curve
(see Equation~[\ref{eq:rmspsd}]).

A ``typical'' PSD is characterized by a cutoff frequency $f_{\rm max}$
at high frequency, and a break frequency $f_{\rm min}$ at low
frequency (to avoid divergence of $\sigma_{\rm rms}^2$). The range
between $f_{\rm min}$ and $f_{\rm max}$ is linked by a power-law
curve, i.e., $P(f) \propto f^{-\alpha}$, and $\alpha$ depends on the
nature of the intrinsic variation of a source (e.g., flicker noise, shot
noise, etc.). 

\section{Structure Function}\label{sec:app:sf}

Simonetti, Cordes \& Heeschen (1985) firstly introduced the concept of 
structure function (SF) into the field of astronomy. The first order SF of
a time series $x(t)$ at a time scale ``$\tau$'' is defined as
\beq 
SF(\tau) = \frac{1}{n}\sum \left [x(t+\tau)-x(t) \right ]^2 \,.
\eeq
In fact, SF($\tau$), as a function of \ts\ $\tau$, is a
measurement of the mean squared flux differences, $x(t+\tau)-x(t)$, of
$N$ pairs with the same time separation $\tau$. For an unevenly
sampled $x(t)$, $n$ is the number of pairs over an
interval of $\tau - \Delta\tau/2 < \tau < \tau + \Delta\tau/2$ for a
specific binning pattern $\Delta\tau$ (equally or logarithmically
binned). The errors on SF($\tau$) are calculated by the standard
deviation of squared flux differences of $n$ pairs in each bin (but
see Simonetti, Cordes \& Heeschen 1985 and Cagnoni, Papadakis \&
Fruscione 2000 for the caveats of using such errors).

A ``typical'' SF($\tau$) is characterized by the minimum and maximum
correlation \tss , $\tau_{\rm min}$ and $\tau_{\rm max}$, and a power
law curve between them, i.e., $SF(\tau) \propto \tau^{\beta}$. Same as PSD,  
$\beta$ depends on the variability nature of a source.
SF is related with the standard
variance and auto-correlation function (ACF) of the \lc\ as
\beq
SF(\tau) = 2 \times \left ( \sigma_{\rm total}^2 - ACF(\tau) \right ) \,.
\eeq 
SF($\tau$) flattens below $\tau_{\rm min}$ and above $\tau_{\rm
max}$, approximating as $2\sigma_{\rm noise}^2$ and $2\sigma_{\rm 
total}^2$, respectively.  

SF gives information similar to PSD. There are simple correspondences
between the parameters derived from SF and PSD  
\begin{eqnarray}
f_{\rm min} = 1/\tau_{\rm max} \,, \\
f_{\rm max} = 1/\tau_{\rm min} \,, \\
\alpha = \beta + 1 \,.
\end{eqnarray}
However, this relation is valid only in the limit $T \rightarrow \infty$
and $\Delta t \rightarrow 0$, this condition definitely requires long
enough observing length and small enough temporal resolution with high
signal-noise ratio. In fact, since the PSD is strongly limited to a small
frequency range $[1/T,1/(2\Delta t]$, these relationships do not hold
any more.  


\clearpage
\begin{deluxetable}{ccccc}
\tabletypesize{\footnotesize}
\tablecolumns{5}
\tablewidth{0pc}
\tablecaption{Journal of Observations}
\tablehead{
\colhead{Observing Time} &\colhead{Duration} 
  &\multicolumn{2}{c}{Net Exposure(ks)} &\colhead{Archive \#} \\
\colhead{(UTC)} &\colhead{(hour)} &\colhead{LECS} &\colhead{MECS}
  &\colhead{}  
}
\startdata
1996/11/20 00:15:58--1996/11/22 13:30:06 &51.2
	&36.29 &106.9 &50016001 \\ 
1997/11/22 16:03:00--1997/11/24 01:35:12 &32.5
	&22.49 &59.49 &50160008 \\
1999/11/04 04:27:27--1999/11/06 16:52:12 &60.4
	&46.12 &104.0 &50880001 \\
\enddata
\label{tab:obs}
\end{deluxetable}

\clearpage
\begin{deluxetable}{rcccccc}
\tablecolumns{7}
\tabletypesize{\footnotesize}
\tablewidth{0pt}
\tablecaption{Excess variance and minimum doubling timescales of
	variability}
\tablehead{
\colhead{band (keV)} &\colhead{cts/s} 
 &\colhead{$\sigma^2_{\rm rms}(10^{-2})$\tablenotemark{a}} 
 &\colhead{$\sigma^2_{\rm rms}(10^{-2})$\tablenotemark{b}}  
 &\colhead{$\sigma^2_{\rm rms}(10^{-2})$\tablenotemark{c}}
 &\colhead{${\rm T_{2}}$(ks)\tablenotemark{a}} 
 &\colhead{${\rm T_{2}}$(ks)\tablenotemark{b}}
}
\startdata
\multicolumn{7}{c}{1996} \\
0.1--0.5&0.34 &1.52$\pm$0.31 &4.08$\pm$0.52 &4.90 &65.4$\pm$12.0
&93.7$\pm$18.6\\
0.5--2  &0.91 &1.35$\pm$0.20 &1.71$\pm$0.37 &2.05 &86.0$\pm$14.5
&104.1$\pm$15.6\\
2--4    &0.72 &1.48$\pm$0.14 &1.79$\pm$0.35 &2.15 &53.5$\pm$12.4
&77.8$\pm$11.1\\
4--10   &0.32 &1.73$\pm$0.20 &3.58$\pm$0.60 &4.30 &35.2$\pm$7.0
&59.7$\pm$10.4\\
\multicolumn{7}{c}{1997} \\
0.1--0.5&0.55 &3.28$\pm$0.54 &5.71$\pm$1.24 &14.28 &27.8$\pm$4.9
&31.0$\pm$4.4\\
0.5--2  &1.40 &6.55$\pm$1.19 &6.28$\pm$1.54 &15.70 &15.2$\pm$1.7
&15.9$\pm$1.5\\
2--4    &0.80 &7.29$\pm$1.19 &6.89$\pm$1.80 &17.23 &9.16$\pm$1.1
&17.5$\pm$2.1\\
4--10   &0.32 &8.34$\pm$1.39 &11.1$\pm$2.18 &27.75 &5.87$\pm$1.0
&14.8$\pm$2.4\\
\multicolumn{7}{c}{1999} \\
0.1--0.5&0.33 &1.60$\pm$0.40 &3.81$\pm$0.62 &... &36.9$\pm$6.5
&107.4$\pm$15.9\\
0.5--2  &0.54 &2.19$\pm$0.31 &2.91$\pm$0.69 &... &33.3$\pm$6.1
&35.2$\pm$5.3\\
2--4    &0.27 &2.75$\pm$0.28 &5.25$\pm$0.80 &... &36.3$\pm$6.7
&51.2$\pm$9.2\\
4--10   &0.11 &4.01$\pm$0.42 &15.2$\pm$1.47 &... &19.4$\pm$3.8
&31.9$\pm$5.2\\
\enddata
\tablenotetext{a}{600~s bin} 
\tablenotetext{b}{5670~s bin}
\tablenotetext{c}{corrected $\sigma^2_{\rm rms}$ of 1996 and 1997 for
 5670~s bin}
\label{tab:rms}
\end{deluxetable}

\clearpage
\begin{deluxetable}{cccc}
\tablecolumns{4}
\tabletypesize{\footnotesize}
\tablewidth{0pt}
\tablecaption{The best fit parameters of NPSD\tablenotemark{a}}
\tablehead{
\colhead{Observation} &\colhead{N (at $10^{-4}$ Hz)\tablenotemark{b}}
 &\colhead{$\alpha$\tablenotemark{b}}
 &\colhead{$\chi^2$(dof)} \\
\colhead{} &\colhead{(Hz$^{-1}$)} &\colhead{} &\colhead{}
}
\startdata
1996 & 0.38$\pm$ 2.13 &2.94$\pm$0.52 &5.14(8) \\
1997 &20.83$\pm$24.72 &1.93$\pm$0.15 &5.99(8) \\
1999 & 0.35$\pm$ 3.01 &3.10$\pm$0.76 &8.64(9) \\
\enddata
\tablenotetext{a}{The power-law model is assumed to be $P(f) = N
	(f/10^{-4})^{-\alpha}$}
\tablenotetext{b}{The errors are 1~$\sigma$}
\label{tab:psd}
\end{deluxetable}

\clearpage
\begin{deluxetable}{cccccc}
\tablecolumns{5}
\tabletypesize{\footnotesize}
\tablewidth{0pt}
\tablecaption{The best fit parameters of structrue funtions}
\tablehead{
\colhead{Observation} &\multicolumn{2}{c}{$\beta$\tablenotemark{a}} 
	   & &\multicolumn{2}{c}{$\tau_{\rm max}$
		(10$^{4}$s)\tablenotemark{a}}  \\
\cline{2-3} \cline{5-6}
\colhead{} &\colhead{0.1--2~keV} &\colhead{2--10~keV}
	& &\colhead{0.1--2~keV}
 	&\colhead{2--10 keV}
}
\startdata
1996 &$1.21^{+0.38}_{-0.38}$ &$1.37^{+0.21}_{-0.24}$ &
     &$9.81^{+0.27}_{-0.63}$ &$8.20^{+0.35}_{-0.34}$  \\
1997 &1.49$\pm$0.30$^b$ &$1.55^{+0.21}_{-0.21}$ &
     &1.10$\pm$0.35$^b$ &$1.28^{+0.38}_{-0.19}$ \\     
1999 &$0.99^{+0.12}_{-0.11}$ &$1.33^{+0.08}_{-0.08}$ &
     &$6.55^{+0.27}_{-0.72}$ &$6.17^{+0.33}_{-0.35}$ \\ 
\enddata
\tablenotetext{a}{The errors are $90\%$ confidence level for one parameter
	of interest}
\tablenotetext{b}{1~$\sigma$ errors due to the SF insensitive to derive the
	$90\%$ confidence level}
\label{tab:sf}
\end{deluxetable}

\clearpage
\begin{deluxetable}{ccc}
\tablecolumns{3}
\tabletypesize{\footnotesize}
\tablewidth{0pt}
\tablecaption{Time Lags (ks) between the 0.1--1.5~keV and 3.5--10~keV
	bands}
\tablehead{
\colhead{Flare} &\colhead{DCF\tablenotemark{a}} 
	&\colhead{MMD\tablenotemark{a}}
}
\startdata
1996 \#2 &6.16$^{+1.38}_{-1.33}$ &5.09$^{+4.98}_{-3.95}$ \\
1997 \#1 &0.41$^{+0.40}_{-0.40}$ &2.07$^{+2.64}_{-1.02}$ \\
1997 \#2 &0.45$^{+0.39}_{-0.29}$ &0.38$^{+0.26}_{-0.46}$ \\
1997 \#3 &1.68$^{+0.52}_{-0.39}$ &.... \\
1999 \#1 &3.50$^{+1.06}_{-1.05}$ &3.03$^{+1.74}_{-1.77}$ \\
1999 \#2 &4.72$^{+2.62}_{-2.61}$ &6.93$^{+4.25}_{-4.13}$ \\
\enddata
\tablenotetext{a}{The errors are $90\%$ confidence level for one parameter
	of interest}
\label{tab:lag}
\end{deluxetable}


\clearpage

\end{document}